\documentclass[a4paper,11pt]{article}
\pdfoutput=1
\usepackage{jheppub}
\usepackage[T1]{fontenc}
\usepackage{tensor}
\usepackage[mathscr]{euscript}
\usepackage{float}
\usepackage{tikz}
\usepackage{mathtools}
\usepackage{soul}
\usepackage{amsthm}
\usepackage[english]{babel}
\usepackage{tikz}
\usetikzlibrary{positioning,calc}
\usetikzlibrary{arrows.meta,decorations.markings}

\newcommand{\C}{\mathbb{C}}
\newcommand{\R}{\mathbb{R}}
\renewcommand{\P}{\mathbb{P}}

\newcommand{\Z}{\mathbb{Z}}
\newcommand{\p}{\partial}
\newcommand{\rd}{\mathrm{d}}
\newcommand{\pbar}{\bar{\p}}
\newcommand{\zbar}{\bar{z}}
\newcommand{\sx}{\mathscr{X}}

\newcommand{\myx}{X}

\theoremstyle{definition}

\theoremstyle{proposition}

\newif\ifdetails
\detailstrue

\title{Spectral flow in AdS$\mathbf{_3}$ revisited}

\author[a]{N. M. McStay,}
\author[b]{R. A. Reid-Edwards,}
\author[c]{S. Seet}

\affiliation[a]{International Centre for Theoretical Sciences-TIFR,\\
Shivakote, Hesaraghatta Hobli, Bengaluru North 560089, India.}
\affiliation[b]{Department of Applied Mathematics \& Theoretical Physics, University of Cambridge,\\
Wilberforce Road, Cambridge CB3 0WA, United Kingdom}
\affiliation[c]{School of Mathematics and Maxwell Institute for Mathematical Sciences, University of Edinburgh}

\emailAdd{nathan.mcstay@icts.res.in}
\emailAdd{rar31@cam.ac.uk}
\emailAdd{sseet@ed.ac.uk}

\abstract{We revisit an old idea that spectral flow in WZW models can be associated with certain $G$-bundles over a worldsheet. This allows us to introduce a novel method for computing correlators of spectrally-flowed vertex operators in WZW models with group $G$ and level $k$, by systematically computing the contribution from a network of line defects. Specialising to the case of $G=SL(2, \mathbb{R})$, as is relevant to AdS$_3$ string theory, we demonstrate that generic $n$-point functions of spectrally-flowed vertex operators at tree level can be rewritten in terms of an $(n+1)$-point function of $n$ unflowed vertex operators and one degenerate insertion. This is to be interpreted as an unflowed $n$-point function in an instanton background. Moreover, we conjecture that this $(n+1)$-point function can be recast as an $n$-point function of unflowed operators and we provide evidence for this. We demonstrate that the $(n+1)$-point function satisfies the same constraints from global spacetime conformal invariance as does an unflowed $n$-point function for a specific choice of spacetime coordinates that we determine. We verify that our results agree with bootstrap results in the literature. This provides the next step towards fully solving the $SL(2,\R)$ WZW model.
}

\begin{document} 
\maketitle
\flushbottom

\pagenumbering{arabic}

\section{Introduction}

Two-dimensional conformal field theory (CFT) lies at the heart of the worldsheet formulation of string theory and yet, formulating and quantising worldsheet sigma models on curved backgrounds is generically a very hard problem. One special class of 2D CFTs which we know how to quantise are WZW models \cite{Wess:1971yu,Witten:1983tw,Witten:1983ar,Knizhnik:1984nr}.\footnote{See \cite{DiFrancesco:1997nk} for a textbook review.} The fundamental fields of these non-linear sigma models describe an embedding into a Lie group, $G$, so are relevant to string theory on group manifolds. The CFT carries an infinite-dimensional affine Kac-Moody algebra at level $k$, $\mathfrak{g}_k$, as a symmetry of the theory, which provides a means of quantisation.

One prominent example is the $\widehat{SL}(2,\R)$ WZW model, since $\widehat{SL}(2,\R)$ describes a copy of real Lorentzian AdS$_3$. The hat refers to the universal cover of $SL(2,\R)$ and in what follows, we will always refer to the universal cover, so we will drop the hat to avoid cluttering notation. This example provides a toy model for studying quantum gravity and holography in low dimensions. It is relevant both in bosonic and superstring theory \cite{Maldacena:1998bw,Evans:1998qu,Giveon:1998ns,deBoer:1998gyt,Kutasov:1999xu,Maldacena:2000hw,Maldacena:2000kv,Maldacena:2001km,Giveon:2001up}, since superstrings propagating on AdS$_3$ supported by pure NS-NS flux are described by the supersymmetric analogue of the $SL(2,\R)$ WZW model.

Many important questions in string theory can be probed through the $SL(2,\R)$ WZW model, since we are not restricted to working perturbatively in $\alpha'$, but have an (in principle) exactly solvable worldsheet theory. The level $k$ of the Kac-Moody current algebra is related to $\alpha'$ via\footnote{Here we refer to the value of $k$ in the bosonic, rather than the supersymmetric, theory since supersymmetry will play no role in this work. In superstring theory, this formula is modified to $k_s = k_b-2 = \frac{R^2_{AdS}}{\alpha'}$, where $k_s$ is the level of the supersymmetric $SL(2,\R)$ WZW model.}
\begin{equation}
    k = \frac{R_{AdS}^2}{\alpha'}\,,
\end{equation}
and one can study the model for any real $k>2$, including in highly stringy regimes \cite{Giveon:2005mi}. One can also ask questions pertaining to black hole physics, since close cousins of the $SL(2,\R)$ WZW model include the cigar geometry $SL(2,\R)/U(1)$, given by a gauged WZW model, which describes a 2D Euclidean black hole \cite{Witten:1991yr} and the 3D BTZ black hole is a quotient of global AdS$_3$ \cite{Banados:1992wn}. The $SL(2,\R)$ WZW model has, moreover, played a vital role in studies of AdS$_3$/CFT$_2$ \cite{Maldacena:1997re}. Enormous progress has been made over the last ten years by applying worldsheet methods to study the duality between pure NS-NS string theories on AdS$_3$ and the conjectured dual 2D CFTs that are related to the symmetric orbifold \cite{Giribet:2018ada,Gaberdiel:2018rqv,Eberhardt:2018ouy,Eberhardt:2019ywk,Eberhardt:2020akk,Gaberdiel:2020ycd,Balthazar:2021xeh,Eberhardt:2021vsx,Dei:2021xgh,Dei:2021yom,Dei:2022pkr,Demulder:2023bux,Eberhardt:2025sbi}.

The $SL(2,\R)$ WZW model is also interesting inherently from a 2D CFT perspective since it is closely related, and yet seemingly richer, than Liouville theory, which has been solved \cite{Dorn:1994xn,Zamolodchikov:1995aa,Teschner:2001rv}. We mean this in the sense that the spectrum and three- and four-point functions on the sphere have been calculated and crossing symmetry has been proven (see \cite{Ribault:2014hia} for a review). There are relatively few examples of interacting, irrational CFTs that have been solved (Liouville theory being one of them) and so one could argue that the $SL(2,\R)$ WZW model is a natural next step in complexity to aim for as a solvable 2D CFT. As with Liouville theory, the $SL(2,\R)$ WZW model contains continuous representations but it additionally contains discrete and so-called \emph{spectrally-flowed} representations, which are not highest weight with respect to the currents of the left- and right-moving affine Kac-Moody algebras. It is chiefly the existence of spectrally-flowed representations that increases the complexity of the $SL(2,\R)$ WZW model compared to other 2D CFTs that have been solved. With this in mind, we would like to revisit spectral flow in WZW models and present novel techniques for computing correlation functions of spectrally-flowed vertex operators.\\

Some of the earliest work on spectral flow automorphisms dates back to the mid 1980s \cite{Schwimmer:1986mf,Felder:1988sd,Lerche:1989uy,Moore:1989yh}. For a group $G$, spectral flow is an outer automorphism of the current algebra $\mathfrak{g}_k$, the generators of which we denote by $J^a_n$. As an outer automorphism, it is distinguished by a shift in the mode numbers of the generators, as opposed to permutations among the generators of $\mathfrak{g}$, which are a feature of automorphisms more generally. In other words, spectral flow automorphisms involve translations in the affine grading of the algebra, $n$, rather than permuting the values of $a$ in $J^a_n$. This shift in $n$ generates a shift in the conformal weights (i.e. the eigenvalues of $L_0$) by a term proportional to a Cartan charge, and this is the origin of the name "spectral flow".\footnote{See Appendix A of \cite{Ridout:2008nh} for a more detailed review of spectral flow automorphisms.}

The techniques we shall discuss will be applicable to general groups $G$, however, we shall use $SL(2,\R)$ as an illustrative example throughout this work. Strictly speaking, we will work with the analytic continuation of the $SL(2,\R)$ WZW model to Euclidean signature, meaning that the Hilbert space of the theory is derived from Lorentzian signature. This is richer than the $H^+_3$ model, which contains only continuous representations and no spectrally-flowed representations.

Spectral flow for $SL(2,\R)$ is labelled by an integer $w \in \mathbb{Z}$. While standard Kac-Moody primary states live in the kernel of $J^a_n$ for $n>0$, a $w$-spectrally-flowed primary state instead lives in the kernel of $\tilde J^a_n$ for $n>0$, where $\tilde J^3_n = J^3_{n}- \delta_{n,0}\frac{kw}{2}$ and $ \tilde J^{\pm}_n := J^{\pm}_{n\pm w}$ (see \eqref{eq:spectral_flow} and \eqref{eq:spectrally_flowed_states}). An analogous spectral flow automorphism exists in the $SU(2)$ WZW model. However, spectral flow in $SU(2)$ maps within the unflowed Hilbert space of the theory. This is not true of $SL(2,\R)$, where spectral flow generates new representations of the loop group. In fact, modular invariance of the one-loop partition function requires that such representations are included \cite{Henningson:1991jc,Maldacena:2000hw,Maldacena:2000kv}.

In the AdS$_3$ literature, the standard construction of spectrally flowed vertex operators dates back to \cite{Fateev,Maldacena:2001km}, where an operator carrying one unit of spectral flow is defined by taking a prescribed coincidence limit of two unflowed vertex operators (see (\ref{eq:spec_flow_limiting_definition_m-basis}) and (\ref{eq:y_basis_spec_flow_limiting_defn})). This means that a $w$-spectrally-flowed vertex operator can in principle be related to the collision of $(w+1)$-many unflowed vertex operators. The evaluation of a correlator of $n$ spectrally-flowed vertex operator insertions (a "spectrally-flowed correlator" for short) can therefore be obtained as a certain limit of higher-point correlators of unflowed vertex operators (an "unflowed correlator"). In practice, this procedure is rarely viable beyond 3 points and low values of spectral flow before the profusion of extra insertions becomes unmanageable, and most recent progress on correlator evaluations follows one of two different tracks.

On the one track, many existing results specialise to corners of parameter space, such as the tensionless limit of string theory \cite{Dei:2020zui,Naderi:2022bus,Dei:2023ivl} or the near-boundary approximation in target space \cite{Hikida:2023jyc,Knighton:2023mhq,Knighton:2024qxd,Sriprachyakul:2024gyl,Knighton:2026wva}. The key simplification in these cases is that the worldsheet theory can be written in terms of free fields, meaning that spectrally-flowed vertex operators can be written as a single local\footnote{Strictly speaking, the vertex operators may be mildly non-local, containing $\delta$ functions whose arguments are finitely many worldsheet derivatives of the worldsheet fields.} insertion. This approach is not sufficient for our interests, since we would like to study the exact $SL(2,\R)$ WZW model for any $k$.

On the other track, one constrains spectrally-flowed correlators in the full $SL(2, \mathbb{R})$ WZW model using bootstrap techniques \cite{Dei:2021xgh,Dei:2021yom,Dei:2022pkr,Iguri:2022eat,Iguri:2022pbp,Bufalini:2022toj,Iguri:2023khc,Iguri:2024yhb,Barone:2025vww}.\footnote{See \cite{Kovensky:2026usc} for a recent review and a more detailed list of historic references.} This includes constraints following from global and local Ward identities (introduced in \cite{Eberhardt:2019ywk}), the KZ equation, reflection symmetry, series identifications, null-vector constraints and OPE factorisation limits. The current state-of-the-art is a complete solution for generic 3-point correlators \cite{Dei:2021xgh,Bufalini:2022toj} and a conjecture for the 4-point correlators \cite{Dei:2021yom}. Solving the WZW model through this bootstrap approach is challenging, since the spectrally-flowed sectors are not affine highest weight and therefore have complicated Ward identities.

One important insight from the bootstrap approach is that the 3- and 4-point spectrally-flowed correlators appear to be related to 3- and 4-point unflowed correlators with modified quantum numbers. This suggests that a formalism exists that can treat correlation functions of flowed and unflowed vertex operators, in some sense, on the same footing. Hints towards such a formalism were present in the foundational paper \cite{Maldacena:2000hw}, in which Maldacena and Ooguri motivate spectral flow in the first instance as a procedure that relates classical solutions of the WZW equations of motion by symmetry transformations that are elements of the loop group of $SL(2, \mathbb{R})_L\times SL(2, \mathbb{R})_R$, which are not continuously connected to the identity. In more geometric terms, one might suspect that allowing for spectral flow is in some sense allowing for the possibility that $g$, the group-valued dynamical field of the WZW model, should be thought of as a section of a non-trivial bundle over the worldsheet.

Studying the WZW model from the perspective that $g$ is a section of a non-trivial bundle predates the foundational Maldacena-Ooguri papers \cite{Felder:1988sd,Moore:1989yh,Nakatsu:1993np,Hori:1994uf,Hori:1994nc}, but this interpretation has seen little application in the AdS$_3$ literature, with \cite{Knighton:2023mhq} providing the only known example as far as we are aware. From this perspective, spectral flow in the WZW model can be thought of in terms of promoting dynamical fields to sections of locally single-valued bundles (as defined in \S\ref{sec:spec_flow_G_bundle}) with curvature concentrated at punctures.

In this paper we leverage this perspective on spectral flow towards the computation of spectrally-flowed correlators, specialising to $SL(2,\R)$ for definiteness. The first main result we obtain is that at genus 0, a generic correlation function of $n$ spectrally-flowed vertex operators in the $y$-basis, denoted $V^{w_i}_{j_i}(y_i;x_i;z_i)$, can be rewritten in a \textit{canonical form} as a dressed correlator of $n$ unflowed insertions and one new insertion at an auxiliary point $u$, which carries $W-1$ units of spectral flow, where $W=\sum_{i=1}^n w_i$. As will be explained in \S\ref{sec: review material}, $w_i$ is the amount of spectral flow, $j_i$ is the $SL(2,\R)$ spin, $z_i$ is a point on the worldsheet, $x_i$ is a point on the boundary $\C\P^1$ and $y_i$ is not a physical quantum number but is a convenient way to encode a family of spectrally-flowed affine primaries. The canonical form of the $n$-point function is
\begin{equation}\label{eq:canonical_form_intro}
    \left\langle \prod_{i=1}^n V^{w_i}_{j_i}(y_i;x_i;z_i) \right\rangle = \left| \mathcal{P} \right|^2\left\langle \Omega_W(\gamma(z);u) \prod_{i=1}^n V_{j_i}(\gamma(z_i);z_i) \right\rangle\,,
\end{equation}
where
\begin{equation}\label{P}
    \mathcal{P}:=\prod_{i=1}^n(z_i-u)^{j_iW} \prod_{i<j} (z_i-z_j)^{-j_iw_j-j_jw_i-\frac{kw_iw_j}{2}}\,,
\end{equation}
and $\Omega_W(\gamma(z);u):=\exp\left(\oint_{z=u} \frac{\rd z}{2 \pi i} \, \gamma(z)J^+(z)\right)V_{\frac{k}{2},\frac{kW}{2}}^{W-1}(u)$. The new target space data is defined in terms of a certain meromorphic function, $\gamma(z)$, whose behaviour at $u$ and $z_i$ is determined by the data of all $n$ insertions $\{z_i,x_i,y_i,w_i\}$. The spectrally-flowed correlator is independent of $u$ and the $\Omega_W( \gamma(z);u)$ insertion can be placed at an arbitrary location. It should be thought of as specifying a certain instanton background of $g$ in which to compute the correlator of unflowed insertions $\prod_{i=1}^n V_{j_i}(\gamma(z_i);z_i)$. This result is derived in \S\ref{sec:spectrally-flowed_correlators_AdS3}.

Building upon this result and the insights from \cite{Dei:2021xgh,Dei:2021yom}, we conjecture that this $(n+1)$-point function can be recast as an $n$-point function of unflowed operators. Specifically, we conjecture that for $W\in 2\Z$,
\begin{equation}\label{eq:conjecture}
    \left\langle \prod_{i=1}^n V^{w_i}_{j_i}(y_i;x_i;z_i) \right\rangle = \left| \mathcal{A} \right|^2\left\langle \prod_{i=1}^n V_{j_i}(\sx_i;z_i) \right\rangle\,,
\end{equation}
where $\mathcal{A}$ is a meromorphic function and $\sx_i$ are a set of spacetime coordinates defined in \eqref{eq:Y_i_defn} that depend on the data $\{z_i,x_i,y_i,w_i,j_i\}$. That is to say, we conjecture that a certain change in target space coordinates, that is adapted to the bundle that $g$ is a section of, trivialises the background to generate an unflowed $n$-point correlator. A proof of this conjecture will appear in a companion paper \cite{McStay:2026xxx}. A similar conjecture can also be made for $W \in  2\Z + 1$, but in this case one of the spins $j_i$ must be replaced by $\frac{k}{2}-j_i$ \cite{Dei:2021xgh,Dei:2021yom}. We will focus on the even $W$ case in this paper for simplicity.

The second main result of this paper is to provide evidence for the conjecture \eqref{eq:conjecture}. For $W\in 2\Z$, we derive a change of coordinates $\gamma(z_i) \mapsto \sx_i$ with respect to which a meromorphic function (that we determine) times the $(n+1)$-point function from \eqref{eq:canonical_form_intro} satisfies the same constraints from the spacetime global conformal invariance as does an $n$-point correlator of unflowed insertions $\langle \prod_{i=1}^n V_{j_i}(\sx_i;z_i)\rangle$. This constitutes the first step of the proof. As an initial check of our claim, we prove the statement in the 2-point case, demonstrating an equivalence between the spectrally-flowed correlator and the dressed unflowed correlator of $V_{j_i}(\sx_i;z_i)$. While the 2-point function is not a new result in itself \cite{Maldacena:2001km,Dei:2021xgh}, it illustrates our argument in a simple example and, we hope, further clarifies our methodology to the reader.\footnote{We emphasise that the methods we apply here are not limited to $SL(2,\R)$, so we hope that such an example will be of benefit to those wanting to apply these methods to other models containing a spectral flow automorphism.} As another check, we verify in \texttt{Mathematica} (checking up to $W=\sum_{i=1}^4w_i \leq 12$) that the cross-ratio of our $\sx_i$ in the 4-point case agrees with the cross-ratio in \cite{Dei:2021yom}.\\

The structure of the paper is as follows. In \S\ref{sec: review material}, we begin with a review of the necessary background material for WZW models, and specifically for $G=SL(2,\R)$, in order to fix our notation and conventions. 

In \S\ref{sec:spec_flow_G_bundle}, we explain in what sense spectral flow can be thought of as a non-trivial bundle over the worldsheet. We show that spectrally-flowed vertex operators can be written in terms of a defect line wrapping two simpler insertions. The defect line is a non-local operator built from a contour integral of the Kac-Moody currents. We explain how these defect lines implement the transition functions of a non-trivial bundle over the worldsheet and we check the equivalence with conventional descriptions of spectral flow.

In \S\ref{sec:spectrally-flowed_correlators_AdS3}, we explain how to practically apply this definition to aid the computation of spectrally-flowed correlators. In brief, the contour integrals that define the defect lines can be deformed around the other insertions. We introduce a new $x(z)$-basis of vertex operators to do this, which is a generalisation of the $y$-basis of \cite{Dei:2021xgh}. On the genus 0 worldsheet, we demonstrate that any spectrally-flowed $n$-point function in the $y$-basis satisfies \eqref{eq:canonical_form_intro}, where $\Omega_W$ is given in the $x(z)$-basis.

In \S\ref{sec: computing correlators}, we constrain the $(n+1)$-point function in \eqref{eq:canonical_form_intro} by solving differential constraints, in a similar manner to the methods of \cite{Dei:2021xgh,Dei:2021yom,Dei:2022pkr}. We derive the change of spacetime variables for which the $(n+1)$-point function satisfies differential equations naturally associated to the spacetime global conformal invariance of an unflowed $n$-point function. To demonstrate parity with the literature, we fully solve the $n=2$ case, and in the $n=4$ case we provide an ancillary \texttt{Mathematica} notebook that verifies agreement between our resulting target space cross-ratio and the result of \cite{Dei:2021yom}.

We conclude in \S\ref{sec: discussion} with a discussion of the results obtained in this work and of future directions. We relegate some details of the calculations in \S\ref{sec: computing correlators} to Appendix \ref{appendix: correlator computations}. Additionally, to elaborate and illustrate how our methods generalise, we consider spectral flow of $G=SU(2,2)$ in Appendix \ref{appendix:SU(2,2)} and discuss higher genus correlators in Appendix \ref{appendix:higher_genus}.

\section{A brief review of AdS$_\mathbf{3}$ string theory}
\label{sec: review material}

In \S\ref{sec:spec_flow_G_bundle} we will give a definition of spectral flow that will apply to any group $G$. Nevertheless, our archetypal example throughout this work will be $G = SL(2,\R)$, as is relevant to string theory on AdS$_3$. Therefore, let us clearly set out our conventions for the (analytically continued) $SL(2,\R)$ WZW model and review the usual definitions of spectral flow that have appeared in the AdS$_3$ literature. A more detailed review (which we follow in part) can be found in \cite{Kovensky:2026usc}.

\subsection{Preliminaries}

A WZW model \cite{Wess:1971yu,Witten:1983tw,Witten:1983ar} is defined as a non-linear sigma model with an embedding field $g:\Sigma \to G$. It possesses a local $G_L \times G_R$ symmetry
\begin{equation}\label{eq:left_right_mult}
    g(z,\zbar) \mapsto h(z)g(z,\zbar) \bar{h}(\zbar)\,,
\end{equation}
where $h$ and $\bar{h}$ are elements of $G$ and $(z,\zbar)$ are local coordinates on $\Sigma$. This symmetry is generated by the currents
\begin{equation}\label{eq:J_definition}
    J = J^at_a = -k\p gg^{-1}\,, \quad J^a = k\,\text{tr}\left(t^a\p g g^{-1}\right)\,,
\end{equation}
which satisfy the OPEs
\begin{equation}\label{eq:J_g_OPEs}
    J^a(z) g(w,\bar w) \sim -\frac{t^a g(w, \bar w)}{z-w}\,,
    \qquad
    \bar J^{a}(\bar z) g(w,\bar w) \sim -\frac{ g(w, \bar w)(t^{\dagger})^{a}}{\bar z-\bar w}\,.
\end{equation}
Note that, irrespective of the choice of representation for $G_L \times G_R$ acting on $g$, the currents $J$ and $\bar{J}$ transform in the adjoint representation of $G$, meaning that $a$ is an adjoint index. These currents generate the affine Kac-Moody algebra $\mathfrak{g}_k$ at level $k$.

For the non-compact group $SL(2,\R)$, $k$ is not quantized but can be any real number. In string theory applications, one typically restricts to $k>2$ for the standard proofs of the no-ghost theorem to apply (see \cite{Evans:1998qu,Maldacena:2000hw} and references therein). We will use the standard basis
\begin{equation}
    t^3 = \frac{1}{2}\begin{pmatrix}
        1 & 0 \\ 0 & -1
    \end{pmatrix}\,, \quad t^+ = \begin{pmatrix}
        0 & 1 \\ 0 & 0
    \end{pmatrix}\,, \quad t^- = \begin{pmatrix}
        0 & 0 \\ -1 & 0
    \end{pmatrix}\,,
\end{equation}
with Killing form
\begin{equation}
    \kappa^{ab} = \begin{pmatrix}
        -\frac{1}{2} &0&0\\
        0&0&1\\
        0&1&0
    \end{pmatrix}\,.
\end{equation}
This is equivalent to $\kappa(A,B) = -\text{tr}(AB)$ for matrices $A$ and $B$ and leads to the following OPEs for the $\mathfrak{sl}(2,\R)_k$ current algebra
\begin{equation}
    \begin{aligned}
        J^+(z)J^-(w) &\sim \frac{k}{(z-w)^2} - \frac{2J^3(w)}{z-w}\,,\\
        J^3(z)J^{\pm}(w) &\sim \frac{\pm J^{\pm}(w)}{z-w}\,,\\
        J^3(z)J^3(w) &\sim \frac{-k}{2(z-w)^2}\,,
    \end{aligned}
\end{equation}
and similarly for the anti-holomorphic sector. We abide by standard conventions for the modes,
\begin{equation}
    J^a_{n} := \oint_{z=0} \frac{\rd z}{2\pi i}\, z^{n} J^a(z)\,,
\end{equation}
where the subscript $z=0$ denotes that the contour is taken to be a small circle of radius $\epsilon$ about $z=0$. We will always choose $\epsilon$ to be sufficiently small so that it does not contain any other insertion points on the worldsheet.

\subsection{The unflowed sector}\label{sec:unflowed_sector}

Kac-Moody primary states (respectively operators) in a $\mathfrak{g}_k$ module are characterised by a set of quantum numbers corresponding to eigenvalues under the Casimirs of $G$ and the elements of a Cartan subalgebra. For $\mathfrak{sl}(2,\mathbb{R})_k$ there is only one independent Casimir, given by  $\frac{1}{2}(t^+t^-+t^-t^+) - t^3t^3$, and we choose $t^3$ to generate the Cartan subalgebra. Kac-Moody primary states for $\mathfrak{sl}(2,\mathbb{R})_k$ are then labelled by eigenvalues $j$ and $m$ where
\begin{equation}\label{eq:j,m_definition}
    \begin{aligned}
        J^{\pm}_0|j,m\rangle &= (m\pm1\mp j)|j,m\pm 1\rangle\,, \quad J^3_0|j,m\rangle = m|j,m\rangle\,,
    \end{aligned}
\end{equation}
such that
\begin{equation}
    \left(\frac{1}{2}(J^+_0J^-_0+J^-_0J^+_0) - (J^3_0)^2\right)|j,m\rangle = -j(j-1)|j,m\rangle\,.
\end{equation}
The full WZW model has affine $\mathfrak{sl}(2,\R)_k$ symmetry on both the left and the right, so the full state is denoted by $|j,m,\bar{m}\rangle$ where $m$ and $\bar{m}$ need not be the same. Often we will work holomorphically and omit the $\bar{m}$ index.

There are two types of $\mathfrak{sl}(2,\R)$ representations that are relevant to string theory on AdS$_3$ \cite{Maldacena:2000hw}:
\begin{enumerate}
    \item \textbf{Discrete representations:} These can be either lowest weight, $\mathcal{D}^+_j$, or highest weight, $\mathcal{D}^-_j$. They are comprised of states
    \begin{equation}
        \begin{aligned}
            \mathcal{D}^+_j &= \{|j,m\rangle : m = j, j+1, j+2, \dots\}\,,\\
            \mathcal{D}^-_j &= \{|j,m\rangle : m = -j, -j-1, -j-2, \dots\}\,,
        \end{aligned}
    \end{equation}
    and unitarity requires $j\in \R^{>0}$.
    \item \textbf{Continuous representations:} These are denoted by $\mathcal{C}^{\alpha}_j$ where $0\leq \alpha < 1$ is the fractional part of the $J^3_0$ eigenvalue.
    \begin{equation}
        \mathcal{C}^{\alpha}_j = \{|j,m\rangle : m = \alpha, \alpha \pm 1, \alpha \pm 2, \dots \}\,.
    \end{equation}
    Unitarity requires $j = \frac{1}{2} + is$ for $s\in \R$. However, the isomorphism between representations with $SL(2,\R)$ spins $j$ and $1-j$ allows one to set $s\geq 0$ without loss of generality.
\end{enumerate}
Both types of representation extend to representations of the affine algebra $\mathfrak{sl}(2,\R)_k$ through the application of lowering operators $J^a_{-n}$.

The vertex operators corresponding to the states $|j,m,\bar{m}\rangle$ are written as $V_{j,m,\bar{m}}(z,\bar{z})$. Since they are eigenstates of $J^3_0$ and $\bar{J}^3_0$ with eigenvalues $m$ and $\bar{m}$, these operators are said to be in the "$m$-basis". Another convenient choice of basis is the so-called ``$x$-basis'', built from operators of the form
\begin{equation}\label{eq:x_basis_sum}
    V_j(x,\bar{x};z,\zbar)=\sum_{m,\bar{m}}x^{m-j}\bar{x}^{\bar{m}-j}\,V_{j,m,\bar{m}}(z,\bar{z}),
\end{equation}
in the discrete representation $\mathcal{D}^+_j$.\footnote{In the continuous representation, $m-\bar{m}$ is an integer but $m+\bar{m}$ can take any real value. This formula is therefore modified to include an integral over $m+\bar{m}$. See, for instance, \cite{Teschner:1997fv} or \S2 of \cite{Kovensky:2026usc} for a modern review.} In what follows, we will often abbreviate these vertex operators to $V_{j,m}(z)$ and $V_j(x;z)$. At $x=0$ in the discrete representation, we identify $V_j(0;z) = V_{j,j}(z)$. More generally,
\begin{equation}
    V_j(x;z)=e^{xJ^+_0}\,V_j(0;z)\,e^{-xJ^+_0}
\end{equation}
holds for any representation and $(x,\bar{x})$ have the holographic interpretation of the insertion point of a vertex operator in the dual CFT.

\subsection{Spectrally-flowed sectors}\label{sec:flowed_sector}

$\mathcal{D}^{\pm}_j$ and $\mathcal{C}^{\alpha}_j$ are not the only representations of the affine algebra $\mathfrak{sl}(2,\R)_k$. Additional representations follow as a result of an outer automorphism on the current algebra known as the \textit{spectral flow}. This acts as
\begin{equation}\label{eq:spectral_flow}
    \sigma^w(J^3_n) = J^3_n + \delta_{n,0}\frac{kw}{2}\,, \quad \sigma^w(J^{\pm}_n) = J^{\pm}_{n\mp w}\,,
\end{equation}
for $w\in \mathbb{Z}$. Spectral flow automorphisms can be defined more generally for semi-simple groups \cite{Lerche:1989uy}; their characteristic feature is that they involve a shift in the mode number i.e. a translation in the grading of the affine algebra. Nevertheless, in many cases we do not need to separately consider flowed and unflowed representations. For instance, for $G=SU(2)$ one finds that spectral flow maps an unflowed primary to an affine descendant. It does not generate new states outside of the original unflowed Hilbert space. This is not true for $G = SL(2,\R)$ where spectral flow maps to physically distinct states.

This was a key observation in \cite{Maldacena:2000hw}, where it was observed that the spectral flow automorphism must be included in the Hilbert space of string theory on AdS$_3$. This removed an unphysical upper bound on the mass of states that is present in the unflowed sector and ensured modular invariance of the 1-loop partition function \cite{Maldacena:2000hw,Henningson:1991jc}.

Referring to $|j,m\rangle$ as an affine primary, we define $[|j,m\rangle]^w$ to be a \textit{spectrally-flowed affine primary}. It satisfies
\begin{equation}\label{eq:spectrally_flowed_states}
    J^a_n[|j,m\rangle]^w = \left[ \sigma^w\left(J^a_n\right) |j,m\rangle \right]^w\,.
\end{equation}
In particular, the analogue of \eqref{eq:j,m_definition} is
\begin{equation}\label{eq:currents_acting_on_flowed_states}
    J^{\pm}_{\pm w}[|j,m\rangle]^w = (m\pm 1\mp j) [|j,m\pm 1\rangle]^w\,, \quad J^3_0[|j,m\rangle]^w = \left(m + \frac{kw}{2}\right)[|j,m\rangle]^w\,. 
\end{equation}
The spectrally-flowed representations $\mathcal{D}^{\pm,w}_j = \sigma^w\left( \mathcal{D}^{\pm}_j \right)$ and $\mathcal{C}^{\alpha,w}_j = \sigma^w\left( \mathcal{C}^{\alpha}_j \right)$ are built from $[|j,m\rangle]^w$. It turns out that not all of these representations are independent because of the $SL(2,\R)$ series identifications,
\begin{equation}
    \mathcal{D}^{\pm,w}_j \cong \mathcal{D}^{\mp,w\pm 1}_{\frac{k}{2}-j}\,.
\end{equation}
For this reason, we do not need to separately consider the flowed highest- and lowest-weight discrete representations.\footnote{A priori, it is plausible that the states living in these representations could appear in the spectrum with degeneracy two. In \cite{Maldacena:2000hw} it was argued that this is not the case by comparing with the Hilbert space $L^2($AdS$_3)$ for quantum mechanics on AdS$_3$, for which the degeneracy is one. This was further supported by computing the thermal one-loop partition function in \cite{Maldacena:2000kv}.} In fact, it was claimed in \cite{Maldacena:2000hw} that the full Hilbert space of the $SL(2,\R)$ WZW model is given by the representations $\mathcal{C}^{\alpha,w}_j \times \mathcal{C}^{\alpha,w}_j$ for $j=\frac{1}{2} + is$ and $\mathcal{D}^{+,w}_j \times \mathcal{D}^{+,w}_j$ for $\frac{1}{2} < j < \frac{k-1}{2}$ with $w\in \mathbb{Z}$.
\\

We can now construct vertex operators describing the states $[|j,m\rangle]^w$. We denote such vertex operators by $V^w_{j,h}(z)$ where
\begin{equation}
    h = m + \frac{kw}{2}\,,
\end{equation}
is the $J^3_0$ eigenvalue. As before, for holographic applications it is more natural to consider vertex operators in the $x$-basis,
\begin{equation}\label{eq:x_translation}
    V^w_{j,h}(x;z) = e^{xJ^+_0}V^w_{j,h}(z)e^{-xJ^+_0}\,.
\end{equation}
These describe coherent states in the worldsheet CFT \cite{Dei:2021xgh}. Unlike the unflowed sector, we maintain the label $h$ on these vertex operators because, even though they don't have a definite $J^3_0$ eigenvalue, they are associated to a definite value of $m$. This is because $J^+_0[|j,m\rangle]^w = \left[J^+_{-w}|j,m\rangle\right]^w$ corresponds to a flowed affine descendant. When working in the $x$-basis, we are free to restrict to $w\geq 0$ because $x\mapsto \frac{1}{x}$ (i.e. changing patch on the boundary $\C\P^1$) maps $w\mapsto -w$. We will always assume $w\geq 0$ in what follows.

Inside correlation functions, $V^w_{j,h}(x;z)$ is inserted at a point $z\in \Sigma$ on the worldsheet. We will then view $e^{-xJ^+_0}$ to be acting on the vacuum (and hence set it to 1), while $e^{xJ^+_0}$ represents a contour integral surrounding the insertion of $V_{j,h}^w(0;z)$. As a result, in what follows we will often write \eqref{eq:x_translation} as 
\begin{equation}\label{x}
    V_{j,h}^w(x;z) = \exp\left(\frac{1}{2 \pi i}\oint_{z'=z} \rd z'\, x J^+(z')\right)V_{j,h}^w(0;z)\,,
\end{equation}
to emphasise that there is a line operator acting on $V_{j,h}^w(0;z)$.

These vertex operators satisfy the OPEs
\begin{equation}\label{eq:x_basis_OPEs}
    \begin{aligned}
        &J^+(z')V^w_{j,h}(x;z) \sim \frac{(h-\frac{kw}{2}+1-j)V^w_{j,h+1}}{(z'-z)^{w+1}} + \sum_{p=1}^{w-1}\frac{J^+_pV^w_{j,h}}{(z'-z)^{p+1}} + \frac{\p_xV^w_{j,h}}{z'-z} + \dots \\
        &(J^3(z') - xJ^+(z'))V^w_{j,h}(x;z)\sim \frac{hV^w_{j,h}}{z'-z} + \dots\,,\\
        &(J^-(z') - 2xJ^3(z') + x^2J^+(z'))V^w_{j,h}(x;z) \sim (z'-z)^{w-1}\left(h-\frac{kw}{2}-1+j\right)V^w_{j,h-1}\\
        &\hspace{7.3cm}+\mathcal{O}((z'-z)^{w})\,,
    \end{aligned}
\end{equation}
where all vertex operators on the right hand side are evaluated at $(x;z)$. Note that these OPEs are defined with respect to $e^{xJ^+_0}J^a(z')e^{-xJ^+_0}$. The terms $J^+_pV^w_{j,h}$ represent many unknowns appearing in the $J^+$ OPE; these unknowns make it much harder to bootstrap the form of spectrally-flowed correlators and are a primary reason why progress has been slow on determining the exact form of $n$-point functions.
\\

It was observed in \cite{Dei:2021xgh} that, for the purposes of computing correlation functions, it is helpful to introduce an additional coordinate $y$ that is analogous to $x$ for the flowed zero-mode $J^+_w$. The $y$-basis is defined by
\begin{equation}
    V^w_{j}(y;x;z) = e^{zL_{-1}}e^{xJ^+_0}e^{yJ^+_w}V^w_{j,h}(0)e^{-yJ^+_w}e^{-xJ^+_0}e^{-zL_{-1}}\,,
\end{equation}
for $h=j + \frac{kw}{2}$. It is important to note that $V^w_{j,h}(0)$ was inserted at $z=0$ because $J^+_w$ and $L_{-1}$ do not commute and hence the ordering matters. These vertex operators do not have definite worldsheet nor spacetime conformal weight. Similarly to \eqref{eq:x_basis_sum},
\begin{equation}\label{eq:y_basis_transform}
    V^w_j(y,\bar{y};x,\bar{x};z,\bar{z})=\sum_{h,\bar{h}}y^{h-\frac{kw}{2}-j}\bar{y}^{\bar{h}-\frac{kw}{2}-j}V^w_{j,h,\bar{h}}(x,\bar{x};z,\zbar)\,,
\end{equation}
in the discrete representation $\mathcal{D}^+_j$. The inverse is given by a Mellin-type transform
\begin{equation}\label{eq:y_basis_Mellin_transform}
    V^w_{j,h,\bar{h}}(x,\bar{x};z,\zbar) = \int \rd^2y\, y^{\frac{kw}{2} + j-h-1} \bar{y}^{\frac{kw}{2} + j-\bar{h}-1} V^w_j(y,\bar{y};x,\bar{x};z,\bar{z})\,,
\end{equation}
which holds for both continuous and discrete representations.

In analogy with \eqref{x}, inside a correlation function we will often rewrite the $y$-basis vertex operator as
\begin{equation}
    V_j^w(y;x;z) = \exp\left(\frac{1}{2 \pi i}\oint_{z'=z} \rd z'\, (x + y(z'-z)^w) J^+(z')\right)V_{j}^w(0;0;z)\,.
\end{equation}
Here, the factor of $(z'-z)^w$ is a consequence of commuting $e^{yJ^+_w}$ past $e^{zL_{-1}}$.

In the $y$-basis, the OPEs \eqref{eq:x_basis_OPEs} are modified to
\begin{equation}\label{eq:y_basis_OPEs}
    \begin{aligned}
        &J^+(z')V^w_{j}(y;x;z) \sim \frac{\p_yV^w_{j}}{(z'-z)^{w+1}} + \sum_{p=1}^{w-1}\frac{J^+_pV^w_{j}}{(z'-z)^{p+1}} + \frac{\p_xV^w_{j}}{z'-z} + \dots \\
        &(J^3(z') - xJ^+(z'))V^w_{j}(y;x;z)\sim \frac{(j+\frac{kw}{2} + y\p_y)V^w_{j}}{z'-z} + \dots\,,\\
        &(J^-(z') - 2xJ^3(z') + x^2J^+(z'))V^w_{j}(y;x;z) \sim (z'-z)^{w-1} (2jy + y^2\p_y)V^w_j\\
        &\hspace{7.5cm} +\mathcal{O}((z'-z)^{w})\,,
    \end{aligned}
\end{equation}
where all vertex operators on the right hand side are evaluated at $(y;x;z)$. 

In \S\ref{subsec: x(z) basis}, we will further generalise the $y$-basis to define what we call the "$x(z)$-basis". The vertex operators will be labelled by a degree $w$ polynomial which describes conjugation by all modes $J^+_n$ for $0\leq n \leq w$.

\subsection{Explicit realisations of vertex operators}

In \S\ref{sec:unflowed_sector} and \S\ref{sec:flowed_sector}, we defined vertex operators at the level of representation theory. We will now discuss how, and to what extent, they can be explicitly realised in terms of fundamental fields.\footnote{In the near-boundary approximation, one can realise the $x$-basis vertex operators using a free field realisation of $\mathfrak{sl}(2,\R)_k$ \cite{Hikida:2023jyc,Knighton:2023mhq}. These methods have been successfully used to compute residues of the exact $n$-point functions \cite{Knighton:2024qxd,Knighton:2026wva}. In this paper, we are interested in exact realisations of the WZW model and so we will not review these near-boundary approximations.}

Let us start by considering states in the $m$-basis. These are most conveniently defined using the parafermionic decomposition \cite{Fateev:1985mm}, where one factorises $SL(2,\R) \cong SL(2,\R)/U(1) \times U(1)$ \cite{Maldacena:2001km,Fateev}. The $U(1)$ is associated to a bosonisation of the Cartan,
\begin{equation}\label{eq:J3_bosonisation}
    J^3(z) = -\sqrt{\frac{k}{2}}\p\phi, \quad \phi(z)\phi(w) \sim -\log(z-w)\,,
\end{equation}
and we define $\Psi_{j,m}$ to be a vertex operator associated to $SL(2,\R)/U(1)$. This is a useful way to describe vertex operators, because the $U(1)$ part of the correlator is free and may be simply evaluated. The parafermionic CFT on the other hand is interacting, making its correlators difficult to compute.

The $m$-basis vertex operators take the form
\begin{equation}
    V^w_{j,h}(z) = \Psi_{j,m}e^{\left(m + \frac{kw}{2}\right)\sqrt{\frac{2}{k}}\phi}(z)\,,
\end{equation}
where we recall that $h = m + \frac{kw}{2}$. It is simple to introduce the spectral flow in this decomposition, because it does not interact with the parafermionic CFT. We deduce that
\begin{equation}\label{eq:spec_flow_limiting_definition_m-basis}
    V^w_{j,h}(z) = \lim_{\epsilon\to 0} \epsilon^{mw} V_{j,m}(z+\epsilon)\Omega_w(z)\,,
\end{equation}
where we define the generalised spectral flow operator \cite{Iguri:2022eat}
\begin{equation}\label{eq:Omega_definition}
    \Omega_w(z) = V^{w-1}_{\frac{k}{2},\frac{kw}{2}} = e^{w\sqrt{\frac{k}{2}}\phi(z)}\,,
\end{equation}
whose parafermionic component is (locally) the identity. $\Omega_w$ is equivalent to applying $w$ units of spectral flow to the identity operator of the full WZW model.

As we have already mentioned, the physically interesting correlation functions correspond to insertions of coherent states given by the $x$-basis, for which the parafermionic decomposition cannot be easily applied. Nevertheless, in the unflowed sector, the vertex operators have a simple expression \cite{Teschner:1997fv},
\begin{equation}\label{eq:unflowed_vertex_op}
    V_j(x;z) = \left[(1,-x)g(z,\zbar)(1,-\bar{x})^T\right]^{-2j}\,.
\end{equation}
That a similar expression does not exist for the flowed sectors can be understood for the following reason. Suppose the spectrally-flowed operator is inserted at $z_i$. Then spectral flow is generated by the outer automorphism \cite{Maldacena:2000hw}
\begin{equation}\label{eq:spectal_flow_automorphism}
    g(z,\zbar) \mapsto U(z)g(z,\zbar)\bar{U}(\zbar)\,, \quad U =  \left(\begin{array}{cc}
        (z-z_i)^{w/2} & 0 \\
        0 & (z-z_i)^{-w/2}
    \end{array}\right)=\exp\left(w\ln(z-z_i)t^3\right)\,,
\end{equation}
where $\bar{U}$ has the same form as $U$ but with $z-z_i$ replaced with $\bar{z}-\bar{z}_i$. This is well-defined on the patch $z\neq z_i$ and satisfies
\begin{equation}\label{eq:J_transformation}
    J=-k\p gg^{-1}\rightarrow -kU\p gg^{-1}U^{-1}-k\p U U^{-1}\,.
\end{equation}
This is consistent with \eqref{eq:spectral_flow} and \eqref{eq:J_definition}, where the term $-k\p UU^{-1}$ provides the necessary shift in the Cartan zero-mode. Non-trivial $x$-dependence can be incorporated via conjugation with $e^{xt^+}$
\begin{eqnarray}
    U(x_i;z)&=&\left(\begin{array}{cc}
        1 & x_i \\
        0 & 1
    \end{array}\right)\left(\begin{array}{cc}
        (z-z_i)^{w_i/2} & 0 \\
        0 & (z-z_i)^{-w_i/2}
    \end{array}\right)\left(\begin{array}{cc}
        1 & -x_i \\
        0 & 1
    \end{array}\right)\nonumber\\
    &=&\exp\left(w_i\ln(z-z_i)(t^3-x_it^+)\right)\,.
\end{eqnarray}
However, this transformation is singular precisely at $z=z_i$ and so the candidate vertex operator "$V^w_{j}(x_i;z_i) = U(x_i;z_i)V_j(x_i;z_i)\bar{U}(\bar{x}_i;\zbar_i)$", analogous to \eqref{eq:unflowed_vertex_op}, is not well-defined. As we shall see in \S\ref{sec:spec_flow_G_bundle}, one really needs to view $g(z,\zbar)$ not as a function $g:\Sigma \to G$, but as a section of a $G$-bundle with a non-trivial transition function in the neighbourhood of spectrally-flowed insertions.

Nevertheless, one can define a spectrally-flowed $x$-basis vertex operator using a definition akin to \eqref{eq:spec_flow_limiting_definition_m-basis}. Including the anti-holomorphic sector,
\begin{equation}\label{eq:x_basis_spec_flow_limiting_defn}
    V^w_{j,h}(x;z) = \lim_{\epsilon,\bar{\epsilon} \to 0} \epsilon^{mw}\bar{\epsilon}^{\bar{m}w} \int \rd^2y\, y^{\frac{kw}{2} +j-h-1}\bar{y}^{\frac{kw}{2}+j-\bar{h}-1}V_{j}(x+y;z+\epsilon)\Omega_w(x;z)\,,
\end{equation}
where $\Omega_w(x;z) = e^{xJ^+_0}V^{w-1}_{\frac{k}{2},\frac{kw}{2}}e^{-xJ^+_0}$. This is the preferred local definition used in the literature for correlation function computations \cite{Maldacena:2001km,Iguri:2022eat}. Recalling \eqref{eq:y_basis_transform}, this formula simplifies in the $y$-basis
\begin{equation}\label{eq:y_basis_spec_flow_limiting_defn}
    V^w_{j}(y;x;z) = \lim_{\epsilon \to 0} \epsilon^{jw}  V_{j}(x+\epsilon^wy;z+\epsilon)\Omega_w(x;z)\,,
\end{equation}
where we choose to only write the holomorphic sector.

\section{\boldmath Locally single-valued flat $G$-bundles and spectral flow}\label{sec:spec_flow_G_bundle}

In this section, we revisit an old idea that spectral flow can be interpreted in terms of certain $G$-bundles over a worldsheet $\Sigma$ \cite{Schwimmer:1986mf,Felder:1988sd,Lerche:1989uy,Moore:1989yh,Nakatsu:1993np,Hori:1994uf,Hori:1994nc}. To the best of our knowledge, this is a definition of spectral flow that has largely evaded the AdS$_3$ literature, except for one application in the near-boundary approximation \cite{Knighton:2023mhq}. Nevertheless, we have found that this framework can be practically applied to constrain the form of spectrally-flowed correlation functions as we will demonstrate in \S\ref{sec:spectrally-flowed_correlators_AdS3} and \S\ref{sec: computing correlators}. It is the goal of this section, therefore, to reintroduce this definition of spectral flow and to phrase it in the modern parlance of line defects. We encourage readers who are primarily interested in how these ideas inform the computation of spectrally-flowed correlators in AdS$_3$ string theory to make note of \eqref{eq: nucleation 2} and to skip ahead to \S\ref{sec:spectrally-flowed_correlators_AdS3}.\\

Let us begin by introducing some definitions. While we will state these definitions using formal mathematics, we trust that the application of these ideas in \S\ref{subsec: nucleation}-\ref{subsec: patches and vertex operators} will be amenable to a physics audience. In the following, we define $\Sigma^\circ = \Sigma\backslash\{z_i\}$ to be a punctured Riemann surface and $G$ a Lie group.
\paragraph{\boldmath Principal $G$-bundle:}
A principal $G$-bundle over $\Sigma^\circ$ is a smooth manifold $E$, equipped with a projection $\pi:E\longrightarrow\Sigma^\circ$ and a smooth right action $E\times G\to E$, denoted
$(p,g)\mapsto p\cdot g$, satisfying:
\begin{itemize}
    \item  $\pi(p\cdot g)=\pi(p)$.

    \item The action is free and transitive on each fibre:
    for any $p,q\in \pi^{-1}(z)$, there exists a unique
    $g\in G$ such that $q=p\cdot g$.

    \item The bundle is locally trivial in a way that respects this action: $E|_U\cong U\times G$ and $(p,g)\cdot h=(p,gh)$ for $U\subset \Sigma^\circ$ a local chart.
\end{itemize}
Thus, each fibre is a copy of $G$.

\paragraph{\boldmath Flat principal $G$-bundle:} Let $(E,\pi)$ be a principal $G$-bundle over $\Sigma^{\circ}$. Let $\omega \in \Omega^1(E,\mathfrak{g})$ (a Lie-algebra-valued one-form on $E$) be a connection on $E$, meaning that $\omega$ respects the right action of $G$ on $E$ and that it identifies vector fields parallel to the fibres of $E$ with the Lie algebra generators of $G$. This specifies how to transport a point in one fibre to a point in another. Locally, we can pullback $\omega$ to $A \in \Omega^1(U,\mathfrak{g})$ for a patch $U\subset \Sigma^{\circ}$, providing a connection on $U$. Then, we define a flat principal $G$-bundle to be a triplet $(E,\pi,\omega)$ where the pullback $A$ is a connection with vanishing curvature on any local patch $U\subset \Sigma^\circ$,
\begin{equation}
    F_A = \mathrm{d} A + A\wedge A = 0\,.
\end{equation}

WZW models are not strictly speaking described by a product of flat principal $G$-bundles, one on the right and one on the left. To see this, recall the $G_L \times G_R$ symmetry \eqref{eq:left_right_mult},
\begin{equation}
    g\mapsto h g \bar{h}\,.
\end{equation}
The kernel of this map on $G$ is isomorphic to the centre of $G$, since the kernel is spanned by $(h,\bar{h})=(c,c^{-1})$ for $c\in Z(G)$, where $Z(G)$ is the centre of $G$. We therefore introduce the effective structure group of the WZW model,
\begin{equation}\label{eq:effective_structure_group}
    H=\frac{G_L\times G_R}{Z(G)}\,.
\end{equation}
If the WZW model was a product of $G$-bundles, it would have a $G\times G$ structure group, defining the possible transition functions on overlaps. Instead, the effective structure group is smaller. With this in mind, we can now explain the bundle structure that defines a WZW model.

\paragraph{\boldmath Locally single-valued flat $G$-bundle:}
We define a locally single-valued flat $G$-bundle\footnote{Let us emphasise that this is not a canonical definition, but one that is sufficient for us to define spectral flow. We use the word "locally" to emphasise that we only constrain holonomies about the punctures. This means we can define a single-valued field $g(z,\zbar)$ in a punctured neighbourhood of $z_i$. If $\Sigma$ is of higher genus, we do not constrain the holonomies around the $A$- and $B$-cycles. Additionally, we will be slightly sloppy and still refer to this as a locally single-valued flat $G$-bundle, even though the definition is actually with respect to a $H$-bundle. This is because most of the time we will only be considering the holomorphic sector of the theory, for which it is more natural to discuss the group $G$ rather than $H$.} over $\Sigma^\circ$ to be a flat principal $H$-bundle where the holonomy around every puncture $z_i$ is trivial. More precisely, let $C_i$ be a small contour in $\Sigma^\circ$ enclosing $z_i$ and let $\rho_A:\pi_1(\Sigma^\circ,p) \to H$ be the representation induced by the connection $A$ for some base point $p\in \Sigma^{\circ}$. Then $\rho_A(C_i) = 1_H$, the identity element of $H$, for all $i=1,\dots, n$. Equivalently, in the lift to $G_L \times G_R$, if the holonomies are $H_i$ and $\bar{H}_i$, then $H_ig\bar{H}_i=g$. This ensures that there exists a single-valued parallel trivialisation on a sufficiently small punctured neighbourhood of each $z_i$. In what follows, we will not impose that such a trivialisation extends regularly across the puncture.\\

We now extend the locally single-valued flat $G$-bundle to be defined over the whole of $\Sigma$. For $g=0$, on a local $\mathbb{C}$ patch of $\Sigma$ where $A\in \Omega^1(\Sigma, \mathfrak{h})$ is the local form of the connection on $E$, the flatness condition becomes
\begin{equation}
    \rd A +A \wedge A = \sum_{i=1}^n p_i \delta^{(2)}(z-z_i) \rd^2 z\,, \quad p_i \in \mathfrak{h}\,.
\end{equation}
As usual, each of the $p_i$s are only defined up to conjugacy.\footnote{If we were working on $\Sigma\cong \C\P^1$, then conjugating by a holomorphic $h(z)\in G$ would constrain $h(z)= \text{constant}$ and therefore every $p_i$ would be conjugated by the same element $h\in G$. However, we are working on a local patch of a punctured Riemann surface. This means that each $p_i$ can be conjugated independently.} The single-valuedness condition means that the exponentiation of each $p_i$ gives the identity element,
\begin{equation}
    e^{p_i} = 1_H\,, \quad \forall i \in \{1, \ldots n\}\,,
\end{equation}
so that the holonomies around each point $\{z_i\}$ are trivial.

Just in words for now, the central claim of this section is the following\footnote{To our knowledge, the first to make a claim like this was \cite{Hori:1994uf,Hori:1994nc}.}:\\

\noindent\fbox{%
\begin{minipage}{\textwidth}
  \textbf{Spectral flow and locally single-valued flat $\mathbf{G}$-bundles:}

In a worldsheet theory which enjoys an affine Kac-Moody symmetry algebra $\mathfrak{g}_k$, evaluating a correlator with a spectrally-flowed insertion at a point $z_i$ is equivalent to declaring that the fundamental worldsheet fields are sections of a locally single-valued flat $G$-bundle (or its associated bundles), with curvature concentrated at $z_i$.
\end{minipage}%
}\\

In order to state the correspondence precisely in equations, we must introduce two prerequisites. In \S\ref{subsec: nucleation} we discuss the concept of nucleation with symmetry generators. We explain that spectrally-flowed vertex operators in theories with an affine Kac-Moody symmetry can be replaced by simpler objects wrapped by a symmetry operator. Then, in \S\ref{subsec: sym gen for trans fn} we will discuss how the use of symmetry generators in correlators is related to specifying the transition functions of $E$, a locally single-valued flat $G$-bundle over the worldsheet.

With these tools in hand, in  \S\ref{subsec: patches and vertex operators} we explain precisely how to think of a spectrally-flowed vertex operator as an unflowed vertex operator written in terms of the local form of the worldsheet field on a particular patch, of a patchwise description of the bundle $E$. Whilst we will primarily focus on AdS$_3$ applications, the ideas presented in this section are directly applicable to other choices of Lie group too. To demonstrate this, we briefly discuss the application to $G=SU(2,2)$ in Appendix \ref{appendix:SU(2,2)}.

\subsection{Nucleation for spectral flow}\label{subsec: nucleation}

Given a local insertion $V$, it is often useful to decompose it into a symmetry generator $L$ wrapping an insertion $V'$,
\begin{equation}
    \langle V(z) \ldots \rangle = \langle L V'(z) \ldots \rangle\,. 
\end{equation}
We refer to this process as "nucleation" (see, for instance, \cite{Heckman:2024obe}). By wrapping the symmetry generator around all other of the insertions and contracting the loop around the point at infinity, one generates relations between correlation functions in the CFT. Analogous methods have been applied in the study of Kramers-Wannier duality \cite{Kramers:1941kn,Frohlich:2004ef}.

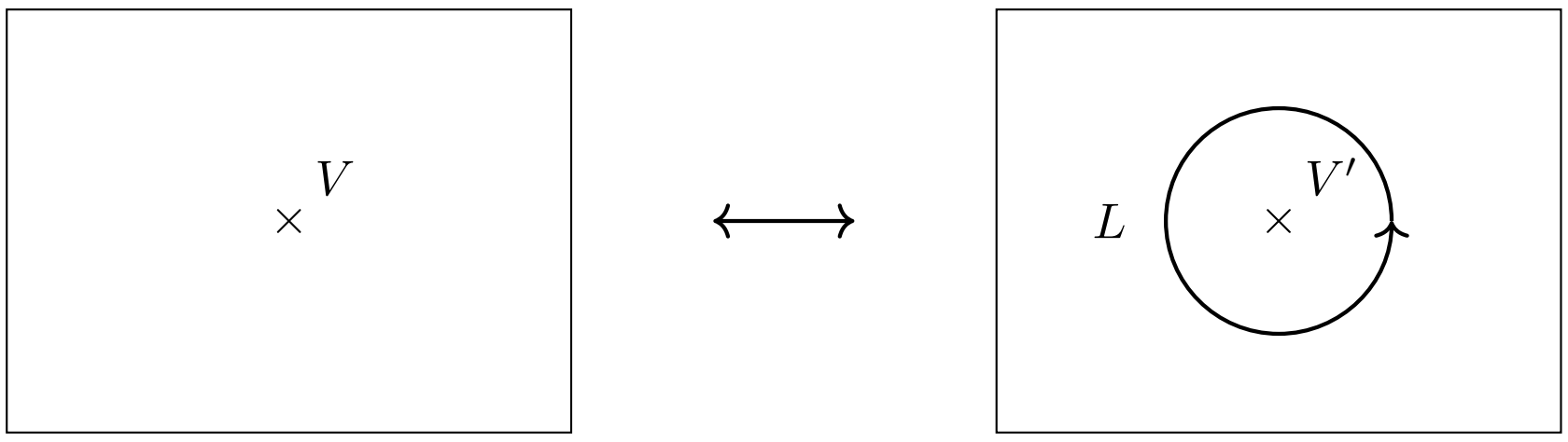
\begin{figure}
\centering
    \begin{tikzpicture}[
  panel/.style={draw, minimum width=4cm, minimum height=3cm},
  every node/.style={font=\small}
]

\node[panel] (left) {};
\node at (left.center) {$\times$};
\node[above right=2pt of left.center] {$V$};

\draw[<->, thick] ($(left.east)+(1,0)$) -- ++(1cm,0);

\node[panel, right=3cm of left] (right) {};
\node at (right.center) (vprime) {$\times$};
\node[above right=2pt of right.center] {$V'$};

\draw[->, thick]
  (vprime) ++(0.8,0)
  arc[start angle=0, end angle=360, radius=0.8cm];

\node at ($(vprime)+(-1.2,0)$) {$L$};

\end{tikzpicture}
\caption{Nucleation is the procedure by which an insertion $V$ is replaced by an insertion $V'$ encircled by a (finite or infinitesimal) symmetry generator $L$. 
}
\end{figure}

For definiteness, we specialise to the case of a WZW model on $\Sigma$ with worldsheet fields $g(z, \bar z)$ valued in $G$. Suppose that the holomorphic left multiplication current $J^+(z)$ is the raising operator on a certain $\mathfrak{g}_k$ module with states $\{|j,m\rangle\}$. A familiar example of nucleation is the decomposition
\begin{equation}
    (m+1-j)|j,m+1 \rangle = J^+_0 |j,m\rangle \quad \leftrightarrow \quad (m+1-j)V_{j,m+1}(0)=\frac{1}{2 \pi i}\oint_{z=0} \rd z\, J^+(z) V_{j,m}(0)\,.
\end{equation}
The operator $\frac{1}{2 \pi i}\oint \rd z\, J^+(z)$ should be thought of as an infinitesimal symmetry generator. By deforming the contour to envelop the other insertions in a correlator, we recover linear relations between CFT correlators. 

Besides deriving linear relations, another use case of nucleation is to simplify a $V$ by decomposing it into an $L V'$, with the hope that $V'$ is something easier to deal with and that the action of $L$ on the other vertex operators is relatively straightforward. A standard example of this is to (somewhat tautologically) undo the $x$-conjugation of an $x$-basis vertex operator which, as we noted in \eqref{x}, is written in terms of symmetry generators as\footnote{Here and in many of the following equations, we suppress the anti-holomorphic sector to simplify notation. When the anti-holomorphic sector plays a more significant role, we will reinstate it.}
\begin{equation}
    V_{j,h}^w(x;z) = \exp\left(\frac{1}{2 \pi i}\oint_{z'=z} \rd z'\, xJ^+(z') \right) V_{j,h}^w(0;z)\,.
\end{equation}
In our case of interest, we will demonstrate that a spectrally-flowed vertex operator $V$ can be replaced in nucleation by a pair of vertex operators and a particular symmetry generator of the Kac-Moody current algebra. 

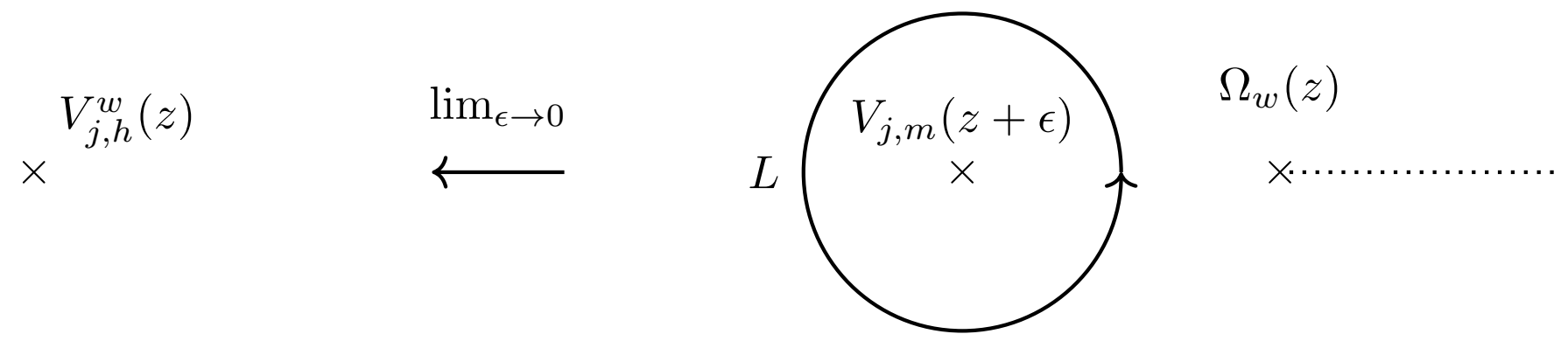
\begin{figure}
\centering
    \begin{tikzpicture}[
  panel/.style={minimum width=4cm, minimum height=3cm},
  every node/.style={font=\small}
]

\node[panel] (left) {};
\node at (left.center) {$\times$};
\node[above right=2pt of left.center] {$V^w_{j,h}(z)$};

\draw[<-, thick] ($(left.east)+(1,0)$) -- ++(1cm,0);
\node at ($(left.east)+(1.5,0.5)$) (arrowlabel){$\lim_{\epsilon\rightarrow 0}$};
\node[panel, right=3cm of left] (right) {};
\node at (right.center) (vprime) {$\times$};
\node[above =2pt of right.center] { $V_{j,m}(z+\epsilon)$};

\draw[->, thick]
  (vprime) ++(1.2,0)
  arc[start angle=0, end angle=360, radius=1.2cm];

\node at ($(vprime)+(-1.5,0)$) {$L$};

\node at ($(vprime)+(2.4,0)$) (vdoubleprime) {$\times$};
\node[above =2pt of vdoubleprime] {$\Omega_w(z)$};
\draw[-, dotted, thick] ($(vdoubleprime.east)+(-0.2,0)$) -- ++(2cm,0);

\end{tikzpicture}
\caption{A spectrally-flowed vertex operator $V^w_{j,h}$ can be expressed in nucleation as the $\epsilon \rightarrow 0$ limit of the right hand panel. $L$ is the symmetry generator obtained by exponentiation of the contour integral of $ w J^3(z') \ln\left(z'-z\right) + c.c$. The branch cut of the logarithm is indicated with a dashed line running right.}
\label{fig:spec_flow_as_nucleation}
\end{figure}

For definiteness, consider the case where we are working with the level $k$ WZW model for $G=SL(2, \mathbb{R})$. We will work in the $m$-basis in this section for simplicity, but a $y$-basis generalisation is discussed in the next section. Recall that a spectrally-flowed $m$-basis vertex operator can be written as \eqref{eq:spec_flow_limiting_definition_m-basis}
\begin{equation}\label{eq: m basis defn of spectral flow}
    V^w_{j,h}(z) = \lim_{\epsilon\rightarrow 0} \epsilon^{mw} V_{j,m}(z+\epsilon) \Omega_w(z)\,,
\end{equation}
where $\Omega_w(z) = V^{w-1}_{\frac{k}{2},\frac{kw}{2}}(z)$. This spectrally-flowed vertex operator can be re-expressed via nucleation as
\begin{equation}\label{eq: nucleation 1}
    V^w_{j,h}(z) =\lim_{\epsilon \rightarrow 0}\exp \left(\frac{w}{2 \pi i}\oint_{|z'-z-\epsilon|=\epsilon/2} \rd z' \, J^3(z') \ln\left(z'-z\right)\right)V_{j,m}(z+\epsilon)\Omega_w(z)\,, 
\end{equation}
as can be checked by direct evaluation of the action of $J^3$ on $V_{j,m}$ --- see Figure \ref{fig:spec_flow_as_nucleation}.

\begin{figure}
\centering
    \begin{tikzpicture}[
  panel/.style={minimum width=4cm, minimum height=3cm},
  every node/.style={font=\small}
]

\node[panel] (left) {};
\node at (left.center)(v1) {$\times$};

\draw[->, thick]
  (v1) ++(0.6,0)
  arc[start angle=0, end angle=360, radius=0.6cm];

\node at ($(v1)+(-0.8,0)$) {$\gamma_1$};

\node at ($(v1)+(1.2,0)$) (v2) {$\times$};
\draw[-, dotted, thick] ($(v2.east)+(-0.2,0)$) -- ++(1cm,0);

\draw[<->, thick] ($(left.east)+(1,0)$) -- ++(1cm,0);
\node at ($(left.east)+(1.5,1.7)$) (arrowlabel){$\oint_{\gamma_1} \frac{\rd z'}{2 \pi i}\, J^3(z') \ln(z'-z) = \oint_{\gamma_2} \frac{\rd z'}{2 \pi i}\, J^3(z') \ln(z'-z) + \int_{\gamma_3} \rd z' \, J^3(z')$};
\node[panel, right=2cm of left] (right) {};
\node at (right.center) (vprime) {$\times$};

\node at ($(vprime)+(1.2,0)$) (vdoubleprime) {$\times$};

\draw[->, thick, line cap=round, line join=round]
  ($(vprime)+(-0.75,0.85)$)

  .. controls ($(vprime)+(-1.00,0.10)$)
       and ($(vprime)+(-0.90,-1.10)$) ..
  ($(vprime)+(-0.20,-1.35)$)

  .. controls ($(vprime)+(0.75,-1.60)$)
       and ($(vdoubleprime)+(1.05,-1.45)$) ..
  ($(vdoubleprime)+(1.40,-0.85)$)

  .. controls ($(vdoubleprime)+(1.70,-0.25)$)
       and ($(vdoubleprime)+(1.68,0.75)$) ..
  ($(vdoubleprime)+(1.25,1.05)$)

  .. controls ($(vdoubleprime)+(0.85,1.35)$)
       and ($(vdoubleprime)+(0.55,1.25)$) ..
  ($(vdoubleprime)+(0.45,0.85)$)

  .. controls ($(vdoubleprime)+(0.30,0.50)$)
       and ($(vdoubleprime)+(0.30,-0.35)$) ..
  ($(vdoubleprime)+(0.05,-0.60)$)

  .. controls ($(vdoubleprime)+(-0.25,-0.85)$)
       and ($(vdoubleprime)+(-0.55,-0.55)$) ..
  ($(vdoubleprime)+(-0.55,-0.05)$)

  .. controls ($(vdoubleprime)+(-0.55,0.45)$)
       and ($(vprime)+(-0.15,0.95)$) ..
  ($(vprime)+(-0.75,0.85)$);

  \draw[->,blue, very thick]
  ($(vdoubleprime)+(0.30,0)$)
  --
  ($(vdoubleprime)+(0.95,0)$);
    \draw[-,blue, very thick]
  ($(vdoubleprime)+(0.950,0)$)
  --
  ($(vdoubleprime)+(1.6,0)$);
\node at ($(vdoubleprime)+(1,0.3)$){$\textcolor{blue}{\gamma_3}$};

\node at ($(vprime)+(-1.2,0)$) {$\gamma_2$};

\draw[-, dotted, thick] ($(vdoubleprime.east)+(-0.2,0)$) -- ++(2cm,0);

\end{tikzpicture}
\caption{The closed contour used in the definition of $L$ can be smoothly deformed. When the contour passes through the branch cut of the logarithm in its integrand (indicated by the dashed line), it deposits a $J^3$ along the blue curve $\textcolor{blue}{\gamma_3}$ where the branch cut was crossed.}
\label{fig:Wilson_line}
\end{figure}
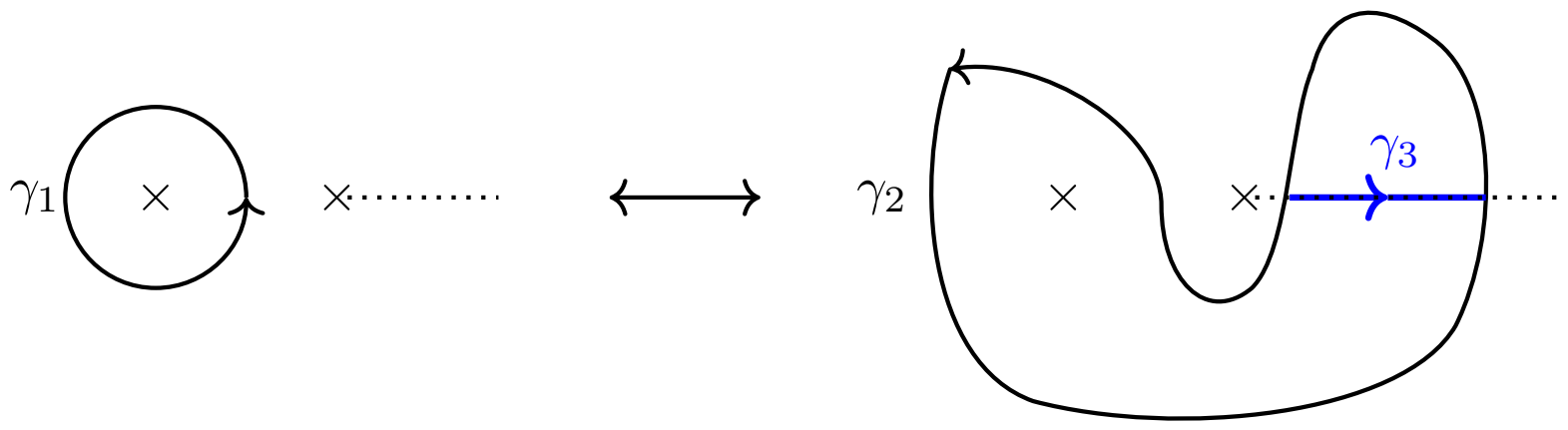

Note that the symmetry parameter $w \ln (z'-z)$ is not holomorphic, since acting with $\bar \partial $ generates a distribution localised to the branch cut of the logarithm. Using Stokes' theorem, this means that deforming the contour to cross the branch cut will lay down a Wilson line $\exp(\int \rd z'\, wJ^3(z'))$ along the segment of the branch cut that was crossed. Intuitively, the $2 \pi i$ jump in value of the logarithm along the branch cut means that the integrand in $L$ is not single-valued, so that pushing the contour of integration across the branch cut picks up an additional contribution of $w J^3$ lying on the segment of the branch cut that was crossed --- see Figure \ref{fig:Wilson_line}.

\begin{figure}
\centering
\begin{tikzpicture}[
  panel/.style={minimum width=4cm, minimum height=3cm},
  every node/.style={font=\small}
]

\node[panel] (left) {};
\node at (left.center) {$\times$};
\node[above left=2pt of left.center] {$V^w_{j,h}(z)$};

\draw[<->, thick] ($(left.east)-(1.5,0)$) -- ++(0.6cm,0);
\node at ($(left.east)+(0,0)$) (arrowlabel){$\lim_{\epsilon\rightarrow 0}$};

\node[panel, right=1cm of left] (right) {};

\node at ($(right.center)+(-0.9,0)$) (vprime) {$\times$};
\node[above=2pt of vprime] {$V_{j,m}(z+\epsilon)$};

\node at ($(right.center)+(0.9,0)$) (vdoubleprime) {$\times$};
\node[above=2pt of vdoubleprime] {$\Omega_w(u-\epsilon)$};

\draw[->, thick]
  ($(right.center)+(2.0,0.25)$)
  arc[
    start angle=0,
    end angle=360,
    x radius=2.0cm,
    y radius=1.65cm
  ];
\draw[dotted] ($(vprime)+(0.3,0)$) -- ($(vdoubleprime)+(-0.3,0)$);

\node at ($(right.center)+(-2.2,1)$) {$L$};

\node at ($(right.east)+(1,0)$) (arrowlabel){$=\lim_{\epsilon\rightarrow 0}$};
\node[panel, right=1.2cm of right] (third) {};

\node at ($(third.center)+(-0.9,0)$) (vthirdleft) {$\times$};
\node[above=2pt of vthirdleft] {$V_{j,m}(z+\epsilon)$};

\node at ($(third.center)+(0.9,0)$) (vthirdright) {$\times$};
\node[above=2pt of vthirdright] {$\Omega_w(u-\epsilon)$};

\draw[->, thick]
  ($(vthirdleft)+(0.2,0)$)
  arc[
    start angle=0,
    end angle=360,
    x radius=0.2cm,
    y radius=0.2cm
  ];

\draw[->, thick]
  ($(vthirdright)+(0.2,0)$)
  arc[
    start angle=0,
    end angle=360,
    x radius=0.2cm,
    y radius=0.2cm
  ];

\draw[->, solid] ($(vthirdleft)+(0.3,0)$) -- ($(vthirdright)+(-0.8,0)$);
\draw[solid] ($(vthirdright)+(-0.9,0)$) -- ($(vthirdright)+(-0.3,0)$);

\node at ($(vthirdleft)+(0,-0.5)$) {$l_1$};
\node at ($(vthirdright)+(0,-0.5)$) {$l_2$};
\node at ($(vthirdleft)+(1,-0.5)$) {$l_3$};

\end{tikzpicture}
\caption{A spectrally-flowed vertex operator $V^w_{jm}$ can also be expressed in nucleation as the $\epsilon \rightarrow 0$ limit of the middle panel rescaled by $(z-u)^{mw}$. $L=\exp\left(\frac{w}{2 \pi i}\oint_{\p U}\rd z' \, J^3(z') \ln\left(\frac{z'-z}{z'-u}\right)\right)$ is the symmetry generator, and the branch cut of the logarithm is indicated with a dashed line running from $z$ to $u$. 
In the third panel, we deformed the contour to envelop each individual insertion, i.e $L = \exp(l_1+l_2+l_3)$ where $l_1 = \frac{w}{2 \pi i}\oint_{z+\epsilon}\rd z' \, J^3(z') \ln\left(\frac{z'-z}{z'-u}\right)$, $l_2 = \frac{w}{2 \pi i}\oint_{u-\epsilon}\rd z' \, J^3(z') \ln\left(\frac{z'-z}{z'-u}\right)$, and $l_3 = w\int^u_z \rd z' \, J^3(z')$.}
\label{fig:dumbell_contour}
\end{figure}
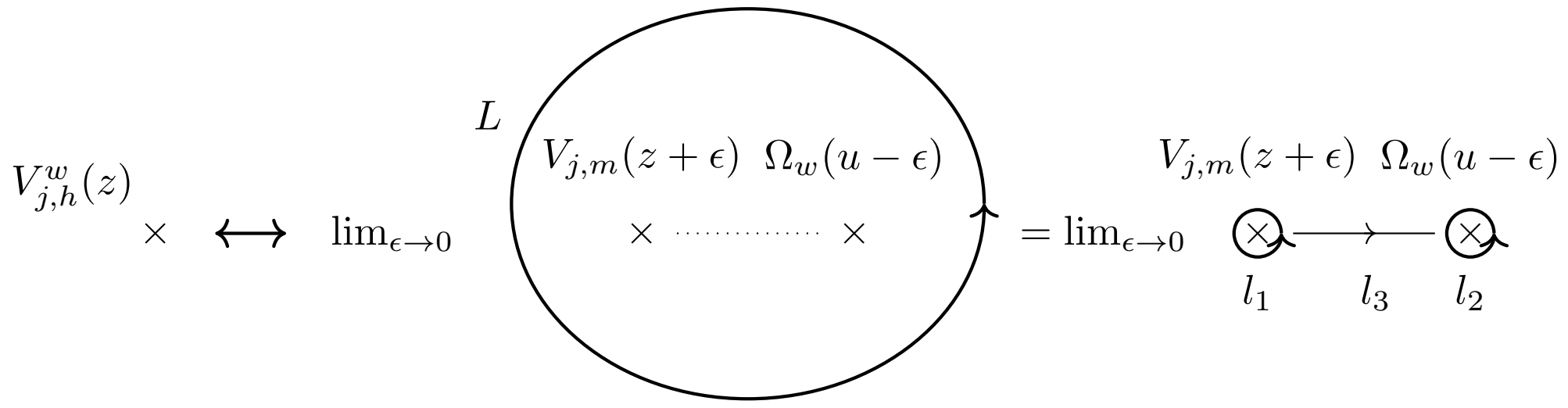

This nucleation is not directly useful for the computation of correlation functions, due to the singular behaviour as the contour of $L$ approaches $\Omega_w(z)$, as well as the fact that deforming $L$ to envelop the rest of the insertions will lay down Wilson lines along the branch cut. A more convenient alternative nucleation is
\begin{equation}\label{eq: nucleation 2}
   \boxed{ \begin{aligned}
        V^w_{j,h}(z) = \exp\left(\frac{w}{2\pi i}\oint_{\partial U} \rd z'\, J^3(z')\ln\left(\frac{z'-z}{z'-u}\right)\right)(z-u)^{m w} \\
        \times \lim_{\epsilon \rightarrow 0} V_{j,m}(z+\epsilon) \Omega_w(u-\epsilon)\,,
    \end{aligned}
        }
\end{equation}
where $u$ is an arbitrary point in $\Sigma$ (away from other insertion points) and the contour integral is taken around the boundary of $U$, a small neighbourhood that contains both $z$ and $u$, as well as the branch cut of the logarithm --- see Figure \ref{fig:dumbell_contour}. In this presentation, the role of $\epsilon$ is as a point-splitting regulator that determines the normalisation of the vertex operator.

In order to see how this works, we can shrink the contour around each vertex to obtain powers of the symmetry parameter evaluated at the vertices. The discontinuity of the logarithm at the branch cut additionally inserts a Wilson line. Recalling \eqref{eq:j,m_definition} and \eqref{eq:currents_acting_on_flowed_states}, we obtain
\begin{multline}\label{eq:verifying_m_basis_contours}
    V^w_{j,h}(z) = \lim_{\epsilon\rightarrow 0}\left( \left((z-u)^{mw}\left(\frac{\epsilon}{z+\epsilon-u}\right)^{mw}\left(\frac{u-\epsilon-z}{-\epsilon}\right)^{kw^2/2}\right)\right.
    \\
    \left.\exp\left(w\int_{z}^u\rd z' J^3(z')\right)V_{j,m}(z+\epsilon)\Omega_w(u-\epsilon)\right)\,.
\end{multline}
Since $J^3(z)\rd z$ is $\rd$-closed, it is locally exact and can be written as the holomorphic derivative of a scalar. This leads to the standard bosonisation in the parafermionic decomposition \eqref{eq:J3_bosonisation}.
The Wilson line can now be written as $\exp\left(w\sqrt{\frac{k}{2}}(-\phi(u)+\phi(z))\right)$. In terms of the field $\phi$, recall that $\Omega_w(u-\epsilon) = e^{w\sqrt{\frac{k}{2}}\phi(u-\epsilon)}$ as in \eqref{eq:Omega_definition}, which is a Wilson line end point (in the $\epsilon\to 0$ limit). Using the Baker-Campbell-Hausdorff (BCH) formula to combine the exponentials, we find that
\begin{multline}
    \exp\left({w\sqrt{\frac{k}{2}}(-\phi(u)+\phi(z))}\right) \exp\left({w\sqrt{\frac{k}{2}}\phi(u-\epsilon)}\right) 
    \\
    =\left(\frac{-\epsilon}{z+\epsilon-u}\right)^{kw^2/2} \exp\left({w\sqrt{\frac{k}{2}}(-\phi(u)+\phi(z)+\phi(u-\epsilon))}\right)\,.
\end{multline}
Substituting this back into the expression for $V^w_{j,h}(z)$, we recover
\begin{equation}\label{eq:verifying_m_basis_contours2}
    \begin{aligned}
             V^w_{j,h}(z) &= \lim_{\epsilon\rightarrow 0}\left( \left((z-u)^{mw}\left(\frac{\epsilon}{z+\epsilon-u}\right)^{mw}\times c.c.\right)\right.
    \\
    &\quad\left.\times \exp\left({w\sqrt{\frac{k}{2}}(-\phi(u)+\phi(z)+\phi(u-\epsilon))}\right)V_{j,m}(z+\epsilon)\right)
    \\
    &=\lim_{\epsilon\rightarrow 0} (\epsilon^{mw} \times c.c.) \left(e^{{w\sqrt{\frac{k}{2}}\phi(z)}}V_{jm}(z+\epsilon)+\mathcal{O}(\epsilon)\right)
    \\
    &=\lim_{\epsilon\rightarrow 0} (\epsilon^{mw} \times c.c.)V_{j,m}(z+\epsilon)\Omega_w(z)\,,
    \end{aligned}
\end{equation}
in agreement with \eqref{eq:spec_flow_limiting_definition_m-basis}. This verifies \eqref{eq: nucleation 2}. Note that an overall multiplicative minus sign cancels between the left- and right-moving sectors and for this reason we write $c.c.$ to denote the complex conjugate. Looking back at \eqref{eq: nucleation 2}, we can now understand why $(z-u)^{mw}\Omega_w(u-\epsilon)$ had to be included. It is a result of the Wilson line obtained from shrinking $L$ around the branch cut that effectively transports $\Omega_w$ from one endpoint to the other, i.e. from $u$ to $z$. The factor of $(z-u)^{mw}$ is then a consequence of the BCH formula.

We stress that the choice of $u$ in \eqref{eq: nucleation 2} is completely free, since the factor of $\Omega_w(u-\epsilon)$ gets precisely cancelled by the $u$-dependent terms generated by $L$.\footnote{In \cite{Giribet:2019new}, it was observed that if $V_{\frac{k}{2},\frac{k}{2}}^{w=-1}$ was inserted inside a correlation function with all states in the $m$-basis, then the correlator was independent of the location of $V_{\frac{k}{2},\frac{k}{2}}^{w=-1}$. Since $\Omega_{1}(z) = V_{\frac{k}{2},\frac{k}{2}}$, $V_{\frac{k}{2},\frac{k}{2}}^{w=-1}$ is the spectral flow operator flowed by $-1$ units. This is equivalent to our statement of $u$-independence at $w=1$, since the line operator in \eqref{eq: nucleation 2} effectively performs $-w$ units of spectral flow at $u$.} In particular, choosing $u = z+\epsilon$ directly recovers \eqref{eq: m basis defn of spectral flow}. This nucleation with the free $u$ parameter is well-suited for use in correlator computations since the symmetry parameter $w \ln \left(\frac{z'-z}{z'-u}\right)$ is holomorphic away from its branch cut, i.e outside of $U$, so that the line $L$ can be deformed around $\Sigma\setminus U$ straightforwardly.

\subsection{Symmetry generators for transition functions}\label{subsec: sym gen for trans fn}
Affine Kac-Moody symmetry operators (such as the line operators we use in nucleation) encode the data of transition functions of holomorphic $G$-bundles on the worldsheet $\Sigma$.

The prototypical example is of a $U(1)$ bundle on $\Sigma = S^2$. Suppose that a worldsheet field $\Phi(z, \bar z)$ on $\Sigma$ is charged, with holomorphic current $J(z)$,
\begin{equation}
    J(z) \Phi(0) \sim \frac{iq \Phi(0)}{z}\,.
\end{equation}
Inside the correlator
\begin{equation}
    \left \langle \exp \left(\frac{1}{2 \pi i}\oint_{\text{equator}}\rd z' \,\,\epsilon(z') J(z') \right) \cdots\right  \rangle\,,
\end{equation}
where the $\cdots$ represent any vertex operator insertions, we claim that $\Phi(z, \bar z)$ is no longer a function but should now be interpreted as a section of a holomorphic $U(1)$ bundle. With the standard two set open cover $U_0$ and $U_1$ being $S^2$ with the north and south poles removed, respectively, the transition functions of the bundle are implemented by the symmetry transformation as
\begin{equation}
    \begin{aligned}
        \Phi_{0}(z, \bar z) =  \exp \left(\frac{1}{2 \pi i}\oint_{z'=z}\rd z'\, \epsilon(z') J(z') \right)\Phi_{1}(z, \bar z) &= e^{iq\epsilon(z)}   \Phi_{1}(z, \bar z)\,, \\
        &\implies f_{01} = e^{iq \epsilon(z)}\,.
    \end{aligned}
\end{equation}
Therefore, if a correlator contains a symmetry generator wrapping the equator, passing a field $\Phi$ through the symmetry generator implements the transition function between the patches. The transition function $f_{01}$ between the patches $U_0, U_1$ is given by the exponentiation of the symmetry parameter. This gives an example of how the data of a line bundle is encoded in a network of symmetry operators. 

These networks are often called topological defect line networks \cite{Chang:2018iay}, due to the fact that the contours of integration of the symmetry operators only need to be defined up to homotopy, so long as they remain in the region where the symmetry parameter is holomorphic. The term "topological" is loaded in our context, however, because it implies null conditions with Virasoro generators, so we will avoid the name topological defect line.

More generally, let us consider the evaluation of a correlator in a theory whose fundamental fields $\Phi(z, \bar z)$ transform in the representation $R$ of $G$ under the action of the affine Kac-Moody currents for $\mathfrak{g}_k$. Given an open cover of the worldsheet $\bigcup_{i}U_i = \Sigma$, for each non-empty intersection $U_i\cap U_j$ a holomorphic transition function $f_{ij}$ valued in $G$ can be implemented by the action of the following symmetry generator
\begin{equation}
    \Phi_j(z, \bar z) = R[f_{ij}]\Phi_i(z, \bar z) = \exp \left(\frac{1}{2 \pi i}\oint_z \text{tr}((\ln f_{ij}) J)\right) \Phi_i(z, \bar z)\, , \quad \Phi_i := \Phi|_{U_i}\,.
\end{equation}
If we prepare a defect line network consisting of symmetry generators such that in order to pass from $U_i$ to $U_j$, a $\Phi$ must go through the symmetry generator that implements $f_{ij}$, this furnishes a realisation of the transition functions of a holomorphic vector bundle over $\Sigma$. In correlator evaluations with such a defect line network present, $\Phi$ no longer has a global description and should not be thought of as a function but rather as a section of a holomorphic vector bundle over $\Sigma$.

\subsection[Spectral flow as a $G$-bundle]{\boldmath Spectral flow as a $G$-bundle}\label{subsec: patches and vertex operators}

We have established a description of spectrally-flowed vertex operators in nucleation as a pair of vertex operators dressed by a symmetry generator --- see \eqref{eq: nucleation 2}. We have also seen that networks of symmetry generators in correlators can be understood as an instruction to promote worldsheet fields from $G$-valued functions to sections of $G$-bundles over the worldsheet. What remains to be explained is a precise characterisation of the bundles that are constructed in this way using spectrally-flowed insertions.

This is best shown with an illustrative example. Let us specialise to the case of interest, which is the WZW model at level $k$ with group $G=SL(2, \mathbb{R})$. Consider the evaluation of a correlator $\mathcal{H}$ that contains $n$ spectrally-flowed vertex operators in the $m$-basis, 
\begin{equation}
 \mathcal{H} = \left\langle \prod_{i=1}^n V^{w_i}_{j_i,h_i}(z_i) \right\rangle   \,.
\end{equation}
Now consider the nucleation as given in \eqref{eq: nucleation 2} and reproduced here
\begin{multline}\label{eq:m-basis-flowed-vertex-operator}
    V^{w_i}_{j_i,h_i}(z_i) = \exp\left(\frac{w_i}{2\pi i}\oint_{\partial U_i} \rd z'\, J^3(z')\ln\left(\frac{z'-z_i}{z'-u_i}\right)\right)(z_i-u_i)^{m_i w_i}  \\\times \lim_{\epsilon \rightarrow 0} V_{j_i,m_i}(z_i+\epsilon) \Omega_{w_i}(u_i-\epsilon)\,,
\end{multline}
where $U_i$ is a neighbourhood of the branch cut that contains $z_i+\epsilon$ and $u_i-\epsilon$. Via the dictionary explained in \S\ref{subsec: sym gen for trans fn}, these symmetry generators encode the data of transition functions of a bundle.

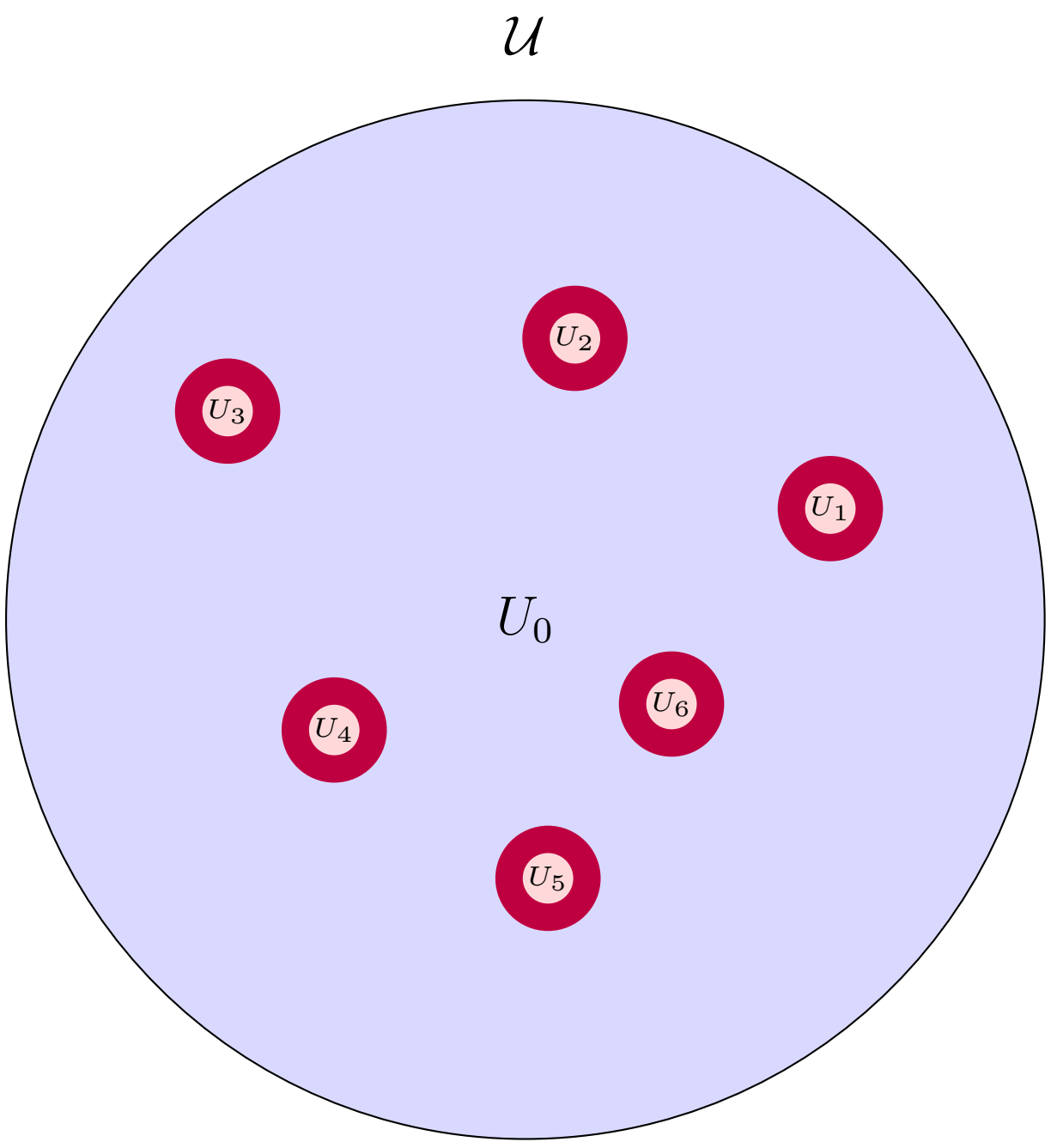
\begin{figure}
    \centering
    \begin{tikzpicture}[every node/.style={font=\small}]
  \def\R{4} 
  \draw[fill=blue!15] (0,0) circle (\R);
  \node[font=\large] at (0,\R+0.5) {$ \mathcal{U} $};
  \node[font=\large] at (0,0) {$ U_0 $};

  \foreach \ang/\rad/\r/\i in {
     20/2.5/0.3/1,
     80/2.2/0.3/2,
     145/2.8/0.3/3,
     210/1.7/0.3/4,
     275/2.0/0.3/5,
     330/1.3/0.3/6
  }{
     \path ({\rad*cos(\ang)},{\rad*sin(\ang)}) coordinate (c\i);
     \draw[draw=purple, line width=6pt, fill=red!15] (c\i) circle (\r);
     \node at (c\i) {\tiny $ U_{\i} $};
  }
\end{tikzpicture}
    \caption{An example of the case of six insertions. The red indicates the overlap regions $U_0 \cap U_i$.}
    \label{fig:covering_Sigma}
\end{figure}

The bundle may be described in the following way. Consider a simply-connected patch $\mathcal{U}$ of $\Sigma$ with the spectrally-flowed vertex operators at $z_i \in \mathcal{U}$. $E|_{\mathcal{U}}$, the restriction of $E$ from $\Sigma$ to $\mathcal{U}$, can be defined with respect to the cover
\begin{equation}
   U_0 \cup \bigcup_{i=1}^n U_i  = \mathcal{U}\,,
\end{equation}
where $U_0$ does not include any of the branch cuts of the $n$ logarithms $\ln \left(\frac{z'-z_i}{z'-u_i}\right)$, and the intersection $U_0 \cap U_i \neq \emptyset$ for all $i$. We sketch such a cover in Figure \ref{fig:covering_Sigma}. The transition functions $f_{0i}$ can be read off from the action of the symmetry generators on $g(z, \bar z)$. 
\begin{figure}
    \centering
\begin{tikzpicture}[
    contour/.style={
        thick,
        postaction={decorate},
        decoration={
            markings,
            mark=at position 0.15 with
                {\arrow{Stealth}}
        }
    },
    insertion/.style={circle,fill,inner sep=1.5pt},
    every node/.style={font=\large}
]

\begin{scope}[scale=1.7]
    \draw[contour] (0,0) circle (1);
    \node at (0,0) {$U_i$};

    \node[insertion,label=above:$g$] at (1.15,0.35) {};
\end{scope}

\node at (2.25,0) {$\rightarrow$};

\begin{scope}[shift={(4.5,0)},scale=1.7]
    \draw[contour]
        (47.5:1)
        arc[start angle=47.5,end angle=377.5,radius=1]

        .. controls (0.90,0.40) and (0.82,0.41) ..
        ({0.55+0.25*cos(-12.5)},
         {0.35+0.25*sin(-12.5)})

        arc[start angle=-12.5,end angle=-282.5,radius=0.25]

        .. controls (0.70,0.573) and (0.72,0.65) ..
        (47.5:1)
        -- cycle;
    \draw[thick,-{Stealth}]
    ({0.55+0.25*cos(-100)},
     {0.35+0.25*sin(-100)})
    arc[start angle=-100,end angle=-145,radius=0.25];
    \node at (0,0) {$U_i$};
    \node[insertion,label=above right:$\,g$]
        at (0.55,0.35) {};
\end{scope}

\node at (6.75,0) {$\rightarrow$};

\begin{scope}[shift={(9,0)},scale=1.7]
    \draw[contour] (0,0) circle (1);
    \node at (0,0) {$U_i$};

    \node[insertion] at (0.55,0.35) {};
    \node at (0.55,0.48) {$g$};

    \draw[contour]
        (0.55,0.35) ++(0:0.25)
        arc[start angle=0,end angle=-360,radius=0.25];
\end{scope}

\end{tikzpicture}
    \caption{The first and third panels show an insertion of $g(z, \bar z)$ where $z \notin U_i$ and $z \in U_i$ respectively, in a correlator evaluation where a line operator implementing a transition function between $U_0$ and $U_i$ is present. In the third panel, $g|_{U_i}$ is expressed in terms of $g|_{U_0}$ wrapped with a symmetry generator (note the reversed orientation). The second panel demonstrates how to think about the transition function as originating from "pushing" an insertion through a line operator, causing it to be wrapped by the line operator with reversed orientation.}
    \label{fig:pushing a g}
\end{figure}
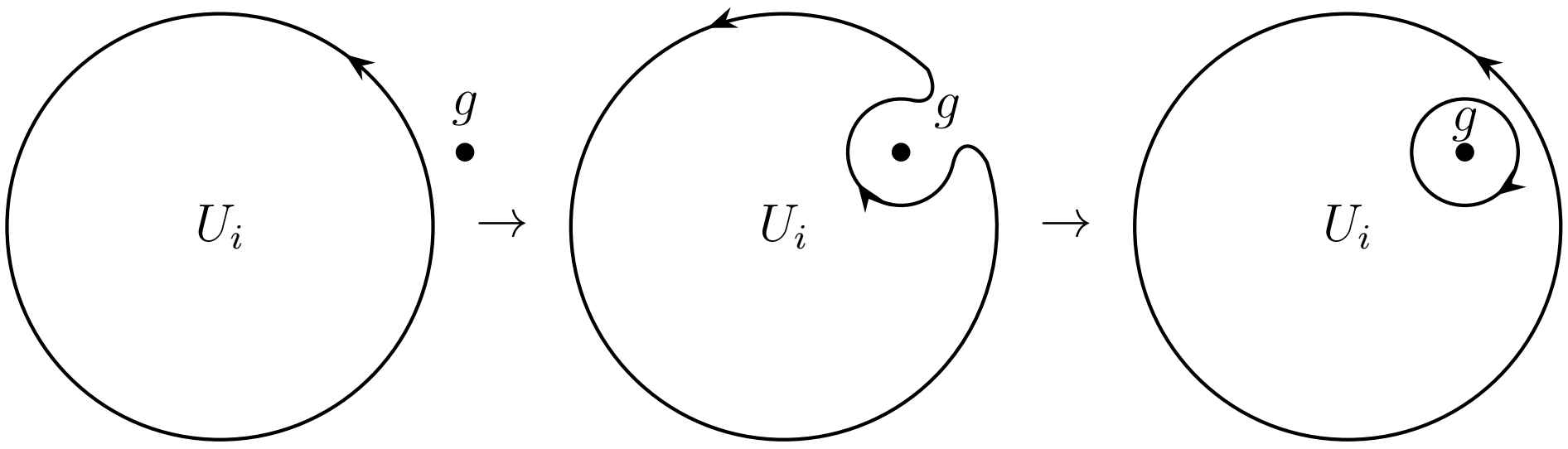
Pushing a $g(z,\bar z)$ insertion through the symmetry generator (see Figure \ref{fig:pushing a g}) associated to the nucleation of $V^{w_i}_{j_i,h_i}(z_i)$, we see that
\begin{equation}\label{eq:spectral_flow_transition_function}
    \begin{aligned}
        g|_{U_i}(z, \bar z) &= \exp\left(-\frac{w_i}{2\pi i}\oint_{z'=z} \rd z'\, J^3(z')\ln\left(\frac{z'-z_i}{z'-u_i}\right)+c.c.\right) g|_{U_0}(z, \bar z) 
    \\
    &= \exp\left(w_i\ln \left(\frac{z-z_i}{z-u_i}\right)t^3\right)g|_{U_0}(z, \bar z) \times (\exp\left(w_i\ln \left(\frac{\zbar-\zbar_i}{\zbar-\bar{u}_i}\right)t^3\right)
    \\
    &=\begin{pmatrix}
        \left(\frac{z-z_i}{z-u_i}\right)^{w_i/2} & 0 \\ 0 & \left(\frac{z-z_i}{z-u_i}\right)^{-w_i/2}
    \end{pmatrix}g|_{U_0}(z, \bar z) \begin{pmatrix}
        \left(\frac{\bar z-\bar z_i}{\bar z-\bar u_i}\right)^{w_i/2} & 0 \\ 0 & \left(\frac{\bar z-\bar z_i}{\bar z-\bar u_i}\right)^{-w_i/2}
    \end{pmatrix}\,.
    \end{aligned}
\end{equation}
Specifically, $g|_{U_i}$ is equivalent to pushing $g|_{U_0}$ through the line operator in \eqref{eq:m-basis-flowed-vertex-operator}, collecting a minus sign for the orientation of the contour. In the second line, we apply \eqref{eq:J_g_OPEs}. The final result for the transition function agrees with \eqref{eq:spectal_flow_automorphism}, except we see that we have also performed spectral flow by $-w_i$ units in a neighbourhood of $z=u$. This is consistent with the fact that the generalised spectral flow operator, $\Omega_w$, which we have inserted at $z=u$, is given by $w$ units of spectral flow acting on the identity.

A local form of the connection, $A$, can be directly constructed from the data of the transition functions. We gauge fix $A=0$ on $U_0$. Then we impose that
\begin{equation}
    t_{0i}\,\rd g|_{U_0}\,\bar{t}_{0i} = \left(\rd + A|_{U_i}\right)g|_{U_i}
\end{equation}
on the overlap, where $t_{0i}$ is the transition function defined by \eqref{eq:spectral_flow_transition_function} as $g|_{U_i} = t_{0i}\,g|_{U_0}\,\bar{t}_{0i}$. Substituting \eqref{eq:spectral_flow_transition_function}, we deduce
\begin{equation}\label{eq: connection with no x}
   (\rd + A)g = \begin{cases}
         \rd g - w_i t^3 g\left(\frac{1}{z-z_i}-\frac{1}{z-u_i}\right)\rd z - w_igt^3\left(\frac{1}{\zbar-\zbar_i}-\frac{1}{\zbar-\bar{u}_i}\right)\rd\zbar \,,  & z \in U_i\,,
         \\
         \rd g\,, & z \in U_0\,.
     \end{cases}
\end{equation}
On $\mathcal{U}$, in this gauge the curvatures are written as
\begin{equation}\label{eq: curvatures for m basis spectrally flowed vertex operators}
    F = \rd A + A \wedge A  = \sum_{i=1}^n 2 \pi i w_i (t^3_R -t^3_L) (\delta^{(2)}(z-z_i) - \delta^{(2)}(z-u_i))\rd^2 z\,,
\end{equation}
where we indicate the left or right multiplication by $t^3$ with the subscript $L/R$ and we use the convention $\rd^2z = \frac{i}{2}\rd z \wedge \rd \zbar$. For integer $w_i$ this is a locally single-valued flat $G$-bundle since\footnote{Parallel transport around $z=z_i$ is generated by the path ordered exponential $\mathrm{P}\exp\left(-\oint_{z=z_i}A\right)$.}
\begin{equation}
    g =e^{ 2 \pi i w_i (t^3_L -t^3_R)} g =  e^{2 \pi i w_i t^3} g e^{-2 \pi i w_i t^3} \,\, \forall g \iff w_i \in \mathbb{Z}\,.
\end{equation}
Note that the effective structure group \eqref{eq:effective_structure_group} in this case is $H = \frac{\widehat{SL}(2,\R) \times \widehat{SL}(2,\R)}{\Z}$, where we emphasise that we are working with the universal cover. The allowed values of the spectral flow are precisely the elements of the fundamental group $\pi_1(H) \cong \Z$ \cite{Maldacena:2000hw,Ribault:2014hia}.

This is the geometrisation of the definition of spectral flow. The conventional characterisation in terms of automorphisms of the mode algebra of $J^a_n$ can be directly recovered from the way the transition functions act on the currents. 
For instance, expressions like $J^a(z, \bar z) = k \text{tr}(t^a \partial g g^{-1})$ should not be thought of as globally defined, but should be transformed between local forms $J^a|_{U_i}$ according to the transition functions. Either by acting on them with the symmetry generators or by substituting in the results for $g$ into \eqref{eq:J_transformation}, we find
\begin{equation}
\begin{aligned}
    J^+|_{U_i}(z) &= \left(\frac{z-z_i}{z-u_i}\right)^{-w_i}J^+|_{U_0}(z)\,, 
    \\
    J^3|_{U_i}(z) &= J^3|_{U_0}(z) + \frac{kw_i}{2} \left(\frac{1}{z-z_i}-\frac{1}{z-u_i}\right)\,, 
    \\ 
    J^-|_{U_i}(z) &= \left(\frac{z-z_i}{z-u_i}\right)^{w_i} J^-|_{U_0}(z)\,.
\end{aligned}
\end{equation}
From this point of view, the mode automorphisms are a result of the transition functions between different local forms of $g|_{U_i}$ as we pass from one patch to another.\\

As a final note, it is important to be careful about operator fusion when attempting to read off the bundle structure encoded by the symmetry generators or when performing computations. Having written down the symmetry generators via nucleation, we can view a correlation function of spectrally-flowed vertex operators as a correlation function with $n$ line defects,
\begin{equation}
    \prod_{i=1}^n \exp\left(\frac{w_i}{2\pi i}\oint_{\partial U_i} \rd z'\, J^3(z')\ln\left(\frac{z'-z_i}{z'-u_i}\right)\right) \,.
\end{equation}
To put them all into the same exponent, one must carefully apply the BCH formula. For instance,
\begin{equation}\label{eq:joining_J3_contours}
    \begin{aligned}
        &\exp\left(\frac{w_i}{2\pi i}\oint_{\partial U_i} \rd z'\, J^3(z')\ln\left(\frac{z'-z_i}{z'-u_i}\right)\right)  \exp\left(\frac{w_j}{2\pi i}\oint_{\partial U_j} \rd z'\, J^3(z')\ln\left(\frac{z'-z_j}{z'-u_j}\right)\right) \\
     &=\left(\frac{(z_i-u_j)(z_j-u_i)}{(z_i-z_j)(u_i-u_j)}\right)^{kw_iw_j/2}
     \\
     &\quad\times \exp\left(\frac{w_i}{2\pi i}\oint_{\partial U_i} \rd z'\, J^3(z')\ln\left(\frac{z'-z_i}{z'-u_i}\right)+\frac{w_j}{2\pi i}\oint_{\partial U_j} \rd z'\, J^3(z')\ln\left(\frac{z'-z_j}{z'-u_j}\right)\right)  \,.
    \end{aligned}
\end{equation}
These sorts of manipulations will be very useful in \S\ref{sec:spectrally-flowed_correlators_AdS3}, allowing us to add and subtract integrands or compose contours freely. We can think of these prefactors as quantum corrections in the evaluation of correlators, since these symmetry generators classically commute.

\section{Spectrally-flowed correlators in AdS$_\mathbf{3}$}\label{sec:spectrally-flowed_correlators_AdS3}

It was established in the previous section that, when a spectrally-flowed vertex operator is inserted inside a correlation function, we can view this as a pair of simpler vertex operators dressed by a symmetry generator that is a line operator surrounding the two insertions. This can be applied quite generically to WZW models of a group $G$. As an illustrative example, we found that 
\begin{multline}
    V^w_{j,h}(z) = \exp\left(\frac{w}{2\pi i}\oint_{\partial U} \rd z'\, J^3(z')\ln\left(\frac{z'-z}{z'-u}\right)\right)(z-u)^{m w} \\\times \lim_{\epsilon \rightarrow 0} V_{j,m}(z+\epsilon) \Omega_w(u-\epsilon)\,,
\end{multline}
for $m$-basis vertex operators in the $SL(2,\R)$ WZW model at level $k$. $U$ is a region containing $z$, $u$ and the branch cut of the logarithm. We would now like to use this technology to constrain the form of correlation functions.

We will be primarily concerned with the evaluation of sphere correlators with spectrally-flowed vertex operators inserted in the $y$-basis since, by comparing \eqref{eq:x_basis_spec_flow_limiting_defn} and \eqref{eq:y_basis_spec_flow_limiting_defn}, it is clear that spectral flow has a simpler description in the $y$-basis. Via the integral transform \eqref{eq:y_basis_Mellin_transform}, these encode the $x$-basis expressions.

In this setting, we will demonstrate that every $n$-point correlator of $y$-basis vertex operators $V^{w_i}_{j_i}(y_i;x_i;z_i)$ can be transformed into an $(n+1)$-point correlator of $n$ unflowed $x$-basis vertex operators and one generalised spectral flow operator \eqref{eq:Omega_definition}, $\Omega_{W}$ for $W = \sum w_i$, inserted at an auxiliary point $u\in \Sigma\backslash\{z_i\}$. We can write this result as
\begin{equation}\label{eq: sketch canonical correlator}
    \left\langle \prod_{i=1}^n V^{w_i}_{j_i}(y_i;x_i;z_i) \right\rangle = \left| \mathcal{P} \right|^2\left\langle \Omega_W(\gamma(z);u) \prod_{i=1}^n V_{j_i}({\gamma}(z_i);z_i)\right\rangle\,,
\end{equation}
for a meromorphic function $\mathcal{P}$ and where $\Omega_W$ and $V_{j_i}$ on the right hand side of (\ref{eq: sketch canonical correlator}) have been dressed by a function we denote by ${\gamma}(z)$.

In order to set up the language for this result, in \S\ref{subsec: x(z) basis} we introduce the $x(z)$-basis for spectrally-flowed vertex operators, which can be understood as a natural generalisation of the $x$- and $y$-bases. This allows us to generalise \eqref{eq: nucleation 2} to the $y$-basis. In \S\ref{sec:canonical_form_of_a_correlator}, we explain how to unwrap the line operators around the rest of the worldsheet to prove \eqref{eq: sketch canonical correlator}. As will be explained, in the $y$-basis it is easiest to nucleate $n$-tuples of spectrally-flowed vertex operators instead of nucleating each spectrally-flowed vertex operator individually. This provides a two-patch description for the $G$-bundle. In simpler language, \S\ref{sec:canonical_form_of_a_correlator} explains a prescription for joining together all of the symmetry generators into one line operator which can readily be contracted around infinity.

\subsection[$x(z)$-basis]{\boldmath $x(z)$-basis}\label{subsec: x(z) basis}

Recall that, for an unflowed operator in the $x$-basis,
\begin{equation}
     V_{j}(x;z) = \exp\left(\frac{1}{2 \pi i}\oint_{z'=z} \rd z'\, x J^+(z')\right) V_{j}(0;z) \,.
\end{equation}
This example can be used to illustrate an important point about nucleation, which is that the line operator $L$ that relates a certain $V$ and $V'$ is not unique if $V'$ is known to be annihilated by some modes of the Kac-Moody currents. For instance, if we were to replace $x$ with any meromorphic function $x(z')$ that is regular in a neighbourhood of $z$ and that satisfies
\begin{equation}
    x(z') = x +\mathcal{O}((z'-z))\,, 
\end{equation}
then the terms of order $\mathcal{O}((z'-z)^1)$ do not contribute to the evaluation because $V_{j}(0;z)$ is a Kac-Moody primary and annihilated by positive modes of $J^+$. Hence,
\begin{equation}
     \exp\left(\frac{1}{2 \pi i}\oint_{z'=z} \rd z' \, x(z') J^+(z')\right) V_{j}(0;z) =  V_{j}(x;z)\,,
\end{equation}
and we see that wrapping an unflowed vertex operator $V_j(0;z)$ with $x(z') J^+(z')$ is only sensitive to the constant term in the near-$z$ expansion of $x(z')$.

When it comes to wrapping spectrally-flowed vertex operators with exponentials of line integrals of $x(z') J^+(z')$, this story will be modified. Since $V^w_{j,h}(0;z)$ is not annihilated by $J^+_1,\ldots , J^+_{w}$, this means that a $w$-spectrally-flowed vertex operator will be sensitive to terms in the near-$z$ expansion of $x(z')$ up to the term of order $(z-z')^{w+1}$. With this in mind, let us introduce the $x(z)$-basis.

\paragraph{\boldmath $x(z)$-basis vertex operator:} Given a meromorphic function $x(z)$ that is non-singular in a neighbourhood of $z_i$, a vertex operator $V^w_j(x(z);z_i)$ is defined as
\begin{equation}\label{eq: defn of x(z) basis}
    V^w_j(x(z);z_i) := \exp\left(\frac{1}{2 \pi i}\oint_{z=z_i} \rd z \,x(z) J^+(z)\right)V^w_{j}(0;z_i)\,. 
\end{equation}
If the expansion coefficients of $x(z)$ around $z_i$ are labelled $\alpha_r$,
\begin{equation}\label{eq:x(z)_expansion}
    x(z) = \alpha_0 + \alpha_1(z-z_i) + \ldots +\alpha_w(z-z_i)^w \quad\quad \text{mod } (z-z_i)^{w+1}\,,
\end{equation}
then the vertex operator $V^w_j(x(z);z_i)$ only depends on the set $\{\alpha_0, \ldots ,\alpha_{w}\}$. Indeed, one can write this as
\begin{equation}\label{eq:x(z)_as_conjugation}
    V^w_j(x(z);z_i) = e^{\sum_{r=0}^w \alpha_r J^+_r}V^w_j(0;z_i)e^{-\sum_{r=0}^w \alpha_r J^+_r}\,.
\end{equation}
We see that $\alpha_0=x$ and $\alpha_w=y$ from the $x$- and $y$-bases, such that we recover the $y$-basis by setting all $\alpha_r=0$ for $1\leq r \leq w-1$. We refer to this as the $x(z)$-basis\footnote{The use of "basis" could appear to be slightly misleading in the following sense. The $x$-basis vertex operators $V^w_{j,h}(x;z)$ provide a basis of spectrally-flowed affine primaries, translated by $x$. The $y$-basis vertex operators $V^w_j(y;x;z)$ are a basis of the same set of spectrally-flowed affine primaries by \eqref{eq:y_basis_transform} and \eqref{eq:y_basis_Mellin_transform}. The $x(z)$-basis therefore parameterises a larger set of vertex operators, including spectrally-flowed affine descendants. For instance, differentiating $V^w_j(x(z);z_i)$ with respect to $\alpha_1$, for a discrete representation with $w\geq 2$, and then setting all $\alpha_r=0$ returns $[J^+_1,V^w_j(0;z_i)]$, which corresponds to the state $[J^+_{1-w}|j,j\rangle]^w$.} because it is equivalent to replacing the constant $x$ by any function $x(z)$ satisfying \eqref{eq:x(z)_expansion}.

It is important to stress that there is a great deal of freedom in the choice of $x(z)$ in $V^w(x(z); z_i)$ since we are always free to add arbitrary meromorphic functions that have a zero of order $w+1$ at $z_i$. That is to say, for any holomorphic $f(z)$ and $g(z)$ such that $g(z_i)\neq 0$,
\begin{equation}
V^{w}_{j}(x(z); z_i) = V^{w}_{j}\left(x(z) + (z-z_i)^{w+1}\frac{f(z)}{g(z)}; z_i\right)\,.
\end{equation}
This defines an equivalence class of meromorphic functions. In practice, the choice of $x(z)$ from this equivalence class should be made with a view towards simplifying the correlator evaluation when we replace vertex operators in nucleation. For instance, it would be counterproductive to add terms where $(z-z_i)^{w+1}f(z)/g(z)$ is singular at infinity, because we would then have to deal with modes of Kac-Moody currents acting at infinity when unwrapping the symmetry generator around the rest of the worldsheet.\\

With these definitions and equivalences in mind, we are ready to supply the analogous nucleation to \eqref{eq: nucleation 2} for the $x(z)$-basis vertex operators. Substituting \eqref{eq: nucleation 2} into \eqref{eq: defn of x(z) basis}, for $x(z)$ meromorphic with no singularities inside $\partial U_i$ (a region containing $z_i$, $u_i$ and the branch cut of the logarithm) we have
\begin{multline}\label{eq:x(z)_basis_separate_contours}
    V^{w_i}_{j_i}(x(z);z_i) = \exp\left(\frac{1}{2 \pi i}\oint_{\partial U_i^+} \rd z \,x(z) J^+(z)\right)
    \\ \times \exp\left(\frac{w_i}{2\pi i}\oint_{\partial U_i} \rd z\, J^3(z)\ln\left(\frac{z-z_i}{z-u_i}\right)\right)(z_i-u_i)^{j_i w_i} \\\times \lim_{\epsilon \rightarrow 0} V_{j_i}(0;z_i+\epsilon) \Omega_{w_i}(u_i-\epsilon)\,,
\end{multline}
where $\partial U^+_i$ is a slightly larger contour that encircles $\partial U_i$. Combining the exponentials, this simplifies to
\begin{multline}\label{eq: x(z) basis nucleation}
    V^{w_i}_{j_i}(x(z);z_i)=\exp\left(\frac{w_i}{2\pi i}\oint_{\partial U_i} \rd z\, (J^3(z)-x(z)J^+(z))\ln\left(\frac{z-z_i}{z-u_i}\right)\right) 
    \\\times(z_i-u_i)^{j_i w_i}  \lim_{\epsilon \rightarrow 0} V_{j_i}(x(z_i+\epsilon);z_i+\epsilon)\Omega_{w_i}(x(z);u_i-\epsilon)\,.
\end{multline}
Here, we have used the fact that the unflowed vertex operator $V_{j_i}$ depends only on the value of $x(z)$ at $z=z_i+\epsilon$. Meanwhile, we have translated $\Omega_{w_i}(z) = V^{w_i-1}_{\frac{k}{2},\frac{kw_i}{2}}(z)$ into the $x(z)$-basis. In other words, it has been conjugated by $\sum_{r=0}^{w_i-1}\alpha_r J^+_r$, where $\alpha_s$ are the coefficients of the first $w_i$ terms in the expansion of $x(z)$ around $u_i-\epsilon$. This reduces to a $y$-basis vertex operator specifically in the case where the near-$z_i$ expansion of $x(z)$ is
\begin{equation}
    x(z) = x_i + y_i(z-z_i)^{w_i} \quad \text{mod } (z-z_i)^{w_i+1}\,,
\end{equation}
provided that $x(z)$ is a meromorphic function with no singularities in $U_i$, a neighbourhood of $z_i$ and $u_i$.

Away from the poles of $x(z)$, the action of the symmetry generator on $g(z,\bar z)$ from $U_i$ being pushed into $U_0$ can be evaluated as follows,
\begin{equation}
    \begin{aligned}
            g|_{U_0}(z, \bar z)&=\exp\left(\frac{w_i}{2\pi i}\oint_{z'=z} \rd z'\, (J^3(z')-x(z')J^+(z'))\ln\left(\frac{z'-z_i}{z'-u_i}\right) \right) g|_{U_i}(z, \bar z)
    \\
    &=\exp\left(-w_i \ln \left(\frac{z-z_i}{z-u_i}\right)(t^3-x(z)t^+)\right) g|_{U_i}(z,\bar z)
     \\
    &=\begin{pmatrix}
        1 & x(z) \\ 0 & 1
    \end{pmatrix}\exp\left(-w_i \ln \left(\frac{z-z_i}{z-u_i}\right)t^3\right)\begin{pmatrix}
        1 & -x(z) \\ 0 & 1
    \end{pmatrix} g|_{U_i}(z,\bar z)
    \\
    &=\begin{pmatrix}
        \left(\frac{z-u_i}{z-z_i}\right)^{w_i/2} & x(z)\left(\left(\frac{z-z_i}{z-u_i}\right)^{w_i/2}-\left(\frac{z-u_i}{z-z_i}\right)^{w_i/2}\right)
        \\
         0 & \left(\frac{z-z_i}{z-u_i}\right)^{w_i/2}
    \end{pmatrix}g|_{U_i}(z, \bar z)\,.
    \end{aligned}
\end{equation}
Together with the antiholomorphic contribution, i.e right multiplication by the Hermitian conjugate of the matrix we just computed, this gives us the transition function of $g$.

When we compute correlators, $x(z)$ must be chosen carefully in \eqref{eq:x(z)_basis_separate_contours}, because when we eventually try to unwrap the contour integral containing $x(z)J^+(z)$ around the rest of $\Sigma$, we will receive contributions from the poles of $x(z)$. We shall choose $x(z)$ to have only simple poles and for it to be regular at $z=\infty$, since this will simplify the calculations.

\subsection{The canonical form of a spectrally-flowed correlator}\label{sec:canonical_form_of_a_correlator}

We now would like to derive what we refer to as the canonical form of a spectrally-flowed correlator. We will show how our definition of spectral flow, in terms of line operators wrapping an unflowed operator and a generalised spectral flow operator, provides a novel method for computing spectrally-flowed correlators in WZW models. We study the particular example of $G = SL(2,\R)$ here, but our methods naturally generalise to other groups. We will begin in \S\ref{sec:m_basis_correlators} with a warm-up example of an $n$-point function in the $m$-basis. The full set of physical correlators requires working in the $x$- or $y$-basis, and we perform this calculation in \S\ref{sec:y_basis_correlators}. In each case, we demonstrate that an $n$-point correlator of spectrally-flowed vertex operators can be rewritten as a meromorphic prefactor multiplying an $(n+1)$-point correlator where $n$ of the vertex operators are unflowed, and one generalised spectral flow operator, $\Omega_W$, living in the $W-1$ spectrally-flowed sector is inserted, where $W =\sum w_i$ is the sum of the number of units of spectral flow in the original correlator.

\subsubsection{\boldmath Correlators in the $m$-basis}\label{sec:m_basis_correlators}

We would like to compute the $n$-point function
\begin{equation}
    \left\langle \prod_{i=1}^n V^{w_i}_{j_i,h_i}(z_i) \right\rangle\,.
\end{equation}
We shall write only the holomorphic part of the correlation function in order to simplify the expressions and we will make use of the contour integral representation \eqref{eq: nucleation 2},
\begin{equation}
\begin{split}
    V^{w_i}_{j_i,h_i}(z_i) &= \exp\left(\frac{w_i}{2\pi i}\oint_{\p U_i} \rd zJ^3(z)\ln\left(\frac{z-z_i}{z-u_i}\right)\right) (z_i-u_i)^{m_i w_i}
    \\
    &\quad\times \lim_{\epsilon \rightarrow 0}V_{j_i,m_i}(z_i+\epsilon) \Omega_{w_i}(u_i-\epsilon)\,.
\end{split}
\end{equation}
There are two ways to proceed. The first is to work iteratively, by first considering the insertion at $z=z_1$. We can reverse the orientation of the contour $\p U_i$ and take it to enclose all other vertex operators at $z_i$ for $i=2, \dots, n$. This leads to contributions such as
\begin{equation}
    \exp\left(-\frac{w_1}{2\pi i}\oint_{z=z_i} \rd zJ^3(z)\ln\left(\frac{z-z_1}{z-u_1}\right)\right) V^{w_i}_{j_i,h_i}(z_i) = \left( \frac{z_i-u_1}{z_i-z_1} \right)^{w_1h_i}V^{w_i}_{j_i,h_i}(z_i)\,.
\end{equation}
One can repeat this process by next reversing the orientation of the contour $\p U_2$ and so on, until the contribution from all $n$ line operators has been computed. In the $m$-basis, this calculation is very simple to do and leads to the correct answer for the correlation function. However, this iterative method is vastly more complicated in the $y$-basis and therefore, with a view to that calculation, let us apply an alternative method.
\\

A more efficient way to calculate the correlator is to join all of the line operators together. That is, we begin with a network of $n$ line operators wrapping regions $U_i$ that enclose the points $z_i$, $u_i$ and a branch cut between them. We would now like to combine all of these into one line operator that encloses the region we called $\mathcal{U}$ in \S\ref{sec:spec_flow_G_bundle}, which contains all $z_i$, $u_i$ and branch cuts. On the sphere, we can take $\mathcal{U}$ to be a patch of $\Sigma$ that omits the point at $z=\infty$.\footnote{Without loss of generality, we assume that no vertex operator is inserted at $z=\infty$.} This is equivalent to defining the $G$-bundle with respect to a two-patch cover.

First define
\begin{equation}\label{eq:R(z)}
    R(z) = \prod_{i=1}^n \left( \frac{z-z_i}{z-u_i} \right)^{w_i}\,.
\end{equation}
Following exactly the same logic as in \eqref{eq:verifying_m_basis_contours}-\eqref{eq:verifying_m_basis_contours2} and keeping only the leading order terms in $\epsilon$, one can show
\begin{equation}\label{eq:m_basis_R(z)_contours}
\begin{aligned}
    &\exp\left(\frac{1}{2\pi i}\oint_{\p U_i} \rd z \,J^3(z)\ln R(z)\right) (z_i-u_i)^{m_i w_i}
    \\
    &\quad\times \lim_{\epsilon \rightarrow 0}V_{j_i,m_i}(z_i+\epsilon) \Omega_{w_i}(u_i-\epsilon) \\
    &= \prod_{j\neq i}\left( \frac{z_i-z_j}{z_i-u_j} \right)^{m_iw_j} \left( \frac{u_i-z_j}{u_i-u_j} \right)^{\frac{kw_iw_j}{2}} \lim_{\epsilon\to 0}\epsilon^{m_iw_i} V_{j_i,m_i}(z_i+\epsilon)\Omega_{w_i}(z_i)\\
    &= \prod_{j\neq i}\left( \frac{z_i-z_j}{z_i-u_j} \right)^{m_iw_j} \left( \frac{u_i-z_j}{u_i-u_j} \right)^{\frac{kw_iw_j}{2}} V^{w_i}_{j_i,h_i}(z_i)\,.
\end{aligned}
\end{equation}
Since the line operator above is now the same for each $i=1, \dots, n$, they can be combined into one contribution. However, as was pointed out in \eqref{eq:joining_J3_contours}, we must carefully apply the BCH formula when we do this:\footnote{In principle, we could have introduced different point-splitting regulators, $\epsilon_i$, for each vertex. Since it has no effect overall, we have chosen to use the same $\epsilon$ for each spectrally-flowed vertex operator.}
\begin{equation}
    \begin{aligned}
        &\left\langle \prod_{i=1}^n V^{w_i}_{j_i,h_i}(z_i) \right\rangle \\
        &= \prod_{i=1}^n (z_i-u_i)^{m_iw_i}\prod_{ i<j}\left( z_i-z_j\right)^{-m_iw_j-m_jw_i} (u_i-u_j)^{kw_iw_j} \prod_{i\neq j} \left( z_i-u_j\right)^{m_iw_j-\frac{kw_iw_j}{2}}\\
        &\quad \times \lim_{\epsilon\to 0} \left\langle \prod_{i=1}^n \exp\left(\frac{1}{2\pi i}\oint_{\p U_i} \rd zJ^3(z)\ln R(z)\right) V_{j_i,m_i}(z_i+\epsilon)\Omega_{w_i}(u_i-\epsilon) \right\rangle\\
        &=\prod_{i,j=1}^n (z_i-u_j)^{m_iw_j}\prod_{ i<j}\left( z_i-z_j\right)^{-m_iw_j-m_jw_i-\frac{kw_iw_j}{2}} (u_i-u_j)^{\frac{kw_iw_j}{2}} \\
        &\quad \times \lim_{\epsilon\to 0} \left\langle  \exp\left(\frac{1}{2\pi i}\oint_{\p \mathcal{U}} \rd zJ^3(z)\ln R(z)\right) \prod_{i=1}^nV_{j_i,m_i}(z_i+\epsilon)\Omega_{w_i}(u_i-\epsilon) \right\rangle\,,
    \end{aligned}
\end{equation}
where we have used
\begin{equation}
\begin{aligned}\label{eq: unifying j3 exponentials}
    &\prod_{i=1}^n \exp\left(\frac{1}{2\pi i}\oint_{\p U_i} \rd zJ^3(z)\ln R(z)\right) \\
    &= \prod_{i<j} \left( \frac{(z_i-u_j)(z_j-u_i)}{(z_i-z_j)(u_i-u_j)} \right)^{\frac{kw_iw_j}{2}} \exp\left(\frac{1}{2\pi i}\oint_{\p \mathcal{U}} \rd zJ^3(z)\ln R(z)\right)\,.
\end{aligned}
\end{equation}
Note that replacing $w_i\ln\left(\frac{z-z_i}{z-u_i}\right)$ by $\ln R(z)$ has no effect on the formula \eqref{eq:joining_J3_contours}, because the non-trivial BCH contribution comes from the neighbourhood of the logarithmic branch cuts contained in $U_i$ and $U_j$.

Since we now only have one line operator, we can close it off at infinity. $J^3(z)$ acts trivially on the vacuum, so the line operator contributes exactly $1$. The $\epsilon\to 0$ limit is now completely regular, with no OPEs to be computed. This implies
\begin{equation}\label{eq:x=0_correlator}
    \begin{split}
        \left\langle \prod_{i=1}^n V^{w_i}_{j_i,h_i}(z_i) \right\rangle &= \prod_{i,j=1}^n(z_i-u_j)^{m_iw_j} \prod_{i<j} (z_i-z_j)^{-m_iw_j-m_jw_i-\frac{kw_iw_j}{2}} (u_i-u_j)^{\frac{kw_iw_j}{2}}\\
        &\quad \times \left\langle \prod_{i=1}^n V_{j_i,m_i}(z_i) \Omega_{w_i}(u_i)  \right\rangle\,.
    \end{split}
\end{equation}
Here, we have not been careful to keep track of the overall phase and have freely exchanged $(z_i-z_j)$ with $(z_j-z_i)$. In the case that $m_i=\bar{m}_i$, this phase will cancel with the anti-holomorphic sector.

We can now revisit a claim made at the end of \S\ref{subsec: nucleation}, which is that $u_i$ is only an auxiliary variable and that our definition of spectral flow does not depend on the specific choice. This can be verified in \eqref{eq:x=0_correlator} explicitly. Working in the parafermionic decomposition where $\Omega_{w_i}(u_i) = e^{w_i\sqrt{\frac{k}{2}}\phi(u_i)}$ and $V_{j_i,m_i}(z_i) = \Psi_{j_i,m_i}e^{m_i\sqrt{\frac{2}{k}}\phi(z_i)}$, the OPE between $\Omega_{w_i}(u_i)$ and any other vertex operator exactly cancels with the $u_i$-dependent part of the pre-factor. We need only consider the $U(1)$ part of the correlator, since the parafermionic contribution at $u_i$ is locally the identity operator. This agrees with the claim made in \cite{Giribet:2019new} that for $m$-basis correlators, the position at which the spectral flow operator is inserted does not matter. Similarly, \eqref{eq:x=0_correlator} agrees precisely with (2.23) of \cite{Ribault:2005ms} in the spectral flow conserving case.

Since there is no dependence on the $u_i$, we can set them to whatever value we choose. In fact, we shall generically take the coincidence limit $u_i\to u$,
\begin{equation}\label{eq:m_basis_canonical_form}
    \begin{aligned}
        \left\langle \prod_{i=1}^n V^{w_i}_{j_i,h_i}(z_i) \right\rangle &= \prod_{i=1}^n (z_i-u)^{m_iW} \prod_{i<j} (z_i-z_j)^{-m_iw_j-m_jw_i-\frac{kw_iw_j}{2}} \\
        &\quad \times \left\langle \Omega_W(u)\prod_{i=1}^n V_{j_i,m_i}(z_i)  \right\rangle\,,
    \end{aligned}
\end{equation}
where $W = \sum_{i=1}^n w_i$. If we further let $u\to \infty$, then we deduce that the unflowed correlator can be thought of as carrying a non-trivial background charge. Via the state-operator correspondence, the insertion of a charged state $\Omega_W(u)$ can be related to a charged out vacuum as $u\to \infty$ via \cite{DiFrancesco:1997nk}
\begin{equation}
    \left\langle -\frac{kW}{2}\right| := \lim_{u\to \infty} u^{2\Delta_W}\langle 0|\Omega_{W}(u)\,,
\end{equation}
where $\Delta_W = \Delta(\Omega_W) = -\frac{kW^2}{4}$. $J^3_0$ charge conservation of the correlator on the RHS of \eqref{eq:m_basis_canonical_form} imposes that $\sum_{i=1}^nm_i = -\frac{kW}{2}$ and hence,
\begin{equation}\label{eq:x=0_n-point_solution}
\boxed{
    \begin{aligned}
        \left\langle \prod_{i=1}^n V^{w_i}_{j_i,h_i}(z_i) \right\rangle = \prod_{i<j} (z_i-z_j)^{-m_iw_j-m_jw_i-\frac{kw_iw_j}{2}}&(\zbar_i-\zbar_j)^{-\bar{m}_iw_j-\bar{m}_jw_i-\frac{kw_iw_j}{2}} \\
        &\times \left\langle \prod_{i=1}^n V_{j_i,m_i}(z_i)  \right\rangle_{-\frac{kW}{2}}\,,
    \end{aligned}
    }
\end{equation}
where we reintroduce the anti-holomorphic dependence and the subscript is to emphasise the charged out vacuum.

\subsubsection{\boldmath Correlators in the $y$-basis}\label{sec:y_basis_correlators}

Let us now determine the canonical form of a spectrally-flowed correlator in the $y$-basis. We will follow a similar methodology to \S\ref{sec:m_basis_correlators}. Our starting point is \eqref{eq:x(z)_basis_separate_contours} and \eqref{eq: x(z) basis nucleation}. Here and in the remainder of the paper, $y_i$ is only permitted to be non-zero if $w_i>0$. This allows us to simultaneously consider cases where some of the vertex operators are unflowed. For any meromorphic function $x(z)$ that is non-singular in $U_i$ and satisfies
\begin{equation}\label{eq: pseudo covering map}
    x(z) = x_i + y_i(z-z_i)^{w_i} \quad \text{mod } (z-z_i)^{w_i+1}
\end{equation}
for all $i=1,\dots, n$, then
\begin{multline}\label{eq:spectral_flow_in_y_basis}
    \begin{aligned}
        V^{w_i}_{j_i}(y_i;x_i;z_i) &= \exp\left(\frac{1}{2 \pi i}\oint_{\partial U_i^+} \rd z \,x(z) J^+(z)\right)
    \\ &\quad \times \exp\left(\frac{w_i}{2\pi i}\oint_{\partial U_i} \rd z\, J^3(z)\ln\left(\frac{z-z_i}{z-u_i}\right)\right) \\
    &\quad \times (z_i-u_i)^{j_i w_i}\lim_{\epsilon \rightarrow 0} V_{j_i}(0;z_i+\epsilon) \Omega_{w_i}(u_i-\epsilon) \\
    &= \exp\left(\frac{w_i}{2\pi i}\oint_{\partial U^+_i} \rd z\, (J^3(z) - x(z)J^+(z)) \ln\left(\frac{z-z_i}{z-u_i}\right) \right) 
    \\
    &\quad \times(z_i-u_i)^{j_i w_i} \lim_{\epsilon \rightarrow 0} V_{j_i}(x(z_i+\epsilon);z_i+\epsilon) \Omega_{w_i}(x(z);u_i-\epsilon)\,.
    \end{aligned}
\end{multline}
Our aim is to combine these line operators, which each wrap $U_i$, into one line operator around $\mathcal{U}$, a region containing all $z_i$, $u_i$ and branch cuts but no poles of $x(z)$. The reason we can do this is because $x_i$ and $y_i$ are encoded by $x(z)$ and its derivatives at $z_i$, and there is a lot of freedom in how we choose $x(z)$. Indeed, if we parameterise $x(z)$ as the ratio of two coprime polynomials $p_1(z)/p_2(z)$, for $\deg p_1 +\deg p_2  \geq \sum_{i=1}^n w_i + n+1$, then it is always possible to construct a $p_1(z)/p_2(z)$ that obeys \eqref{eq: pseudo covering map}. We will also assume without loss of generality that $z=\infty$ is a regular point of $x(z)$ such that $\deg p_1 \leq \deg p_2$ and moreover that $x(z)$ only contains simple poles (i.e. that $p_2(z)$ has simple roots).\footnote{To see why this is always possible, note that we can fix the location of the poles arbitrarily, evaluate $\partial^nx(z)$, and then solve for the coefficients of the numerator polynomial via a linear system. The degree of the numerator polynomial is bounded by the number of denominator poles, but by choosing to increase the number of poles we can always ensure that there are sufficiently many degrees of freedom that the linear system can be solved.} We therefore parameterise $x(z)$ by its value at infinity, $x_{\infty}$, the location of its $m$-many simple poles, $\lambda_a$, and their residues $c_a$,
\begin{equation}\label{eq:gamma_defn}
x(z) = x_{\infty} + \sum_{a=1}^m \frac{c_a}{z-\lambda_a}   \,. \end{equation}
It is simple to combine the $n$ line operators appearing in the first line of \eqref{eq:spectral_flow_in_y_basis} since they all have the same integrand,
\begin{equation}
    \begin{aligned}
        \prod_{i=1}^n V^{w_i}_{j_i}(y_i;x_i;z_i) &= \prod_{i=1}^n V^{w_i}_{j_i}(x_i + y_i(z-z_i)^{w_i};z_i)\\
        &= \exp\left(\frac{1}{2\pi i}\oint_{\p \mathcal{U}} \rd z \,x(z) J^+(z)\right) \prod_{i=1}^n V^{w_i}_{j_i}(0;z_i)\\
        &= \exp\left(\frac{1}{2\pi i}\oint_{\p \mathcal{U}} \rd z \,x(z) J^+(z)\right)\\
        &\quad \times \prod_{i=1}^n\exp\left(\frac{w_i}{2\pi i}\oint_{\partial U_i} \rd z\, J^3(z)\ln\left(\frac{z-z_i}{z-u_i}\right)\right) \\
    &\quad \times (z_i-u_i)^{j_i w_i}\lim_{\epsilon \rightarrow 0} V_{j_i}(0;z_i+\epsilon) \Omega_{w_i}(u_i-\epsilon)\,.
    \end{aligned}
\end{equation}
In order to combine the $n$ line operators associated to $J^3$, we first need them all to have the same integrand. We will once again use the function $R(z)$ from \eqref{eq:R(z)} as we did in the $m$-basis. As in \eqref{eq:m_basis_R(z)_contours},
\begin{equation}
    \begin{aligned}
        &\prod_{i=1}^n V^{w_i}_{j_i}(y_i;x_i;z_i)\\ &=\exp\left(\frac{1}{2\pi i}\oint_{\p \mathcal{U}} \rd z \,x(z) J^+(z)\right) \prod_{i=1}^n \exp\left(\frac{1}{2\pi i}\oint_{\partial U_i} \rd z\, J^3(z)\ln R(z)\right)\\
        &\quad \times (z_i-u_i)^{j_i w_i} \prod_{j\neq i}\left( \frac{z_i-u_j}{z_i-z_j} \right)^{j_iw_j} \left( \frac{u_i-u_j}{u_i-z_j} \right)^{\frac{kw_iw_j}{2}} \lim_{\epsilon \rightarrow 0} V_{j_i}(0;z_i+\epsilon) \Omega_{w_i}(u_i-\epsilon)\,.
    \end{aligned}
\end{equation}
Now the line operators associated to $J^3$ can be put into the same exponent and pulled outside the product, picking up powers of cross-ratios just as we did in \eqref{eq: unifying j3 exponentials}. Up to an overall phase that will cancel with the anti-holomorphic sector, the result is
\begin{equation}
    \begin{aligned}
        &\prod_{i=1}^n V^{w_i}_{j_i}(y_i;x_i;z_i)\\ &=\exp\left(\frac{1}{2\pi i}\oint_{\p \mathcal{U}^+} \rd z \,x(z) J^+(z)\right) \exp\left(\frac{1}{2\pi i}\oint_{\partial \mathcal{U}} \rd z\, J^3(z)\ln R(z)\right)\\
        &\quad \times \prod_{i, j}(z_i-u_j)^{j_iw_j}\prod_{i<j}(z_i-z_j)^{-j_iw_j-j_jw_i-\frac{kw_iw_j}{2}} (u_i-u_j)^{\frac{kw_iw_j}{2}} \\
        &\quad \times \prod_{i=1}^n \lim_{\epsilon \rightarrow 0} V_{j_i}(0;z_i+\epsilon) \Omega_{w_i}(u_i-\epsilon)\,,
    \end{aligned}
\end{equation}
where $\p \mathcal{U}^+$ is a contour that lies just outside of $\p \mathcal{U}$. 

Now, let us insert the above inside a correlation function
\begin{equation}
    \mathcal{H} = \left\langle \prod_{i=1}^n V^{w_i}_{j_i}(y_i;x_i;z_i)  \right\rangle\,.
\end{equation}
The first step is to shrink the contour of $e^{\frac{1}{2\pi i}\left(\oint_{\p \mathcal{U}^+} \rd z \, x(z) J^+(z)\right)}$ around $\Sigma \setminus \mathcal{U}$. Since $J^+(z)$ is regular in this region, including at $z=\infty$, we will collect contributions only from the poles of $x(z)$. This means we may make the replacement
\begin{equation}
    \exp\left(\frac{1}{2\pi i}\oint_{\p \mathcal{U}^+} \rd z \, x(z) J^+(z)\right)  = \exp\left(-\sum_{a=1}^m c_a J^+(\lambda_a)\right)\,,
\end{equation}
where we use \eqref{eq:gamma_defn} and the minus sign follows from reversing the orientation of the contour. Next, we must shrink the contour associated to $J^3$ around $\Sigma\setminus \mathcal{U}$. There is no pole at infinity in the symmetry parameter, so the only contribution comes from commutators with $J^+(\lambda_a)$:
\begin{equation}
    \begin{aligned}
        &\exp\left(-\sum_{a=1}^m c_a J^+(\lambda_a)\right)\exp\left(\frac{1}{2\pi i}\oint_{\partial \mathcal{U}} \rd z\, J^3(z)\ln R(z)\right) \\
        &= \exp\left(\frac{1}{2\pi i}\oint_{\Gamma} \rd z\, J^3(z)\ln R(z)\right)\exp\left(-\sum_{a=1}^m \frac{c_a}{R(\lambda_a)} J^+(\lambda_a)\right)\,,
    \end{aligned}
\end{equation}
where $\Gamma$ is a contour that surrounds all $z_i$, $u_i$, $\lambda_a$ and branch cuts. The contour $\Gamma$ can now be shrunk around infinity, where it returns zero. We deduce
\begin{multline}
    \mathcal{H} = \prod_{i, j}(z_i-u_j)^{j_iw_j}\prod_{i<j}(z_i-z_j)^{-j_iw_j-j_jw_i-\frac{kw_iw_j}{2}} (u_i-u_j)^{\frac{kw_iw_j}{2}}  \\
   \left\langle \exp\left(-\sum_{a=1}^m \frac{c_a}{R(\lambda_a)} J^+(\lambda_a)\right)
     \prod_{i=1}^n \lim_{\epsilon\to 0} V_{j_i}(0;z_i+\epsilon) \Omega_{w_i}(u_i-\epsilon) \right\rangle
\end{multline}
Recall that the $\epsilon\rightarrow 0$ limit was introduced to get the correct normalisation for spectrally-flowed operators in the point-splitting between the symmetry generator with $J^3 \ln(\ldots)$ and the vertex operators in nucleation. Now that the $J^3 \ln(\ldots)$ symmetry generator is gone (as was also the case in the $m$-basis) the limit can be taken without any hesitation, leaving us with
\begin{multline}
    \mathcal{H} = \prod_{i, j}(z_i-u_j)^{j_iw_j}\prod_{i<j}(z_i-z_j)^{-j_iw_j-j_jw_i-\frac{kw_iw_j}{2}} (u_i-u_j)^{\frac{kw_iw_j}{2}}  \\
   \left\langle \exp\left(-\sum_{a=1}^m \frac{c_a}{R(\lambda_a)} J^+(\lambda_a)\right)
      \prod_{i=1}^n  V_{j_i}(0;z_i) \Omega_{w_i}(u_i) \right\rangle\,.
\end{multline}
From the arguments of \S\ref{subsec: nucleation} we know that we can choose each $u_i$ to be any value we wish, so let us again take the coincident limit $u_i \rightarrow u$ as we did in \S\ref{sec:m_basis_correlators}. We find
\begin{multline}
    \mathcal{H} = \prod_{i=1}^n(z_i-u)^{j_iW}\prod_{i<j}(z_i-z_j)^{-j_iw_j-j_jw_i-\frac{kw_iw_j}{2}}  \\
   \left\langle \exp\left(-\sum_{a=1}^m \frac{c_a}{R(\lambda_a)} J^+(\lambda_a)\right) \Omega_{W}(u)
      \prod_{i=1}^n  V_{j_i}(0;z_i)  \right\rangle\,,
\end{multline}
where $W := \sum_{i=1}^n w_i$.

Our final step is to now evaluate the one remaining line operator. Defining\footnote{Let us emphasise that $\gamma(z)$ is not to be identified with an embedding coordinate of $H_3^+$ in Poincar\'{e} coordinates. For us, it is simply a meromorphic function without a direct target space interpretation.}
\begin{equation}\label{eq:gamma_tilde_defn}
    \gamma(z) := x_{\infty} + \sum_{a=1}^m \frac{c_a/R(\lambda_a)}{z-\lambda_a}
\end{equation}
we can write
\begin{equation}
    \exp\left(-\sum_{a=1}^m \frac{c_a}{R(\lambda_a)}J^+(\lambda_a)\right) = \exp \left( \frac{1}{2\pi i} \oint_{\p \mathcal{U}} \rd z \, \gamma(z)J^+(z) \right)\,,
\end{equation}
where, as before, the minus sign follows from the change in orientation. Acting on all of the vertex operators, it will place the $V_{j_i}$ in the $x$-basis and $\Omega_W$ in the $x(z)$-basis. Reintroducing the anti-holomorphic sector, the final result is
\begin{equation}\label{eq:n_point_x_basis_formula}
    \boxed{\begin{aligned}
        &\left\langle \prod_{i=1}^n V^{w_i}_{j_i}(y_i;x_i;z_i) \right\rangle =\\
        &\prod_{i=1}^n|z_i-u|^{2j_iW}\prod_{i<j}|z_i-z_j|^{-2j_iw_j-2j_jw_i-kw_iw_j} \left\langle \Omega_{W}({\gamma}(z);u)\prod_{i=1}^n  V_{j_i}({\gamma}(z_i);z_i)  \right\rangle
    \end{aligned}}
\end{equation}
where $W = \sum_{i=1}^n w_i$. Let us recall the various definitions here. $\Omega_W$ is defined in \eqref{eq:Omega_definition}, the $x(z)$-basis is defined in \eqref{eq: defn of x(z) basis} and the function ${\gamma}(z)$ is defined using \eqref{eq: pseudo covering map}, \eqref{eq:gamma_defn} and \eqref{eq:gamma_tilde_defn}. We also elaborate on ${\gamma}(z)$ below. \eqref{eq:n_point_x_basis_formula} is the canonical form of a spectrally-flowed correlator. It can be put back into the $x$-basis by performing the integrals from \eqref{eq:y_basis_Mellin_transform}. It says that any correlator of spectrally-flowed vertex operators in the $y$-basis can be recast as a meromorphic function times an $(n+1)$-point function of $n$ unflowed operators and one degenerate operator, a generalised spectral flow operator in the $x(z)$-basis. Similarly to \eqref{eq:x=0_n-point_solution}, we can think of this as an unflowed correlator in a non-trivial background.\\

To improve our understanding of the above result, it will be useful to be more explicit about $\gamma(z)$, because with this presentation the $x_i$ and $y_i$ dependence of the correlator is somewhat hidden. Consider the meromorphic function
\begin{equation}
   \frac{x(z)}{R(z)}=\frac{x(z) (z-u)^W}{\prod_{i=1}^n(z-z_i)^{w_i}} = \left(x_{\infty}+\sum_{a=1}^m \frac{c_a}{z-\lambda_a}\right)\frac{(z-u)^W}{\prod_{i=1}^n(z-z_i)^{w_i}}\,.
\end{equation}
The singularities are simple poles at $\lambda_a$ and order $w_i$ poles at each $z_i$. Its residues around each $\lambda_a$ fully agree with the residues of $\gamma(z)$ around each $\lambda_a$. Now define
\begin{equation}
    p_i(z) = \text{PP}_{z=z_i} \left(\frac{(z-u)^W}{\prod_{i=1}^n(z-z_i)^{w_i}}\right)\,,
\end{equation}
where PP indicates the principal part,\footnote{In \S\ref{sec:2-point_functions} we give an explicit example of this construction for 2-point functions.} i.e the singular terms in the Laurent expansion around $z_i$,
\begin{multline}
    p_i(z) = \frac{1}{(z-z_i)^{w_i}}\frac{(z_i-u)^W}{\prod_{j\neq i}(z_i-z_j)^{w_j}} + \frac{1}{(z-z_i)^{w_i-1}} \frac{\partial}{\partial z_i}\left(\frac{(z_i-u)^W}{\prod_{j\neq i}(z_i-z_j)^{w_j}}\right) + \ldots 
    \\ +\frac{1}{z-z_i} \frac{1}{(w_i-1)!} \frac{\partial^{w_i-1}}{\partial z_i^{w_i-1}}\left(\frac{(z_i-u)^W}{\prod_{j\neq i}(z_i-z_j)^{w_j}}\right)\,.
\end{multline}
Then $ \gamma(z)$ is explicitly given by
\begin{equation}\label{eq:gamma_explicit_realisation}
     \gamma(z) = \frac{x(z) (z-u)^W}{\prod_{i=1}^n(z-z_i)^{w_i}}-\sum_{i=1}^n x_i p_i(z)\,.
\end{equation}
By construction, the right hand side only has poles at the $\lambda_a$ with the same residues as ${\gamma}(z)$. Since $x(z) = x_i + y_i(z-z_i)^{w_i} + \dots$, any potential poles at $z_i$ exactly cancel. Moreover, by comparing \eqref{eq:gamma_defn} and \eqref{eq:gamma_tilde_defn}, we see that we have equality at $z=\infty$.

We know that \eqref{eq:n_point_x_basis_formula} only depends on the values ${\gamma}(z_i)$ and the coefficients in the Taylor expansion of ${\gamma}(z)$ near $z=u$ up to order $W-1$. These can be readily expressed in terms of the original data $\{y_i,x_i,z_i,w_i\}$ of the spectrally-flowed correlator as well as the auxiliary variable $u$. The Taylor expansion coefficients are given by expanding
\begin{equation}
    {\gamma}(z) \equiv -\sum_{i=1}^n x_i p_i(z) \quad \text{mod } (z-u)^{W}\,,
\end{equation}
so that the $ \gamma(z)$ argument of $\Omega_W$ can be replaced by $-\sum_{i=1}^n x_i p_i(z)$. To evaluate ${\gamma}(z_i)$, we must use the identity
\begin{equation}
    R(z)^{-1} = \frac{(z-u)^W}{\prod_{i=1}^n (z-z_i)^{w_i}} = 1 + \sum_{i=1}^n p_i(z)\,,
\end{equation}
which again follows by matching all singularities and the constant at $z=\infty$. Then, we determine that
\begin{equation}\label{eq:gamma_tilde_at_zi}
    \begin{split}
        {\gamma}(z_i) &= \lim_{z\to z_i}\left( \frac{x(z)}{R(z)} - x_ip_i(z) \right) - \sum_{j\neq i}x_jp_j(z_i) \\
        &= \lim_{z\to z_i}\left( (x_i + y_i(z-z_i)^{w_i} + \dots)\left(1+\sum_{j=1}^n p_j(z)\right) - x_ip_i(z) \right) - \sum_{j\neq i}x_jp_j(z_i) \\
        &= x_i + \sum_{j\neq i}(x_i-x_j)p_j(z_i) + y_i\frac{(z_i-u)^W}{\prod_{j\neq i} (z_i-z_j)^{w_j} }\,.
    \end{split}
\end{equation}

\subsection{Instanton background}\label{sec:instanton_background}

We have demonstrated that generic $n$-point spectrally-flowed correlators in the $G=SL(2,\mathbb{R})$ WZW model at level $k$ can be written as a meromorphic prefactor multiplying an $(n+1)$-point correlator, where $n$ of the insertions are unflowed vertex operators and there is one generalised spectral flow operator, $\Omega_W$. As we have mentioned already, one should interpret this result as an $n$-point unflowed correlator in a non-trivial background which, specifically, we would like to think of as an instanton background.\footnote{See \cite{Nakatsu:1993np} for a similar claim in the context of topologically twisted, gauged WZW models.}

The instantonic interpretation of the $(n+1)$-point correlator is easiest to see in the $m$-basis calculation as in \eqref{eq:x=0_n-point_solution}, where the statement of the canonical form reads
\begin{equation}\label{eq:m_basis_instanton}
    \begin{aligned}
        \left\langle \prod_{i=1}^n V^{w_i}_{j_i,h_i}(z_i) \right\rangle &= \prod_{i=1}^n (z_i-u)^{m_iW} \prod_{i<j} (z_i-z_j)^{-m_iw_j-m_jw_i-\frac{kw_iw_j}{2}} \\
        &\quad \times \left\langle e^{W \sqrt\frac{k}{2}\phi(u) }\prod_{i=1}^n V_{j_i,m_i}(z_i)  \right\rangle\,
        \\
        & = \prod_{i<j} (z_i-z_j)^{-m_iw_j-m_jw_i-\frac{kw_iw_j}{2}} \left\langle \prod_{i=1}^n V_{j_i,m_i}(z_i)  \right\rangle_{-\frac{kW}{2}}\,,
    \end{aligned}
\end{equation}
where we write only the holomorphic sector and in the first equality we have replaced $\Omega_W(u)$ with an equivalent expression through bosonisation. We use independence on the choice of $u$ to set $u=\infty$ in the final line.

To see why this can be called an instanton background, consider the following toy model in which it is well-established that a sum over instanton number arises. We consider a theory of charged symplectic bosons $Y_A$ and $Z^A$ for $A \in 1, \ldots, 4$ with $Y_A \in K(\mathbb{CP}^1)$, the canonical bundle over $\C\P^1$, and a gauged $GL(1)$ symmetry current,\footnote{Strictly speaking, one should also include fermions to ensure that this gauged $GL(1)$ symmetry is not anomalous. This does not affect our discussion.}
\begin{equation}
    S = \int_{\mathbb{CP}^1} \rd^2 z \,\left(Y_A\bar \partial Z^A + a (Y_A Z^A)\right)\,.
\end{equation}
These are the left-movers of the bosonic matter content of the closed string formulation of the Berkovits-Witten twistor string \cite{Witten:2003nn,Berkovits:2004hg}. The instanton number $W \in\mathbb{Z}$ is conventionally introduced (see, for instance, \S7.2 of \cite{Cachazo:2005ga} ) as a twisting of the line bundle on the worldsheet in which $Y_A$ and $Z^A$ take values by a holomorphic line bundle $\mathcal{L} \cong \mathcal{O}(W)$,
\begin{equation}
\begin{aligned}
    Y_A \in K \quad \quad &\xrightarrow{\text{twist}} \quad \quad Y_A \in K \otimes \mathcal{L}^{-1}\,,
    \\
    Z^A \in \mathcal{O} \quad \quad &\xrightarrow{\text{twist}} \quad \quad Z^A \in \mathcal{L}\,,
\end{aligned}
\end{equation}
where $\mathcal{O}$ is the trivial bundle over $\C\P^1$ and $K$ is the canonical bundle. The covariant derivative acting on $Z^A$ has a connection term whose curvature integrated over ${\mathbb{CP}^1}$ evaluates to $W$. Working on the patch of $\mathbb{CP}^1$ where $z \neq u$ for some reference point $u$, we can choose a gauge for the connection such that the twisting is implemented by the following covariant derivative
\begin{equation}
    Y_A \bar \partial Z^A \rightarrow Y_A \bar D Z^A = Y_A \left(\bar \partial+ \frac{W}{\pi}\bar \partial \ln(z-u)\right) Z^A\,,
\end{equation}
where the location of $u$ is a gauge choice and $\frac{W}{\pi}\p\pbar \ln (z-u) = W\delta^{(2)}(z-u)$.

However, this twisting can alternatively be interpreted as turning on a background for the gauge field that sources a background charge for the $GL(1)$ current in the following way:
\begin{equation}
    a \rightarrow a + \frac{W}{\pi }\bar \partial \ln (z-u)\,.
\end{equation}
We choose to adopt this perspective, that the instanton number refers to the non-triviality of the $GL(1)$ bundle that $Y_A$ and $Z^B$ are valued in. Specifically, the background gauge field now has a non-zero Chern class
\begin{equation}
    \int_{\mathbb{CP}^1} \rd^2z\, \left(\frac{W}{\pi}\p\bar \partial \ln(z-u)\right) = W\,.
\end{equation}
Defining the bosonisation of the $GL(1)$ current as $Y_A Z^A = -\sqrt{\frac{k}{2}}\partial \phi$ (where $k=8$ in this example), we see that the twisted action can be rewritten as
\begin{equation}
    \begin{aligned}
        \exp(S_{\text{twisted}}) &= \exp\left(\int \rd^2 z \,\, Y_A \left(\bar \partial + a + \frac{W}{\pi}\bar \partial \ln (z-u)\right) Z^A \right) 
    \\
    &=\exp\left(\int \rd^2 z \, \left[ Y_A \left(\bar \partial + a\right) Z^A +  W\sqrt{\frac{k}{2}} \delta^2(z-u) \phi(z) \right]\right)
    \\
    &= \exp\left(S_{\text{untwisted}} + W\sqrt{\frac{k}{2}} \phi(u)\right)\,,
    \end{aligned}
\end{equation}
where in the second line we performed an integration by parts on $\p\phi$ and substituted $\frac{1}{\pi}\p\pbar\ln(z-u) =\delta^{(2)}(z-u)$. In this way we see that introducing a source for the scalar $\phi$ that bosonises a $GL(1)$ current, as we did in \eqref{eq:m_basis_instanton}, is equivalent to working in an instanton background for the fields charged under the $GL(1)$. This is the sense in which we choose to interpret the $(n+1)$-point function as an $n$-point function of unflowed vertex operators in an instanton background, where the curvature of the bundle is concentrated at $u$. In this context, one can think of $\Omega_W(u)$ as an instanton number changing operator \cite{siegel2004untwistingtwistorsuperstring}.

\section{Computing correlators at genus zero}\label{sec: computing correlators}

In \S\ref{sec:spectrally-flowed_correlators_AdS3}, we established our key result that any $g=0$ $n$-point function of spectrally-flowed vertex operators in the (analytically continued) $SL(2,\R)$ WZW model can be rewritten as the product of a meromorphic function times an $(n+1)$-point function consisting of $n$ unflowed vertex operators and one generalised spectral flow operator, $\Omega_W$, sitting in the $W-1=\sum_{i=1}^nw_i - 1$ spectrally-flowed sector. This gave the interpretation that spectrally-flowed correlation functions are equivalent to unflowed correlation functions in an instanton background.

We would now like to take this one step further and begin to argue that the $(n+1)$-point function can be reduced to an unflowed $n$-point function, which we denote by $\mathcal{F}.$ Since the correlation functions of the unflowed sector have been solved \cite{Teschner:1997fv,Teschner:1997ft,Teschner:1999ug,Teschner:2001gi}, this gives a pathway towards solving the full $SL(2,\R)$ WZW model. The similarities between flowed 2- and 3-point functions and their unflowed counterparts were noted in \cite{Dei:2021xgh}. It was also conjectured in \cite{Dei:2021yom} that the spectrally-flowed 4-point function could be written in terms of an unflowed 4-point function. We will not fully prove our claim here and show how $(n+1)$-point functions containing $\Omega_W$ can be recast as $n$-point functions. Instead, we will provide the next step in this direction.

Specifically, we will show that the correlation function
\begin{equation}
    \mathcal{G} = \left\langle \Omega_W(\myx(z);u)\prod_{i=1}^n V_{j_i}(X_i;z_i) \right\rangle\,,
\end{equation}
for $W\in 2\Z$ and any $\myx(z)$ can be written as a certain meromorphic function in $z_i,X_i, u$ and $\alpha_r$ (where $\alpha_r$ are the Taylor coefficients of $\myx(z)$ around $z=u$) multiplying a function $\tilde{\mathcal{G}}$ --- see \eqref{eq:H_to_G_tilde}. For an appropriate choice of spacetime coordinates that we denote by $\mathscr{X}_i$, $\tilde{\mathcal{G}}$ is constrained to satisfy the same differential constraints as those imposed on an unflowed $n$-point function by spacetime global conformal invariance. Significantly, if our conjecture that the spectrally-flowed $n$-point function can be written in terms of an unflowed $n$-point function is correct, then this constitutes a derivation of the spacetime cross-ratios that appear in the unflowed $n$-point function (for $W\in 2\mathbb{Z}$). Our derivation is valid under the assumption that certain polynomials known as Pad\'{e} approximants exist --- see \S\ref{sec:spacetime_global_conf_inv}.

We have chosen the name $\tilde{\mathcal{G}}$ here, rather than $\mathcal{F}$, which we save for our conjectural unflowed $n$-point function. After analysing spacetime global conformal invariance as described above, if one can additionally show that $\tilde{\mathcal{G}}$ is a meromorphic function (that is independent of $\sx_i$) times some function $\mathcal{F}$ which satisfies the expected constraints on an unflowed correlator from global worldsheet conformal invariance and (for $n\geq 4$) the KZ equations, then $\mathcal{F}$ is guaranteed to live in the space of conformal blocks of the unflowed $n$-point functions. These additional constraints will be analysed in \cite{McStay:2026xxx}. For now, after studying spacetime conformal invariance for $n$-point functions with $W\in 2\mathbb{Z}$, we content ourselves with analysing global worldsheet conformal invariance only in the case of the 2-point function. For generic operator insertions (i.e. away from series identifications) such correlation functions are known to be non-vanishing only for $w_1=w_2$ and, hence, $W\in 2\mathbb{Z}$ is guaranteed.\footnote{For $n$-point functions, it is sufficient to initially study the case of $W\in 2\mathbb{Z}$, since correlation functions with $W\in 2\Z + 1$ can then be constructed using the series identifications.}

Ultimately, our analysis will provide an alternative derivation of the flowed 2-point function by extracting $\mathcal{F}$. Our derivation is, unequivocally, a computationally inefficient method for computing this 2-point function and shorter derivations have appeared elsewhere \cite{Maldacena:2001km,Dei:2021xgh}. Nevertheless, our calculation will serve two purposes. Firstly, it will verify our claims in earlier sections and show that our methods can accurately be applied to compute spectrally-flowed correlation functions. Secondly, and more importantly, our method of computation is constructive. In \cite{Dei:2021xgh}, the presence of the unflowed correlator inside flowed 2- and 3-point functions was observed a posteriori. For the 4-point case in \cite{Dei:2021yom}, the unflowed correlator was observed through examples when studying the Ward, KZ and null equations. By contrast, the methods developed here and in \cite{McStay:2026xxx} will analytically extract the meromorphic function multiplying $\mathcal{F}$ and find appropriate coordinates in which $\mathcal{F}$ is related to an unflowed $n$-point function.\\

To orient the reader, our calculations in this section proceed as follows. We view the correlation function $\mathcal{G}$ as a function of $z_i$, $X_i$, $u$ and $\alpha_r$. In \S\ref{sec:null_constraints}, we study the null constraints associated to $\Omega_W(\myx(z);u)$ which are \eqref{eq:null_constraints}. These provide $W$ linearly independent differential constraints on $\mathcal{G}$ which we explicitly determine. They involve derivatives with respect to $X_i$ and $\alpha_r$. In \S\ref{sec:spacetime_global_conf_inv}, we then study the constraints from the spacetime global conformal invariance of $\mathcal{G}$. These provide three additional linearly independent differential equations, which are also built from derivatives with respect to $X_i$ and $\alpha_r$. We then determine a linear combination of these constraints that removes all $\alpha_r$ derivatives, leaving three differential equations, written solely with respect to $X_i$ derivatives. After an appropriate change of variables $X_i \mapsto \sx_i$, we solve these constraints and show that $\mathcal{G}$ may be written as a meromorphic function times $\tilde{\mathcal{G}}$, where $\tilde{\mathcal{G}}$ satisfies the constraints from global spacetime conformal invariance on an unflowed $n$-point function with spacetime coordinates $\sx_i$. In \S\ref{sec:2-point_functions}, we then restrict to the case $n=2$ and specialise to the locus $\myx(z)={\gamma}(z)$ and $X_i = \gamma(z_i)$ from \eqref{eq:n_point_x_basis_formula}. We further constrain $\tilde{\mathcal{G}}$ using worldsheet global conformal invariance, extract the unflowed 2-point function $\mathcal{F}$ and hence derive the spectrally-flowed 2-point function in the $y$-basis.

\subsection{Null constraints}\label{sec:null_constraints}

The key result that we will use in order to constrain $\mathcal{G}$ is that the generalised spectral flow operator, $\Omega_W$, satisfies
\begin{equation}
    J^-(z)\Omega_W(0;u) = \mathcal{O}\big((z-u)^W \big)\,.
\end{equation}
Since $\Omega_W = V^{W-1}_{k/2,kW/2}$, one might expect this OPE to vanish only up to order $W-1$. However, $J^-_{-1}\left| \frac{k}{2},\frac{k}{2} \right\rangle$ defines a null state, such that the OPE vanishes up to order $W$. Defining 
\begin{equation}
\myx(z)= \sum_{r=0}^{W-1}\alpha_r (z-u)^r\,,
\end{equation}
such that $(\myx J^+)_0:=\frac{1}{2\pi i}\oint \rd z\, \myx(z) J^+(z) = \sum_{r=0}^{W-1}\alpha_rJ^+_r$, we can transform $\Omega_W$ into the $x(z)$-basis using \eqref{eq:x(z)_as_conjugation},
\begin{equation}
    \Omega_W(\myx(z);u) = e^{\sum_{r=0}^{W-1}\alpha_rJ^+_r}\Omega_W(0;u)e^{-\sum_{r=0}^{W-1}\alpha_rJ^+_r}\,.
\end{equation}
One can check that
\begin{equation}
    \begin{split}
        J^-(\myx(z);z) := e^{(\myx J^+)_0} J^-(z) e^{-(\myx J^+)_0} = J^-(z) - 2\myx(z)J^3(z) + \myx(z)^2J^+(z) + k\p \myx(z)\,,
    \end{split}
\end{equation}
satisfies the null relation
\begin{equation}
    J^-(\myx(z);z)\Omega_W(\myx(z);u) = \mathcal{O}\big((z-u)^W\big)\,.
\end{equation}
This gives us $W$ null constraints $\mathscr{N}_r=0$ on $\mathcal{G}$, where $0\leq r\leq W-1$ and
\begin{equation}\label{eq:null_constraints}
    \mathscr{N}_r:=\frac{1}{2\pi i}\oint_{z=u} \rd z\, \frac{1}{(z-u)^{r+1}} \left\langle J^-(\myx(z);z)\Omega_W(\myx(z);u) \prod_{i=1}^n V_{j_i}(X_i;z_i) \right\rangle\,.
\end{equation}

These null constraints can be recast as differential equations using the Ward identities. Explicitly, the Ward identities are
\begin{equation}\label{eq: 2 pt function spec flowed Ward identities}
    \left\langle J^a(z) \Omega_W(\myx(z);u) \prod_{i=1}^nV_{j_i}(X_i;z_i)  \right\rangle = \left[ \sum_{i=1}^n\frac{D^a_i}{z-z_i} + \sum_{r=0}^{W-1} \frac{\mathscr{D}^a_r}{(z-u)^{r+1}} \right] \mathcal{G}\,,
\end{equation}
where
\begin{equation}\label{eq: D1}
    \begin{split}
        D^+_i = \p_{X_i}\,, \quad D^3_i = X_i\p_{X_i} + j_i\,, \quad D^-_i = X_i^2\p_{X_i} + 2j_iX_i\,,
    \end{split}
\end{equation}
along with 
\begin{equation}\label{eq: D2}
    \begin{split}
        \mathscr{D}_r^+ &= \p_{\alpha_r}\,,\\
        \mathscr{D}^3_r &= \frac{kW}{2}\delta_{r,0} + \sum_{s=0}^{W-1-r}\alpha_s\p_{\alpha_{s+r}}\,,\\
        \mathscr{D}^-_r &= kW \alpha_0 \delta_{r,0} + \sum_{\substack{s,s'\geq 0\\r+s+s'< W}} \alpha_s \alpha_{s'}\p_{\alpha_{r+s+s'}}\,,
    \end{split}
\end{equation}
for $0\leq r \leq W-1$. 

Using (\ref{eq: 2 pt function spec flowed Ward identities}), the null constraints (\ref{eq:null_constraints}) take the form of $W$ linear differential equations (one for each $r$)
\begin{equation}\label{eq: linear DE}
    \sum_{s=0}^{W-1}L_{r}^{\alpha_s}\p_{\alpha_s} \mathcal{G} + \sum_{i=1}^n L_{r}^{X_i} \p_{X_i}\mathcal{G} + L_r\mathcal{G} = 0\,,
\end{equation}
for some functions $L_{r}^{\alpha_s}$, $L_{r}^{X_i}$ and $L_r$ of $\{X_i,\alpha_r,z_i, u\}$, with $0\leq r \leq W-1$. These functions are explicitly given by (see Appendix \ref{appendix:deriving_null_constraints} for details)
\begin{equation}\label{eq:null_constraints_coefficients}
    \begin{split}
        L_{r}^{\alpha_s} &= [(z-u)^r(\zeta-u)^s] \frac{(\myx(z)-\myx(\zeta))^2}{z-\zeta}\,,\\
        L_{r}^{X_i} &= [(z-u)^r] \frac{(X_i-\myx(z))^2}{z-z_i}\,, \\
        L_r &= k\alpha_{r+1}(r+1-W) + [(z-u)^r] \sum_{i=1}^n\frac{2j_i(X_i-\myx(z))}{z-z_i} \,.
    \end{split}
\end{equation}
We use the notation $[(z-u)^r]F(z)$ to denote the coefficient of $(z-u)^r$ in the Taylor expansion of $F(z)$ around $z=u$. In the formula for $L_{r}^{\alpha_s}$, we perform a Taylor expansion in $z$ around $z=u$ and a Taylor expansion in $\zeta$ around $\zeta = u$, i.e. $L_{r}^{\alpha_s}$ is the coefficient of $(z-u)^r(\zeta-u)^s$ in a Taylor expansion. This is well-defined for the following reason. $\myx(z)-\myx(\zeta) = \sum_{r=1}^{W-1}\alpha_r[(z-u)^r-(\zeta-u)^r]$ is divisible by $(z-\zeta)$. Therefore, $\frac{(\myx(z)-\myx(\zeta))^2}{z-\zeta}$ is a multinomial in $z$ and $\zeta$, for which the Taylor expansions are well-defined.

The $W$ null constraints $\mathscr{N}_r=0$ constitute $W$ equations (one for each $r$) that can be used, in principle, to solve for each of the $\p_{\alpha_s}\mathcal{G}$ in (\ref{eq: linear DE}) in terms of $\mathcal{G}$ and $\p_{X_i}\mathcal{G}$. We shall see that other physical constraints, such as those from global spacetime/worldsheet conformal invariance, are naturally written as linear combinations of terms involving $\p_{\alpha_s}\mathcal{G}$ \emph{and} $\p_{X_i}\mathcal{G}$. By using the null constraints to write the $\p_{\alpha_s}\mathcal{G}$ terms as functions of $\p_{X_i}\mathcal{G}$ and ${\cal G}$, we arrive at expressions for these physical constraints expressed as differential equations that involve derivatives with respect to the $X_i$ \emph{only}. It is these resultant differential equations that we want to relate to the standard constraints on an $n$-point function of unflowed vertex operators.

\subsection{Spacetime global conformal invariance}\label{sec:spacetime_global_conf_inv}

The correlation function $\mathcal{G}$ is also constrained by spacetime global conformal invariance,
\begin{equation}
    \left( \mathscr{D}^a_0 + \sum_{i=1}^nD^a_i \right)\mathcal{G}=0\,.
\end{equation}
It will be helpful to isolate the terms involving derivatives with respect to the $\alpha_s$. Using (\ref{eq: D1}) and (\ref{eq: D2}), we can write these constraints as
\begin{equation}\label{eq:G_global_st_conf_inv}
    \sum_{s=0}^{W-1}v_s^a\p_{\alpha_s}\mathcal{G} + H^a\mathcal{G} = 0\,,
\end{equation}
where\footnote{The $v^a_s$ are the Taylor coefficients of $v^+(z) = 1$, $v^3(z) = \myx(z)$ and $v^-(z) = \myx(z)^2$ up to order $(z-u)^W$.} $v^+_s = \delta_{s0}$, $v^3_s = \alpha_s$, $v^-_s = \sum_{r=0}^s \alpha_r\alpha_{s-r}$ and
\begin{equation}
    H^+=\sum_{i=1}^n \p_{X_i}, \quad H^3= \frac{kW}{2} + \sum_{i=1}^n(X_i\p_{X_i} + j_i), \quad H^- = kW\alpha_0 + \sum_{i=1}^n(X_i^2\p_{X_i} + 2j_iX_i)\,.
\end{equation}
By substituting $\p_{\alpha_s}\mathcal{G}$ from the null constraints \eqref{eq: linear DE} into (\ref{eq:G_global_st_conf_inv}), one obtains a differential equation that only involves derivatives with respect to $X_i$. Let us now construct this differential equation.
\\

We do not need to solve for each $\p_{\alpha_s}\mathcal{G}$ individually. Instead, we seek vectors $\lambda^a_r$ such that $\sum_{r=0}^{W-1}\lambda_r^aL_{rs} = v^a_s$. Multiplying (\ref{eq: linear DE}) by $\lambda^a_r$, summing over $r$ and then using (\ref{eq:G_global_st_conf_inv}), gives
\begin{equation}\label{eq:reduced_null_constraints}
    H^a\mathcal{G} - \sum_{r=0}^{W-1}\lambda^a_r \left( \sum_{i=1}^n L_{r}^{X_i}\p_{X_i} + L_r\right) \mathcal{G} = 0\,.
\end{equation}
This is the desired linear differential equation involving only derivatives with respect to $X_i$. $L_{r}^{X_i}$ and $L_r$ may still have a functional dependence on the $\alpha_s$, but the role of the $\alpha_s$ has been reduced to that of a set of parameters in (\ref{eq:reduced_null_constraints}). In particular, one could now evaluate them on the locus $\myx(z)=\gamma(z)$ from \eqref{eq:n_point_x_basis_formula}.

In arriving at (\ref{eq:reduced_null_constraints}), we assumed that the $\lambda^a_r$ exist --- how do we, in practice, find such a set of vectors? A helpful trick is to instead define an associated function $\lambda^a(z) = \sum_{r=-1}^{W-1} \lambda^a_{r} (z-u)^{W-r-1}$, which is a polynomial of degree at most $W$. Then,
\begin{equation}\label{eq:lambda_times_L_formula}
    \begin{split}
        \text{res}_{z=u} \left( \frac{\lambda^a(z)}{(z-u)^W}\frac{(\myx(z)-\myx(\zeta))^2}{z-\zeta} \right) &= \sum_{r=-1}^{W-1}\lambda^a_r [(z-u)^r] \frac{(\myx(z)-\myx(\zeta))^2}{z-\zeta} \\
        &= \sum_{r=0}^{W-1}\sum_{s=0}^{W-1} \lambda^a_r L_{r}^{\alpha_s} (\zeta-u)^s + \mathcal{O}((\zeta-u)^W)\,.
    \end{split}
\end{equation}
Note that, in passing to the second line, the auxiliary variable $\lambda^a_{-1}$ drops out because $\frac{(\myx(z)-\myx(\zeta))^2}{z-\zeta}$ is a multinomial in $z$ and $\zeta$. Therefore, if we find functions $\lambda^a(z)$ such that the residue on the LHS picks out the Taylor expansions of $v^a(\zeta)$ around $\zeta=u$ up to order $W$, then we will have determined the desired $\lambda^a_r$.

\subsubsection{Pad\'{e} approximants}

Our construction for $\lambda^a(z)$ depends on the existence of four polynomials $A,$ $B$, $C$ and $D$ that solve the finite approximation problem
\begin{equation}\label{eq:Pade_approximant}
\begin{aligned}
 A\myx-B &=\mathcal{O}\left((z-u)^{W}\right)\,, &\deg A &=W/2\,, &[(z-u)^{W/2}]A &=1\,, &\deg B &<W/2\,,\\
 C\myx-D &=\mathcal{O}\left((z-u)^{W}\right)\,, &\deg D &=W/2\,, &[(z-u)^{W/2}]D &=1\,, &\deg C &<W/2\,.
\end{aligned}
\end{equation}
Specifically, these polynomials describe Pad\'{e} approximants of $\myx(z)$ near $z=u$.\footnote{A Pad\'{e} approximant for a function $f(z)$ of order $[m/n]$ about $z=u$ is a rational function $g(z) = \frac{g_1(z)}{g_2(z)}$ that is a ratio of two polynomials of degrees $\deg g_1 = m$ and $\deg g_2 = n$ such that $g_2(z)f(z) - g_1(z) = \mathcal{O}((z-u)^{m+n+1})$. We will assume that $g_2(u)\neq 0$, since this is relevant to the case where $\myx(u)$ is well-defined. When it exists, the Pad\'{e} approximant for fixed $m$ and $n$ is unique. Here, $(B,A)$ is of order $m<W/2$ and $n=W/2$, while $(D,C)$ is of order $m=W/2$ and $n<W/2$.}

It is a straightforward linear algebra problem to construct such polynomials. For instance, one requires that the coefficients of $(z-u)^t$ in $A\myx(z)$ vanish for $W/2\leq t < W$ and the coefficients in $C\myx(z)$ vanish for $W/2< t < W$, while the coefficient of $(z-u)^{W/2}$ must be one. Define $\tau$ to be a matrix with entries $\tau_{rs} = \alpha_{W/2+r-s}$ (which are the coefficients of $\myx(z)$) for $0\leq r,s < W/2$. The linear system of equations to solve is
\begin{equation}
    \sum_{s=0}^{W/2-1}\tau_{rs}a_s = -\alpha_r\,, \qquad \sum_{s=0}^{W/2-1}\tau_{rs}c_s = \delta_{r0}\,,
\end{equation}
for each $r$, where $A(z) = (z-u)^{W/2} + \sum_{s=0}^{W/2-1} a_s(z-u)^s$ and $C(z) = \sum_{s=0}^{W/2-1} c_s(z-u)^s$. Then $B(z)$ and $D(z)$ are determined by the first $W/2$ terms in the expansions of $A\myx(z)$ and $C\myx(z)$ about $z=u$, respectively. The explicit solution to this linear system is not important for the present work, so we will simply assume that $A$, $B$, $C$ and $D$ solve \eqref{eq:Pade_approximant}. Under this assumption, it is a simple exercise to show that
\begin{equation}\label{eq:polynomial_det}
    AD(z) - BC(z) = (z-u)^W\,.
\end{equation}
The desired polynomials $\lambda^a(z)$ have a simple solution in terms of these Pad\'{e} approximants. One finds that
\begin{equation}\label{eq:lambda_solution}
    \lambda^+=C^2\,,\qquad \lambda^3=-AC\,,\qquad
    \lambda^-=A^2\,.
\end{equation}
This is proven in Appendix \ref{appendix:deriving_lambda} using \eqref{eq:lambda_times_L_formula}.

\subsubsection{Effective spacetime coordinates}

We can now eliminate all $\alpha_r$ derivatives from the null constraint equations and derive differential equations in $X_i$. To do so, we must find an explicit expression for the differential operator
\begin{equation}
    \sum_{r=0}^{W-1}\lambda^a_r  \left( \sum_{i=1}^n L_{r}^{X_i}\p_{X_i} + L_r \right)\,,
\end{equation}
in (\ref{eq:reduced_null_constraints}) in terms of the data $\{X_i,z_i,u\}$. This can be achieved through the following formulae where we recall \eqref{eq:null_constraints_coefficients}. Firstly,
\begin{equation}\label{eq:lambda A}
    \sum_{r=0}^{W-1}\lambda^a_r L_{r}^{X_i} = \text{res}_{z=u} \left( \frac{\lambda^a(z)}{(z-u)^W}\frac{(X_i-\myx(z))^2}{z-z_i} \right).
\end{equation}
Next, we write $L_r = L_r^{(1)} + L_r^{(2)}$ where
\begin{equation}
    L_r^{(1)} = [(z-u)^r] \sum_{i=1}^n \frac{2j_i(X_i-\myx(z))}{z-z_i}\,,
\end{equation}
and
\begin{equation}
    \begin{split}
        L_r^{(2)} = k\alpha_{r+1}(r+1-W) &= [(z-u)^r] \left( k\p \myx(z)  - \frac{kW(\myx(z)-\alpha_0)}{z-u}  \right)\,.
    \end{split}
\end{equation}
Then,
\begin{equation}\label{eq:lambda CD}
    \begin{split}
        \sum_{r=0}^{W-1}\lambda^a_r L_{r}^{(1)} &= \sum_{i=1}^n\text{res}_{z=u} \left( \frac{\lambda^a(z)}{(z-u)^W}\frac{2j_i(X_i-\myx(z))}{z-z_i} \right)\,,\\
        \sum_{r=0}^{W-1} \lambda^a_r L_r^{(2)} &=  \text{res}_{z=u} \left( \frac{\lambda^a(z)}{(z-u)^{W}} \left( k\p \myx(z)  - \frac{kW(\myx(z)-\alpha_0)}{z-u}  \right) \right) \\
        &= \text{res}_{z=u} \left( -\frac{k\p\lambda^a(z)}{(z-u)^{W}} \myx(z)  +kW \alpha_0\frac{\lambda^a(z)}{(z-u)^{W+1}} \right) \,,
    \end{split}
\end{equation}
where we have performed an integration by parts in passing to the final line. Explicitly substituting \eqref{eq:lambda_solution} into (\ref{eq:lambda A}) and (\ref{eq:lambda CD}), one finds that
\begin{equation}\label{eq:lambda everything}
    \begin{split}
        \sum_{r=0}^{W-1} \lambda^+_r L_{r}^{X_i} &= 1 - \frac{(C(z_i)X_i - D(z_i))^2}{(z_i-u)^W}\,,\\
        \sum_{r=0}^{W-1} \lambda^3_r L_{r}^{X_i} &= X_i + \frac{(A(z_i)X_i - B(z_i))(C(z_i)X_i - D(z_i))}{(z_i-u)^W}\,,\\
        \sum_{r=0}^{W-1} \lambda^-_r L_{r}^{X_i} &= X_i^2 - \frac{(A(z_i)X_i - B(z_i))^2}{(z_i-u)^W}\,,\\
        \sum_{r=0}^{W-1} \lambda^+_r L_{r}^{(1)} &= -\sum_{i=1}^n \frac{2j_i(C(z_i)^2X_i - D(z_i)C(z_i))}{(z_i-u)^W}\,,\\
        \sum_{r=0}^{W-1} \lambda^3_r L_{r}^{(1)} &= \sum_{i=1}^n \frac{2j_i(A(z_i)C(z_i)X_i - B(z_i)C(z_i))}{(z_i-u)^W}\,,\\
        \sum_{r=0}^{W-1} \lambda^-_r L_{r}^{(1)} &= \sum_{i=1}^n \left( 2j_iX_i - \frac{2j_i(A(z_i)^2X_i - A(z_i)B(z_i))}{(z_i-u)^W} \right) \,,\\
        \sum_{r=0}^{W-1} \lambda^+_r L_{r}^{(2)} &= 0\,,\\
        \sum_{r=0}^{W-1} \lambda^3_r L_{r}^{(2)} &= \frac{kW}{2}\,,\\
        \sum_{r=0}^{W-1} \lambda^-_r L_{r}^{(2)} &= kW\alpha_0\,.
    \end{split}
\end{equation}
These can be readily verified using \eqref{eq:poly_residue2}.

We are finally in a position to write down the three constraints \eqref{eq:reduced_null_constraints} explicitly. Defining
\begin{equation}\label{eq:M_N_defn}
 M_i=A(z_i)X_i-B(z_i)\,,\qquad
 N_i=C(z_i)X_i-D(z_i)\,,
\end{equation}
we can use (\ref{eq:lambda everything}) to write (\ref{eq:reduced_null_constraints}) as
\begin{equation}\label{eq:X constraints}
    \begin{split}
        \sum_{i=1}^n \frac{N_i^2\p_{X_i}+2j_iC(z_i)N_i}{(z_i-u)^W}\mathcal{G} &= 0\,,\\
        \sum_{i=1}^n\frac{M_iN_i\p_{X_i}
              +j_i[A(z_i)N_i+C(z_i)M_i]}{(z_i-u)^W} \mathcal{G} &= 0\,,\\
        \sum_{i=1}^n\frac{M_i^2\p_{X_i}+2j_iA(z_i)M_i}{(z_i-u)^W} \mathcal{G} &= 0\,,
    \end{split}
\end{equation}
which are, respectively, the equations for $a=+, 3$ and $-$. These can be further simplified via a change of variable from $X_i$ to a new \emph{effective spacetime coordinate} $\sx_i$, where
\begin{equation} \label{eq:Y_i_defn}
 \sx_i=\frac{N_i}{M_i}\,,\qquad
 \mathcal{J}_i=\frac{\p\sx_i}{\p X_i}
       =\frac{(z_i-u)^W}{M_i^2}\,,\qquad
 \mathcal{G}=\left(\prod_{i=1}^n\mathcal{J}_i^{j_i}\right)\tilde{\mathcal{G}}(\sx_i;z_i,u)\,,
\end{equation}
where \eqref{eq:polynomial_det} is used in evaluating the Jacobian, ${\cal J}_i$, and the last equation in (\ref{eq:Y_i_defn}) can be taken as the definition of $\widetilde{\cal G}$.\footnote{Strictly speaking, we should also include dependence on the $\alpha_r$ in $\tilde{\mathcal{G}}$. We have suppressed their dependence since the differential equations under consideration do not contain $\alpha_s$ derivatives any more and we therefore view them as parameters. On the locus of $\myx(z)=\gamma(z)$ from \eqref{eq:n_point_x_basis_formula}, the $\alpha_s$ are all determined by $x_i$, $z_i$, $u$ and $w_i$.} At fixed $z_i$ and $u$, the derivative of ${\cal G}$ with respect to $X_i$ may be written as
\begin{equation}\label{eq:derivative}
\partial_{X_i}\mathcal{G}
=
\left(\prod_{\ell=1}^{n}\mathcal J_\ell^{j_\ell}\right)
\left[
\mathcal J_i\partial_{\mathscr X_i}
-\frac{2j_iA(z_i)}{M_i}
\right]\tilde{\mathcal{G}}\,.
\end{equation}
Substituting (\ref{eq:derivative}) into the three constraints (\ref{eq:X constraints}), and using
\eqref{eq:polynomial_det}, gives
\begin{equation}\label{eq:G_tilde_spacetime_invariance}
\begin{aligned}
\sum_{i=1}^{n}
\left(\mathscr X_i^2\partial_{\mathscr X_i}
+2j_i\mathscr X_i\right)\tilde{\mathcal{G}}&=0\,,\\
\sum_{i=1}^{n}
\left(\mathscr X_i\partial_{\mathscr X_i}+j_i\right)
\tilde{\mathcal{G}}&=0\,,\\
\sum_{i=1}^n\partial_{\mathscr X_i}\tilde{\mathcal{G}}&=0\,,
\end{aligned}
\end{equation}
for $a=+$, 3 and $-$ respectively. These are the spacetime global Ward
identities for an unflowed $n$-point function in the
coordinates \(\mathscr X_i\).\footnote{$\sx_i$ has vanishing conformal weight when evaluated on the locus of \eqref{eq:n_point_x_basis_formula}. To see this, note that $X_i = {\gamma}(z_i)$ has vanishing conformal weight as can be seen from \eqref{eq:gamma_tilde_at_zi}. The M\"{o}bius transformation in \eqref{eq:M_N_defn} and \eqref{eq:Y_i_defn} used to define $\sx_i$ preserves this fact.} For this reason, we refer to the $\sx_i$ as \textit{effective spacetime coordinates}.\\

Let us pause momentarily to interpret this result. Consider the $n$-point function
\begin{equation}
    \mathcal{H} = \left\langle \prod_{i=1}^n V^{w_i}_{j_i}(y_i;x_i;z_i) \right\rangle\,,
\end{equation}
for $W=\sum_{i=1}^n w_i \in 2\mathbb{N}$. Under the assumption that the Pad\'{e} approximants \eqref{eq:Pade_approximant} exist, we have shown that
\begin{equation}\label{eq:H_to_G_tilde}
\boxed{
    \mathcal{H} = \prod_{i=1}^n\frac{|z_i-u|^{4j_iW}}{|M_i|^{4j_i}} \prod_{i<j} |z_i-z_j|^{-2j_iw_j-2j_jw_i-kw_iw_j} \tilde{\mathcal{G}}(\mathscr{X}_i;z_i;u)\,,
    }
\end{equation}
where we use \eqref{eq:n_point_x_basis_formula} and \eqref{eq:Y_i_defn} and we reinstate the anti-holomorphic dependence. Specifically, we evaluate this equation on the locus where $\myx(z)=\gamma(z)$ and $X_i = \gamma(z_i)$ are determined by the function ${\gamma}(z)$ as defined in \S\ref{sec:y_basis_correlators}. Hence, if it is indeed the case that any tree level spectrally-flowed $n$-point function can be written as a meromorphic function times a tree level unflowed $n$-point function, then we conjecture that the spacetime coordinates of the unflowed vertex operators are the $\sx_i$. That is to say, we have constructed the map from the spacetime cross-ratios of the flowed vertex operators to the spacetime cross-ratios of the unflowed vertex operators.

We can perform an initial sanity check of our conjecture by comparing it to the 4-point conjecture of \cite{Dei:2021yom}. It is known that 4-point functions can only be non-vanishing provided $W \geq 2\max_i w_i - 2$. The edge case of $W=2\max_i w_i - 2$ was observed to be somewhat singular in \cite{Dei:2021yom}, so let us therefore concentrate on $W\in 2\mathbb{N}$, with $W \geq 2\max_i w_i$. We set $(x_1,x_2,x_3,x_4)=(0,1,\infty,x)$, $(z_1,z_2,z_3,z_4)=(0,1,\infty,z)$ and we work at generic values of $j_i$, $x$ and $z$ such that we are away from any isolated poles of the full correlator. Then we can compare our cross-ratio
\begin{equation}
    \chi = \frac{\sx_{41}\sx_{23}}{\sx_{43}\sx_{21}}\,,
\end{equation}
where $\mathscr{X}_{ij}=\sx_i-\sx_j$ to their cross-ratio --- see equation (3.7) of \cite{Dei:2021yom}. We have explicitly performed this check in \texttt{Mathematica}, symbolically in $x$, $z$ and $y_i$, for all 531 cases where $W\leq 12$ and found exact agreement. Specifically, in the attached ancillary notebook we evaluate $X_i$ and $\myx(z)$ on the locus determined by ${\gamma}(z)$ in \eqref{eq:n_point_x_basis_formula}, construct the Pad\'{e} approximants, evaluate the cross-ratio of the $\sx_i$ and compare this to the answers of \cite{Dei:2021yom}. We use the \texttt{Mathematica} notebook \texttt{4ptf\_ancillary.nb}
accompanying \cite{Dei:2021yom} in order to calculate their cross-ratios.

In all of the cases we checked using \texttt{Mathematica}, the Pad\'{e} approximants satisfying \eqref{eq:Pade_approximant} for $\myx(z)=\gamma(z)$ existed. Let us conclude this section with a brief sketch of why one might expect this to be the case for $W\in 2\mathbb{N}$ and $W\geq 2\max_i w_i$. It was observed in \cite{Dei:2021yom} that the 4-point function had poles at points where $x_4=x$ and $z_4=z$ were chosen such that a genus zero covering map $\Gamma(z)$ of degree $\frac{W}{2}-1$ exists, where $\Gamma(z) = x_i + \mathcal{O}((z-z_i)^{w_i})$. Let us therefore show that the Pad\'{e} approximants of $\gamma(z)$ of the desired orders in \eqref{eq:Pade_approximant} do not exist if a covering map $\Gamma(z)$ exists. We will not discuss the converse argument here.

Without loss of generality, let us choose the $z_i$ to be finite and distinct, and the same for $x_i$. This implies that, if $\Gamma(z)$ exists, then $\Gamma(z) = \frac{P(z)}{Q(z)}$ for polynomials $P$ and $Q$ of degree $\frac{W}{2}-1$ by the Riemann-Hurwitz formula. We moreover assume that $\Gamma(z)$ is regular at $z=u$, since we can choose $u$ to be any auxiliary point. Next, recall that the Pad\'{e} approximants only depend on the Taylor expansion of $\gamma(z)$ around $z=u$ up to order $(z-u)^W$. This means that, for the sake of constructing Pad\'{e} approximants of $\myx(z)=\gamma(z)$, we can replace $x(z)$ in \eqref{eq:gamma_explicit_realisation} by $\Gamma(z)$. Specifically,
\begin{equation}
     \gamma(z) \equiv \hat{\gamma}(z) \quad \text{mod } (z-u)^W\,, \qquad \hat{\gamma}(z)= \frac{\Gamma(z) (z-u)^W}{\prod_{i=1}^n(z-z_i)^{w_i}}-\sum_{i=1}^n x_i p_i(z) \,.
\end{equation}
This is not an exact equality because in \S\ref{sec:y_basis_correlators}, we required $x(z) = x_i + y_i(z-z_i)^{w_i} + \dots$; the data $y_i$ are not encoded in $\Gamma(z)$. Nevertheless, the Taylor expansions of $\gamma(z)$ and $\hat{\gamma}(z)$ agree near $z=u$ up to order $(z-u)^W$.

Additionally, $\sum_{i=1}^n x_ip_i(z) = \frac{S(z)}{\prod_{i=1}^n(z-z_i)^{w_i}}$ where $S$ is a polynomial of degree at most $W-1$. Hence,
\begin{equation}
    \hat{\gamma}(z) = \frac{P(z)(z-u)^W - Q(z)S(z)}{Q(z)\prod_{i=1}^n(z-z_i)^{w_i}}\,.
\end{equation}
By construction, $\gamma(z)$ and $\hat{\gamma}(z)$ are regular at $z=z_i$, implying $\hat{P}(z) = \frac{P(z)(z-u)^W - Q(z)S(z)}{\prod_{i=1}^n(z-z_i)^{w_i}}$ is a polynomial of degree at most $\frac{W}{2}-1$. Therefore, $\hat{\gamma}(z) = \frac{\hat{P}(z)}{Q(z)}$ is a ratio of polynomials of degree at most $\frac{W}{2}-1$. Now suppose that \eqref{eq:Pade_approximant} holds for $\myx(z)=\hat{\gamma}(z)$. Then
\begin{equation}
    A\hat{P} - BQ = \mathcal{O}\left((z-u)^W\right)\,, \quad C\hat{P} - DQ = \mathcal{O}\left((z-u)^W\right)\,.
\end{equation}
Both of these combinations are polynomials of degree at most $W-1$, so the left hand sides must identically vanish. Eliminating $\hat{P}$ leads to $Q(AD-BC) = 0$. However, this is a contradiction because $Q(z)$ is not identically zero and \eqref{eq:polynomial_det} implies that $(AD-BC)(z)=(z-u)^W$ is also not identically zero. We deduce that the Pad\'{e} approximants do not exist in this case.

\subsection{2-point functions}\label{sec:2-point_functions}

We now restrict our attention to the case of $n=2$. Our aim is to further constrain \eqref{eq:H_to_G_tilde} and extract from $\tilde{\cal G}$ an unflowed 2-point function, when evaluated on the locus determined by ${\gamma}(z)$. To do so, we must constrain the dependence of $\tilde{\cal G}$ on the worldsheet coordinates. Although the formula for the flowed 2-point function in the $y$-basis is known \cite{Dei:2021xgh}, this subsection is included to fully demonstrate how our method can be used to rewrite spectrally-flowed correlation functions as a meromorphic function times their unflowed counterpart. We also hope that by seeing the application of the ideas presented in the previous section to a simple, concrete example, it will help to elucidate the proposed construction.
\\

\subsubsection{\boldmath Effective spacetime coordinates at $n=2$}

To begin, we explicitly perform all of the steps above for $n=2$, illustrating the method. For generic spins $j_i$ (i.e. away from series identifications), the 2-point function is only non-vanishing provided $w_1=w_2 = w$, which guarantees that $W=2w$ is even. Next, to simplify our computations, we choose to set $x_1=0$ and $x_2=1$, but we shall keep $z_1$ and $z_2$ arbitrary, since they do not significantly change the complexity of the algebra below. We will repeatedly make use of the following identities and definitions introduced in \S\ref{sec:canonical_form_of_a_correlator}:
\begin{equation}
    \begin{aligned}
        R(z) &= \frac{(z-z_1)^w(z-z_2)^w}{(z-u)^{2w}}\,,\\
        p_i(z) &= \text{PP}_{z=z_i}\left( \frac{(z-u)^{2w}}{(z-z_1)^w(z-z_2)^w} \right) := \frac{P_i(z)}{(z-z_i)^w}\,,\\
        R(z)^{-1} &= 1 + p_1(z) + p_2(z)\,,\\
        {\gamma}(z) &\equiv -p_2(z) \text{ mod } (z-u)^{2w}\,,\\
        {\gamma}(z_1) &= -p_2(z_1) + y_1\frac{(z_1-u)^{2w}}{(z_1-z_2)^w}\,,\\
        {\gamma}(z_2) &= 1+p_1(z_2) + y_2\frac{(z_2-u)^{2w}}{(z_2-z_1)^w}\,.
    \end{aligned}
\end{equation}
Following (\ref{eq:gamma_tilde_at_zi}), the final two equations determine the insertion points of the unflowed vertex operators inside $\mathcal{G}$. Similarly, the polynomial $\myx(z) \equiv -p_2(z)$ mod $(z-u)^{2w}$. The Pad\'{e} approximants are then explicitly given by
\begin{equation}
\begin{aligned}
    A &= (z-z_2)^w\,, &  B &= -P_2(z)\,,\\
    C &= (z-z_1)^w - (z-z_2)^w\,, & D &= P_1(z) + P_2(z) + (z-z_1)^w\,.
\end{aligned}
\end{equation}
It is a simple exercise to check that these satisfy the properties stated in \eqref{eq:Pade_approximant}. The effective spacetime coordinates, introduced in (\ref{eq:Y_i_defn}), are
\begin{equation}
\begin{aligned}
    \sx_1 &= \frac{-z_{12}^w{\gamma}(z_1) -P_1(z_1) - P_2(z_1)}{z_{12}^w{\gamma}(z_1) +P_2(z_1)} = -1-\frac{1}{z_{12}^wy_1}\,,\\
    \sx_2 &= \frac{z_{21}^w{\gamma}(z_2) -P_1(z_2) - P_2(z_2)-z_{21}^w}{P_2(z_2)} = -1 + z_{21}^wy_2\,,
\end{aligned}    
\end{equation}
where we use the fact that $P_1(z_1) = \frac{(z_1-u)^{2w}}{z_{12}^w}$ and $P_2(z_2) = \frac{(z_2-u)^{2w}}{z_{21}^w}$. Specifically, $M_1 = y_1(z_1-u)^{2w}$ and $M_2 = \frac{(z_2-u)^{2w}}{z_{21}^w}$ which from \eqref{eq:H_to_G_tilde} implies
\begin{equation}
    \mathcal{H}
      =\tilde{\mathcal{P}}\,
         \tilde{\mathcal{G}}(\mathscr{X}_i;z_i)\,, \quad \tilde{\mathcal{P}} = (-1)^{2wj_2}y_1^{-2j_1}
       z_{12}^{w(j_2-j_1)-kw^2/2}\,.
    \label{eq:H-A-F}
\end{equation}
Note that explicit powers of $z_i-u$ exactly cancelled in the prefactors. Therefore, the independence of $\mathcal{H}$ on $u$ implies that $\tilde{\mathcal{G}}$ has no remaining $u$-dependence.\\

\subsubsection{\boldmath Global worldsheet conformal invariance at $n=2$}

We will now use the invariance of $\mathcal{H}$ under global worldsheet conformal transformations to constrain $\tilde{\mathcal{G}}$ appearing in (\ref{eq:H-A-F}). In the $y$-basis, vertex operators do not have definite worldsheet conformal weight. Therefore, in the worldsheet Ward identities, one replaces the eigenvalue $\Delta_i$ by the differential operator
\begin{equation}
    \widehat{\Delta}_i
       =
       \Delta_i
       -w\left(j_i+y_i\partial_{y_i}\right)
       -\frac{kw^2}{4} 
       =-\frac{j_i(j_i-1)}{k-2} -w\left(j_i+y_i\partial_{y_i}\right)
       -\frac{kw^2}{4} \,.
    \label{eq:y-basis-worldsheet-weight}
\end{equation}
The global worldsheet Ward identities read \cite{Dei:2021xgh}
\begin{subequations}
\begin{align}
       \sum_{i=1}^{2}
       \left.\partial_{z_i}\right|_y
       \mathcal{H} &= 0\,, \label{eq:worldsheet_translation}\\
       \sum_{i=1}^{2}
       \left(
          z_i\left.\partial_{z_i}\right|_y
          +\widehat{\Delta}_i
       \right)\mathcal{H} &= 0\,,\label{eq:worldsheet_dilation}\\
       \sum_{i=1}^{2}
       \left(
          z_i^2\left.\partial_{z_i}\right|_y
          +2z_i\widehat{\Delta}_i
       \right)\mathcal{H} &= 0\,.\label{eq:worldsheet_special_conf}
\end{align}
\end{subequations}
We use the notation $|_y$ to emphasise that both $y_1$ and $y_2$ are held fixed (as well as $x_1$ and $x_2$). However, in constraining $\tilde{\mathcal{G}}(\sx_i;z_i)$, we instead need to keep $\mathscr{X}_i$ fixed. From
\begin{align}
    \left.\partial_{z_1}\mathscr{X}_1\right|_y
       &= -\frac{w}{z_{12}}(\mathscr{X}_1+1)\,,
    &
    \left.\partial_{z_1}\mathscr{X}_2\right|_y
       &= \frac{w}{z_{12}}(\mathscr{X}_2+1)\,,
    \\
    \left.\partial_{z_2}\mathscr{X}_1\right|_y
       &= \frac{w}{z_{12}}(\mathscr{X}_1+1)\,,
    &
    \left.\partial_{z_2}\mathscr{X}_2\right|_y
       &= -\frac{w}{z_{12}}(\mathscr{X}_2+1)\,,
\end{align}
the chain rule gives
\begin{equation}
    \begin{split}
        \left.\partial_{z_1}\right|_y
          &=
          \left.\partial_{z_1}\right|_{\mathscr{X}}
          -\frac{w}{z_{12}}(\mathscr{X}_1 + 1)\p_{\mathscr{X}_1} +\frac{w}{z_{12}}(\mathscr{X}_2 + 1)\p_{\mathscr{X}_2}\,,
       \\
       \left.\partial_{z_2}\right|_y
          &=
          \left.\partial_{z_2}\right|_{\mathscr{X}}
          +\frac{w}{z_{12}}(\mathscr{X}_1 + 1)\p_{\mathscr{X}_1} -\frac{w}{z_{12}}(\mathscr{X}_2 + 1)\p_{\mathscr{X}_2}\,.
    \end{split}
\end{equation}
We must also differentiate the prefactor $\tilde{\mathcal{P}}$,
\begin{equation}
    \left.\partial_{z_1}\right|_y\tilde{\mathcal{P}}
       =
       \frac{w(j_2-j_1)-kw^2/2}{z_{12}}\tilde{\mathcal{P}}\,,
    \qquad
    \left.\partial_{z_2}\right|_y\tilde{\mathcal{P}}
       =
       -\frac{w(j_2-j_1)-kw^2/2}{z_{12}}\tilde{\mathcal{P}}\,.
\end{equation}
Thus,
\begin{equation}
\begin{split}
    \tilde{\mathcal{P}}^{-1}\left.\partial_{z_1}\right|_y\mathcal{H}
       &=
       \left[
          \left.\partial_{z_1}\right|_{\mathscr{X}}
           -\frac{w}{z_{12}}(\mathscr{X}_1 + 1)\p_{\mathscr{X}_1} +\frac{w}{z_{12}}(\mathscr{X}_2 + 1)\p_{\mathscr{X}_2}
          +\frac{w(j_2-j_1)-kw^2/2}{z_{12}}
       \right]\tilde{\mathcal{G}}\,,
    \\
    \tilde{\mathcal{P}}^{-1}
    \left.\partial_{z_2}\right|_y\mathcal{H}
       &=
       \left[
          \left.\partial_{z_2}\right|_{\mathscr{X}}
           +\frac{w}{z_{12}}(\mathscr{X}_1 + 1)\p_{\mathscr{X}_1} -\frac{w}{z_{12}}(\mathscr{X}_2 + 1)\p_{\mathscr{X}_2}
          -\frac{w(j_2-j_1)-kw^2/2}{z_{12}}
       \right]\tilde{\mathcal{G}}\,.
    \label{eq:conjugated-z2-derivative}
\end{split}
\end{equation}
Adding these two equations and using \eqref{eq:worldsheet_translation} immediately gives
\begin{equation}
    (\partial_{z_1}|_{\mathscr{X}}+\partial_{z_2}|_{\mathscr{X}})\tilde{\mathcal{G}}=0\,.
    \label{eq:G_tilde_worldsheet_translation}
\end{equation}
For worldsheet dilations and special conformal transformations, we also need to compute $\p_{y_i}$ derivatives. At fixed worldsheet positions, the change of coordinates induces a Jacobian
\begin{equation}
    \frac{\partial(\mathscr{X}_1,\mathscr{X}_2)}
         {\partial(y_1,y_2)}
    =
    \begin{pmatrix}
       \displaystyle\frac{1}{z_{12}^{w}y_1^2} & 0\\[6pt]
       0 & z_{21}^{w}
    \end{pmatrix}\,,
    \label{eq:y-to-Y-Jacobian}
\end{equation}
on a chart where $y_1\neq 0$. Hence,
\begin{equation}
    \begin{split}
            \tilde{\mathcal{P}}^{-1}
       y_1\partial_{y_1}\mathcal{H}
       &=
       \left[
          -2j_1-(\mathscr{X}_1+1)
                    \partial_{\mathscr{X}_1}
       \right]\tilde{\mathcal{G}}\,,
    \\
    \tilde{\mathcal{P}}^{-1}
       y_2\partial_{y_2}\mathcal{H}
       &=
       (\mathscr{X}_2+1)
                    \partial_{\mathscr{X}_2}\tilde{\mathcal{G}}\,.
    \end{split}
\end{equation}
In particular,
\begin{equation}
    \tilde{\mathcal{P}}^{-1}
    \sum_{i=1}^{2}y_i\partial_{y_i}\mathcal{H}
       =\left[-2j_1 -(\mathscr{X}_1+1)
                    \partial_{\mathscr{X}_1} + (\mathscr{X}_2+1)
                    \partial_{\mathscr{X}_2}\right]\tilde{\mathcal{G}}\,.
    \label{eq:conjugated-y-Euler}
\end{equation}
Therefore, (\ref{eq:conjugated-z2-derivative}) implies
\begin{equation}
\begin{aligned}\label{eq:1}
    &\tilde{\mathcal{P}}^{-1}
    \sum_{i=1}^{2}z_i
       \left.\partial_{z_i}\right|_y\mathcal{H}
    = \\
    &\Bigg[ z_1\partial_{z_1}|_{\mathscr{X}} + z_2\partial_{z_2}|_{\mathscr{X}} - w(\mathscr{X}_1+1)\p_{\mathscr{X}_1} + w(\mathscr{X}_2+1)\p_{\mathscr{X}_2}
       +w(j_2-j_1)-\frac{kw^2}{2}
    \Bigg]\tilde{\mathcal{G}}\,,
    \end{aligned}
\end{equation}
while (\ref{eq:y-basis-worldsheet-weight}) and (\ref{eq:conjugated-y-Euler}) imply
\begin{equation}\label{eq:2}
\begin{aligned}
    \tilde{\mathcal{P}}^{-1}
    \sum_{i=1}^{2}\widehat{\Delta}_i\mathcal{H}
    &= \\
    \Bigg[
       \Delta_1+&\Delta_2
       +w(\mathscr{X}_1+1)
         \partial_{\mathscr{X}_1} - w(\mathscr{X}_2+1)
        \partial_{\mathscr{X}_2}
       +w(j_1-j_2)-\frac{kw^2}{2}
    \Bigg]\tilde{\mathcal{G}}\,.
\end{aligned}
\end{equation}
(\ref{eq:1}) and (\ref{eq:2}) combine to give
\begin{equation}
    \left( z_1\partial_{z_1}|_{\mathscr{X}}+ z_2\partial_{z_2}|_{\mathscr{X}}
       +\Delta_1+\Delta_2-kw^2
    \right)\tilde{\mathcal{G}}=0\,,
    \label{eq:G_tilde_worldsheet_dilation}
\end{equation}
from \eqref{eq:worldsheet_dilation}. \eqref{eq:worldsheet_special_conf} similarly implies
\begin{align}
    0={}&
    \Bigg[
       z_1^2\partial_{z_1}|_{\sx}+z_2^2\partial_{z_2}|_{\sx}
       +2z_1\Delta_1+2z_2\Delta_2
       -kw^2(z_1+z_2)
       \nonumber\\
       &\hspace{1cm}
       +wz_{12}
       \left\{
          \sum_{i=1}^{2}
             (\mathscr{X}_i+1)\partial_{\mathscr{X}_i}
          +j_1+j_2
       \right\}
    \Bigg]\tilde{\mathcal{G}}\,.
    \label{eq:F-worldsheet-special-before-reduction}
\end{align}
The final term in braces is the sum of the spacetime
translation and dilation generators from \eqref{eq:G_tilde_spacetime_invariance} which explicitly vanish. Hence, we derive a third worldsheet constraint
\begin{equation}
    \left( z_1^2\partial_{z_1}|_{\mathscr{X}}+ z_2^2\partial_{z_2}|_{\mathscr{X}}
    +2z_1\Delta_1+2z_2\Delta_2
       -kw^2(z_1+z_2)
    \right)\tilde{\mathcal{G}}=0\,.
    \label{eq:G_tilde_worldsheet_special}
\end{equation}
If we perform the redefinition
\begin{equation}\label{eq:G_tilde_to_F}
       \tilde{\mathcal{G}}(\mathscr{X}_1,\mathscr{X}_2;z_1,z_2)=z_{12}^{kw^2}\mathcal{F}(\mathscr{X}_1,\mathscr{X}_2;z_1,z_2) \,,
\end{equation}
then equations \eqref{eq:G_tilde_worldsheet_translation},
\eqref{eq:G_tilde_worldsheet_dilation} and
\eqref{eq:G_tilde_worldsheet_special} become
\begin{equation}
    \begin{aligned}
       \sum_{i=1}^{2}\partial_{z_i}\mathcal{F} &= 0\,,
       \\
       \sum_{i=1}^{2}
             \left(z_i\partial_{z_i}+\Delta_i\right)\mathcal{F} &= 0\,,
       \\
       \sum_{i=1}^{2}
             \left(z_i^2\partial_{z_i}+2z_i\Delta_i\right)\mathcal{F} &= 0\,,
    \end{aligned}
\end{equation}
which are the equations of \emph{worldsheet} global conformal invariance. $\mathcal{F}$ also satisfies \eqref{eq:G_tilde_spacetime_invariance}, the equations of \emph{spacetime} global conformal invariance. In other words, $\mathcal{F}$ satisfies the system of differential equations that uniquely determines the unflowed 2-point function, with external spins $j_i$, target-space positions $\mathscr{X}_i$ and worldsheet positions $z_i$. This is a particular instance of the conjecture \eqref{eq:conjecture}. Restoring the antiholomorphic factors, up to normalisation it must be that \cite{Maldacena:2001km}
\begin{align}
    \mathcal{F}
       (\mathscr{X}_1,\mathscr{X}_2;z_1,z_2)
    &= \mathcal{N}(w,k,j_1)
    \left\langle
       V^0_{j_1}(\mathscr{X}_1;z_1)
       V^0_{j_2}(\mathscr{X}_2;z_2)
    \right\rangle
    \nonumber\\
    &=
    \mathcal{N}(w,k,j_1)\frac{
       \delta(j_1+j_2-1)
          \delta^{(2)}(\mathscr{X}_1-\mathscr{X}_2)
       +
       B(j_1)\delta(j_1-j_2)
          |\mathscr{X}_1-\mathscr{X}_2|^{-4j_1}
    }{
       |z_{12}|^{4\Delta_1}
    }\,.
    \label{eq:quoted-unflowed-two-point}
\end{align}
The relative coefficient between the two terms is fixed via reflection symmetry. After substituting in the known values of $\mathscr{X}_i$ and all of the prefactors of $\mathcal{F}$, one arrives at 
\begin{equation}
\boxed{
    \begin{aligned}
        &\left\langle V^w_{j_1}(y_1;0;z_1)V^w_{j_2}(y_2;1;z_2) \right\rangle \\
        &= \mathcal{N}(w,k,j_1) |z_{12}|^{-4\Delta_1+2w(j_1+j_2)+kw^2}  \bigg[ B(j_1)\delta(j_1-j_2) |(-1)^w + z_{12}^{2w}y_1y_2|^{-4j_1} \\
        &\hspace{2cm}+ |z_{12}|^{-2w}|y_1|^{-2j_1}|y_2|^{-2j_2} \delta(j_1+j_2-1) \delta^{(2)}((-1)^w + z_{12}^{2w}y_1y_2)  \bigg]
    \end{aligned}
    }
\end{equation}
In fact, we can go further and fix this normalisation directly from our contour integral prescription. By construction, our contour integral definition of spectral flow is consistent with the definition in the $m$-basis. We showed in \eqref{eq:x=0_n-point_solution} that
\begin{equation}
    \begin{aligned}
        \langle V^{w}_{j_1,h_1}(z_1)V^{-w}_{j_2,h_2}(z_2) \rangle = (z_1-z_2)^{(m_1 - m_2)w + \frac{kw^2}{2}}&(\zbar_1-\zbar_2)^{(\bar{m}_1 - \bar{m}_2)w + \frac{kw^2}{2}} \\
        &\times \langle V_{j_1,m_1}(z_1)V_{j_2,m_2}(z_2) \rangle\,.
    \end{aligned}
\end{equation}
We can set $m_1=\bar{m}_1=j_1$ and $m_2=\bar{m}_2=-j_2$ to deduce the $x$-basis 2-point function at $x_1=0$ and $x_2=\infty$. Since this holds for any $w$, $k$ or $j_i$, this fixes $\mathcal{N}(w,k,j_1) = 1$. Hence, we exactly recover the expected result for the flowed 2-point function in the $y$-basis \cite{Dei:2021xgh}.\footnote{(4.12) in \cite{Dei:2021xgh} states the 2-point function at $z_1=x_1=0$ and $z_2=x_2=\infty$. For a more direct comparison with our stated formula, equation (4.50) of \cite{Kovensky:2026usc} explicitly includes the $x$- and $z$-dependence.}

\section{Discussion}\label{sec: discussion}

By interpreting spectral flow in terms of non-trivial bundles over the worldsheet, we explained in \S\ref{sec:spec_flow_G_bundle} how to think of spectrally-flowed vertex operators in WZW models as line operators dressing simpler objects --- see \eqref{eq: nucleation 2}. Specialising to $SL(2,\mathbb{R})$, the case of interest for applications to AdS$_3$ string theory, we demonstrated in \S\ref{sec:spectrally-flowed_correlators_AdS3} that a correlator of $n$ spectrally-flowed vertex operators can be rewritten as a meromorphic prefactor multiplying a correlator of $n$ unflowed vertex operators in an instantonic background --- see \eqref{eq:n_point_x_basis_formula}. We refer to this as the canonical form of a spectrally-flowed correlator. We showed in \S\ref{sec: computing correlators} that we can take a certain redefinition of the target space data to obtain spacetime coordinates, $\sx_i$, in which the correlator obeys the same spacetime conformal invariance constraints as does an $n$-point correlator of unflowed vertex operators in a trivial background --- see \eqref{eq:H_to_G_tilde}.

Our results constitute a major step towards establishing a direct relation between spectrally-flowed $n$-point correlators and unflowed $n$-point correlators which we see as the natural method for solving the $SL(2,\R)$ WZW model. The techniques explained here provide a constructive algorithm with which to relate flowed and unflowed vertex operators and can be used to construct the canonical form of a correlator by deforming the line operators that implement spectral flow. Our manipulations of the line operators lead to a transformation in the spacetime coordinates, $x_i$, of the vertex operators, mapping them to points $\gamma(z_i)$ for a particular meromorphic function $\gamma(z)$ defined in \S\ref{sec:y_basis_correlators}. We conjectured and provided evidence that a further transformation from $\gamma(z_i) \mapsto \sx_i$ (as defined in \S\ref{sec: computing correlators}) would place the unflowed correlator into a trivial background and we will discuss this further in \cite{McStay:2026xxx}. We hope that the introduction of these techniques will allow for a new perspective with which to approach calculations of interest in the $SL(2,\R)$ WZW model, complementing the mature bootstrap programme, as well as other theories with spectral flow automorphisms.

Let us briefly comment on the connection between our work and that of the bootstrap programme, especially \cite{Dei:2021xgh,Dei:2021yom,Dei:2022pkr}. Redefining the cross-ratio of the spacetime coordinates such that a 4-point flowed correlator obeys the spacetime global conformal constraints of a 4-point unflowed correlator was first done in \cite{Dei:2021yom}. Their precise calculations were specific to 4-point functions and their expressions were derived using properties of covering maps. It is reassuring that our expression for the 4-point cross-ratio seemingly agrees with theirs, as we have verified in \texttt{Mathematica}. However, we have provided a more geometric understanding of the required redefinitions, especially in our intermediate steps where we studied the line operators that implement spectral flow to derive the unflowed $n$-point function in an instanton background. After this, we applied similar methods to \cite{Dei:2021xgh,Dei:2021yom,Dei:2022pkr} by studying the differential constraints on the correlator. Yet, even in this step, the Pad\'{e} approximants and the fact that our construction applies for all $n$ make the underlying geometry easier to parse.

Another key feature of our work is that our methodology can be applied to more general groups than $SL(2,\R)$. Both the definition of spectral flow in terms of line operators given in \S\ref{sec:spec_flow_G_bundle} and the manipulation of these contours inside a correlation function in \S\ref{sec:spectrally-flowed_correlators_AdS3} generalise for the following reason. Given a Lie group $G$, we can diagonalise its Cartan subalgebra into generators $H^I$. These can be bosonised analogously to $J^3$ in \eqref{eq:J3_bosonisation} and then all of the formulae we have used here are applicable. To help demonstrate this, we elaborate on the case of $G=SU(2,2)$ in Appendix \ref{appendix:SU(2,2)}. Similarly, one anticipates a natural generalisation of our work to higher genus Riemann surfaces and we sketch this in Appendix \ref{appendix:higher_genus}.

We shall conclude with a discussion of several future directions as well as some open questions that follow from our work.

\paragraph{\boldmath Solving the $SL(2,\R)$ WZW model:} In a companion paper \cite{McStay:2026xxx} we will prove our conjecture \eqref{eq:conjecture} by demonstrating that spectrally-flowed correlators can be written as a certain meromorphic function multiplying an unflowed correlator. We believe that the tools developed in this work are sufficient to solve this problem, especially since we have identified candidate spacetime coordinates for the unflowed correlator. In fact, we believe that we already have the final statement and a detailed proof will appear in the companion paper. This is an exciting state of affairs since the unflowed sector is described by the $H^+_3$ model, which has been solved \cite{Teschner:1997fv,Teschner:1997ft,Teschner:1999ug,Teschner:2001gi}. Indeed, we hope that our methods can not only be used to derive the 4-point function, but that they may also be of use for proving crossing symmetry.

\paragraph{When does spectral flow act non-trivially on the spectrum?} We have presented a systematic way to think about spectral flow in correlation functions. However, we have not been able to use our methods to determine when spectral flow acts non-trivially on the Hilbert space of a theory. Put in other words, if and when must spectrally-flowed states be included in a worldsheet theory and when can a worldsheet theory be consistently truncated to its unflowed sector?

The classic example of this is the $SU(2)$ WZW model for which spectral flow exists but maps within the unflowed Hilbert space. More formally, spectral flow is an endomorphism on the space of unitary irreducible representations for $\mathfrak{su}(2)_k$ and we expect that this fact is related to topological statements about $G$-bundles over Riemann surfaces of compact, simply connected Lie groups. It would be satisfying if classic results like this had a sharp statement in the language we have introduced. Perhaps one can use the language of non-trivial $G$-bundles to deduce more generally when spectral flow acts non-trivially in theories with affine symmetry. It may well be necessary to study higher genus correlation functions (see Appendix \ref{appendix:higher_genus}) to make progress on this topic, since the necessity of spectral flow in the $SL(2,\R)$ WZW model is most directly seen from modular invariance of the one-loop partition function \cite{Henningson:1991jc,Maldacena:2000hw,Maldacena:2000kv}.

One interesting setting that we can now connect to both conceptually and at the calculational level is the closed-string formulation of the Berkovits action \cite{Berkovits:2004hg} for Berkovits-Witten twistor string theory for $\mathcal{N}=4$ super Yang-Mills. As we suggested in \S\ref{sec:instanton_background}, a correlation function in a $GL(1)_{\mathbb{C}}$ instantonic background can be instead thought of as computing a spectrally-flowed correlator with respect to the $U(1)$ generator. This perspective is precedented in the twistor variable formulation of the tensionless limit of type IIB string theory on AdS$_3 \times $S$^3 \times $T$^4$ \cite{Dei:2020zui,McStay:2023thk,Knighton:2023mhq,McStay:2024dtk}. The matter content of this theory is reminiscent of the matter fields of the Berkovits twistor string action, and spectral flow is known to be necessary for the complete string spectrum. Can the methods developed here, therefore, provide a new perspective on correlation functions in the twistor string?

It would be interesting more generally to investigate to what extent the methods here can be applied to other models. We emphasise that the method of implementing spectral flow in correlation functions via line operators should not be specific to WZW models. Provided one has a holomorphic current analogous to $J^3(z)$ in the $SL(2,\R)$ WZW model, then line integrals imposing transition functions can be constructed. Indeed, some of the earliest work on spectral flow automorphisms in the literature \cite{Schwimmer:1986mf,Lerche:1989uy,Moore:1989yh} refers to $N=2$ SCFTs, where spectral flow is generated by a holomorphic $U(1)$ and relates NS and R sectors. It would be fascinating if the methods provided here had a useful application in this context.

\paragraph{AdS$_{\mathbf{3}}$/CFT$_{\mathbf{2}}$:} It is natural to ask about the implications of our work for holography. For bosonic string theory, one must couple the CFT to the compact sector, the $bc$ ghost system and then perform an integral over the moduli space of (punctured) Riemann surfaces. For superstring theory in the RNS formalism, the $\mathcal{N}=1$ supersymmetric WZW model $\mathfrak{sl}(2,\R)^{(1)}_k$ is equivalent to a bosonic $\mathfrak{sl}(2,\R)_{k+2}$ WZW model combined with three free fermions. However, our results cannot yet be applied as directly as in the bosonic case because one must insert $(n-2)$ picture-changing operators inside the correlator. We anticipate that the easiest way to circumvent this issue, at least in the NS sector at genus zero, is to work with super Riemann surfaces \cite{DHoker:1988pdl,Witten:2012bh}. Spectral flow should have an analogous description in terms of line operators in the $\mathfrak{sl}(2,\R)_k^{(1)}$ WZW model on a super Riemann surface and one need only consider $n$-point functions of spectrally-flowed affine primaries as we have done in this work. The contribution from picture-changing operators would be repackaged into an integral over the moduli space of super Riemann surfaces. Recently, it was found that this approach was especially useful for computing residues of superstring theory on AdS$_3$ \cite{Knighton:2026wva}.

It has been known for a long time that the correlation functions of string theory on AdS$_3$ contain many bulk poles as the spins, $j_i$, of the vertex operators are varied \cite{Maldacena:2001km}. Physically, these are a consequence of worldsheet instantons \cite{Seiberg:1999xz,Maldacena:2001km}\footnote{There are additionally LSZ-type poles \cite{Aharony:2004xn} corresponding to discrete representations. We will not discuss these poles here.} and they form a two-parameter lattice that can be thought of in analogy with Liouville theory.\footnote{A one-parameter family of poles can be computed in Liouville theory by working perturbatively in the interaction term, but the duality $b\mapsto \frac{1}{b}$ implies that additional poles exist. A similar statement holds for AdS$_3$ string theory.}  In \cite{Dei:2022pkr}, these bulk poles were studied. When trying to calculate the residues of these poles in 4-point functions, Dei and Eberhardt encountered unknown identities of covering maps (which they numerically verified) that were required in order to match with the same residues in the dual CFT. It is plausible that, with the greater geometric understanding that follows from our method, it may now be easier to match the residues with those of the dual CFT. More generally, the residues of a one-parameter family of the bulk poles have been computed exactly for $n$-point functions \cite{Hikida:2023jyc,Knighton:2023mhq,Knighton:2024qxd,Knighton:2026wva,Iguri:2026lxp} but the full two-parameter family has not been determined. If our conjecture \eqref{eq:conjecture} is true, then spectrally-flowed correlators can be related to the unflowed sector. Via the $H^+_3$/Liouville correspondence \cite{Teschner:1997ft,Ribault:2005wp,Hikida:2007tq}, one may then be able to compute residues for the full two-parameter lattice of poles in AdS$_3$.

\paragraph{Acknowledgements} It is a pleasure to thank Vit Sriprachyakul for useful conversations and Tim Adamo and Bob Knighton for comments on an early version of the draft. The work of NM was supported by the Department of
Atomic Energy, Government of India, under project no. RTI4001. RR is supported by STFC consolidated grant UKRI4224 and the generosity of the Avery-Wong Foundation. The work of SS was supported by the Simons Collaboration on Celestial Holography CH-00001550-1 and the Maxwell Institute for Mathematical Sciences.

\paragraph{Generative AI declaration} GPT 5.6 Terra was used in the preparation of the figures. GPT 5.6 Sol and GPT 6 Astra were used to assist with calculations, writing \texttt{Mathematica} code and searching the literature. We have verified all calculations and code that were assisted by AI. The arguments presented are the sole work of the authors and we take full responsibility for the results presented.

\appendix

\section{Computations for spectrally-flowed correlators at genus zero}\label{appendix: correlator computations}

In \S\ref{sec: computing correlators}, we studied spectrally-flowed $n$-point functions at tree level. We constrained the form of
\begin{equation}
        \mathcal{G} = \left\langle \Omega_W(\myx(z);u)\prod_{i=1}^n V_{j_i}(X_i;z_i) \right\rangle\,,
\end{equation}
for $W\in 2\Z$ to show that it may be rewritten (holomorphically) as \eqref{eq:Y_i_defn}
\begin{equation}
    \mathcal{G} = \prod_{i=1}^n \frac{(z_i-u)^{j_iW}}{M_i^{2j_i}}\tilde{\mathcal{G}}(\sx_i;z_i;u)\,.
\end{equation}
In the case of $n=2$, we further constrained $\tilde{\mathcal{G}}$ to extract an unflowed 2-point function in \eqref{eq:G_tilde_to_F}. It is the aim of this appendix to elaborate on these derivations, providing steps in the algebraic manipulation that were not explained in the main text. In particular, we explain equations \eqref{eq:null_constraints_coefficients} and \eqref{eq:lambda_solution}.

\subsection{Deriving the null constraints}\label{appendix:deriving_null_constraints}

In \S\ref{sec:null_constraints} we began by studying the constraints on $\mathcal{G}$ coming from the insertion of $\Omega_W$, which were given in \eqref{eq:null_constraints}. We referred to these as null constraints and claimed that they were given by \eqref{eq:null_constraints_coefficients}. Let us first verify these expressions for $L_{r}^{\alpha_s}$, $L_{r}^{X_i}$ and $L_r$.

The null constraints $\mathscr{N}_r = 0$ are given by the vanishing of the first $W$ Taylor coefficients of
\begin{equation}
    \begin{split}
        &\left[ \sum_{i=1}^n\frac{D^-_i-2\myx(z)D_i^3 + \myx(z)^2D^+_i}{z-z_i} + \sum_{r=0}^{W-1} \frac{\mathscr{D}_r^- - 2\myx(z)\mathscr{D}^3_r + \myx(z)^2\mathscr{D}^+_r}{(z-u)^{r+1}} + k\p \myx(z) \right] G\\
        &\sim \mathcal{O}((z-u)^W)\,,
    \end{split}
\end{equation}
as $z\to u$. The differential operators $D_i^a$ and $\mathscr{D}^a_r$ are defined in \eqref{eq: D1} and \eqref{eq: D2}, respectively. As we did in the main text, we use the notation $[(z-u)^r]F$ to denote the coefficient of $(z-u)^r$ in the Taylor expansion of $F(z)$ around $z=u$. Collecting derivatives with respect to $X_i$, we have
\begin{equation}
    L_{r}^{X_i} = [(z-u)^r] \frac{(X_i-\myx(z))^2}{z-z_i}\,.
\end{equation}
The scalar contribution is
\begin{equation}
    \begin{split}
        L_r &= [(z-u)^r] \left( \sum_{i=1}^n \frac{2j_i(X_i-\myx(z))}{z-z_i} - \frac{\myx(z)kW}{z-u} + \frac{\alpha_0kW}{z-u} + k\p \myx(z) \right) \\
        &= k\alpha_{r+1}(r+1-W) + [(z-u)^r] \sum_{i=1}^n \frac{2j_i(X_i-\myx(z))}{z-z_i} \,.
    \end{split}
\end{equation}
Lastly, collecting the derivatives with respect to $\alpha_s$,
\begin{equation}
    \begin{split}
        L_{r}^{\alpha_s} = [(z-u)^r]\left( \frac{\myx(z)^2}{(z-u)^{s+1}} -2\myx(z)\sum_{m=0}^s \frac{\alpha_{s-m}}{(z-u)^{m+1}} + \sum_{m=0}^s\sum_{m'=0}^{s-m} \frac{\alpha_{m'}\alpha_{s-m-m'}}{(z-u)^{m+1}} \right)\,.
    \end{split}
\end{equation}
We can simplify this last expression by introducing an auxiliary complex coordinate $\zeta$. Since $\myx(z) = \sum_{r=0}^{W-1}\alpha_r(z-u)^r$ is a polynomial in $z$, $\frac{(\myx(z)-\myx(\zeta))^2}{z-\zeta}$ is a multinomial in $z$ and $\zeta$. In particular, the denominator cancels with a zero in the numerator. One finds that
\begin{equation}
    \begin{split}
        &[(\zeta-u)^s]\frac{(\myx(z)-\myx(\zeta))^2}{z-\zeta}\\
        &= \frac{\myx(z)^2}{(z-u)^{s+1}} -2\myx(z)\sum_{m=0}^s \frac{\alpha_{s-m}}{(z-u)^{m+1}} + \sum_{m=0}^s\sum_{m'=0}^{s-m} \frac{\alpha_{m'}\alpha_{s-m-m'}}{(z-u)^{m+1}}\,.
    \end{split}
\end{equation}
Hence, we can write
\begin{equation}
    L_{r}^{\alpha_s} = [(z-u)^r(\zeta-u)^s]\frac{(\myx(z)-\myx(\zeta))^2}{z-\zeta}\,.
\end{equation}
This verifies all of the coefficients in the null constraints.

\subsection[Deriving $\lambda^a_r$]{\boldmath Deriving $\lambda^a_r$}\label{appendix:deriving_lambda}

Having constructed the null constraints and the constraints from spacetime global conformal invariance, our next step was to find linear combinations of these with no $\alpha_s$ derivatives. Our aim was to find three vectors $\lambda^a_r$ such that $\sum_{r=0}^{W-1}\lambda^a_r L_{r}^{\alpha_s} = v_s^a$ where $v^+_a = \delta_{s0}$, $v^3_s = \alpha_s$ and $v^-_s = \sum_{r=0}^s \alpha_r\alpha_{s-r}$. These are the coefficients in the Taylor expansions of $v^+(z)=1$, $v^3(z) = \myx(z)$ and $v^-(z) = \myx(z)^2$ up to order $(z-u)^W$. The proposed method in \eqref{eq:lambda_times_L_formula} was to construct polynomials $\lambda^a(z) = \sum_{r=-1}^{W-1}\lambda^a_r (z-u)^{W-r-1}$ such that 
\begin{equation}\label{eq:lambda_residue}
    \begin{split}
        \text{res}_{z=u} \left( \frac{\lambda^a(z)}{(z-u)^W}\frac{(\myx(z)-\myx(\zeta))^2}{z-\zeta} \right) 
        &= \sum_{r=0}^{W-1}\sum_{s=0}^{W-1} \lambda^a_r L_{r}^{\alpha_s} (\zeta-u)^s + \mathcal{O}((\zeta-u)^W)\,.
    \end{split}
\end{equation}
We choose the $\lambda^a(z)$ to be polynomials whose residue on the LHS agrees with the Taylor expansions of $v^a(\zeta)$ around $\zeta=u$ up to order $W$. We claimed that the solution to this problem was
\begin{equation}
    \lambda^+=C^2\,,\qquad \lambda^3=-AC\,,\qquad
    \lambda^-=A^2\,,
\end{equation}
where $A$ and $C$ are Pad\'{e} approximants \eqref{eq:Pade_approximant}. Let us verify that this is indeed the case.

We will make use of the following identity. Suppose that $Q(z) = \sum_{m=0}^{W}q_m (z-u)^m$ is a polynomial of degree $W$. Then,
\begin{equation}\label{eq:poly_residue}
    \begin{split}
        \text{res}_{z=u}\left(\frac{Q(z)}{(z-u)^{W}(z-\zeta)}\right) &= \text{res}_{z=u}\left(\frac{Q(z)}{(z-u)^{W}}\sum_{n\geq0} (-1)^{n}\frac{(z-u)^n}{(u-\zeta)^{n+1}}\right)\\
        &= -\sum_{m=0}^{W}\sum_{n\geq 0}\frac{1}{(\zeta-u)^{n+1}}  \text{res}_{z=u}\left( (z-u)^{m+n-W} q_m\right)\\
        &= - \sum_{m=0}^{W-1} \frac{1}{(\zeta-u)^{W-m}} q_m\\
        &= q_{W} - \frac{Q(\zeta)}{(\zeta-u)^{W}}\,.
    \end{split}
\end{equation}
This identity trivially generalises to the case where $Q(z,\zeta) = \sum_{m=0}^W q_m(\zeta)(z-u)^m$ is a polynomial in $z$ and the $q_m(\zeta)$ are functions of $\zeta$. In that case,
\begin{equation}\label{eq:poly_residue2}
    \begin{split}
        \text{res}_{z=u}\left(\frac{Q(z,\zeta)}{(z-u)^{W}(z-\zeta)}\right) 
        &= q_{W}(\zeta) - \frac{Q(\zeta,\zeta)}{(\zeta-u)^{W}}\,.
    \end{split}
\end{equation}\\

First consider $\lambda^+=C^2$. Using \eqref{eq:lambda_residue}, one calculates
\begin{equation}
    \begin{split}
        \text{res}_{z=u}&\left( \frac{C(z)^2}{(z-u)^W} \frac{(\myx(z)-\myx(\zeta))^2}{z-\zeta} \right)\\
        &= \text{res}_{z=u}\left( \frac{1}{(z-u)^W} \frac{(D(z)-C(z)\myx(\zeta))^2}{z-\zeta} \right)\,, \\
        &= \left[ (z-u)^W \right] (D(z)-C(z)\myx(\zeta))^2 - \frac{(D(\zeta)-C(\zeta)\myx(\zeta))^2}{(\zeta-u)^W}\,, \\
        &= 1 - \mathcal{O}\left((\zeta-u)^W\right)\,.
    \end{split}
\end{equation}
In the first line we used that $C(z)\myx(z)=D(z) + \mathcal{O}\left((z-u)^W\right)$ from the definition of the Pad\'{e} approximants. In the second line we used \eqref{eq:poly_residue2} and in the final line we used that $D(z)$ is monic in $(z-u)$ of degree $W/2$ and that $C(\zeta)\myx(\zeta)-D(\zeta) = \mathcal{O}\left((\zeta-u)^W\right)$. Therefore $\lambda^+ = C^2$ agrees with the Taylor expansion of $v^+(z)=1$ up to the desired order.

For $\lambda^- = A^2$, one calculates
\begin{equation}
    \begin{split}
        \text{res}_{z=u}&\left( \frac{A(z)^2}{(z-u)^W} \frac{(\myx(z)-\myx(\zeta))^2}{z-\zeta} \right)\\
        &= \text{res}_{z=u}\left( \frac{1}{(z-u)^W} \frac{(B(z)-A(z)\myx(\zeta))^2}{z-\zeta} \right)\,, \\
        &= \left[ (z-u)^W \right] (B(z)-A(z)\myx(\zeta))^2 - \frac{(B(\zeta)-A(\zeta)\myx(\zeta))^2}{(\zeta-u)^W}\,, \\
        &= \myx(\zeta)^2 - \mathcal{O}\left((\zeta-u)^W\right)\,.
    \end{split}
\end{equation}
In the first line we used that $A(z)\myx(z)=B(z) + \mathcal{O}\left((z-u)^W\right)$ from the definition of the Pad\'{e} approximants. In the second line we used \eqref{eq:poly_residue2} and in the final line we used that $A(z)$ is monic in $(z-u)$ of degree $W/2$ and that $A(\zeta)\myx(\zeta)-B(\zeta) = \mathcal{O}\left((\zeta-u)^W\right)$. Therefore $\lambda^- = A^2$ agrees with the Taylor expansion of $v^-(z)=\myx(z)^2$ up to the desired order.

The final case of $\lambda^3 = -AC$ proceeds analogously using the same steps as appeared for $\lambda^+$ and $\lambda^-$. We pair $A$ with one factor of $\myx(z)-\myx(\zeta)$ in \eqref{eq:lambda_residue} and $C$ with the other factor.

\section{\boldmath Spectral flow for $G=SU(2,2)$}\label{appendix:SU(2,2)}

We claimed in the main text that the constructions presented here generalise to other groups. Hence, we will illustrate another example. Of particular interest is the supergroup $PSU(2,2|4)$, which plays an important role as the numerator group in the coset construction of the $AdS_5\times S^5$ Green-Schwarz string. The bosonic subgroup of $PSU(2,2|4)$ is $\frac{SU(2,2)\times SU(4)}{\Z_4}$ and its maximal compact subgroup is $S(U(2)\times U(2))\simeq \frac{SU(2)\times SU(2)\times U(1)}{\Z_2}$. Spectral flow in the $SU(2,2)$ WZW model is generated by the three Cartan generators $t^I$ ($I=1,2,3$) of $S(U(2)\times U(2))$ which, in a convenient basis, have the form
\begin{equation}\label{eq:Cartan_basis}
    t^1=\frac{1}{2}\text{diag}(1,-1,0,0)\,,    \qquad     t^2=\frac{1}{2} \text{diag}(0,0,1,-1)\,, \qquad  t^3=\frac{1}{2}\text{diag}(1,1,-1,-1)\,.
\end{equation}
$t^1$ and $t^2$ are the Cartan generators of the two $SU(2)$ factors in the maximal compact subgroup and $t^3$ generates the $U(1)$. In addition to the Cartan generators are twelve other generators; six of these can be thought of as raising operators, akin to $J^+$ in $SL(2,\R)$ and six lowering operators akin to $J^-$. Collectively we denote these operators by $E^{\alpha}$, where $\alpha$ denotes one of the twelve root vectors of $SU(2,2)$. In the Cartan-Weyl basis the algebra includes
\begin{equation}
    [H^I,H^J]=0\,,    \qquad  [H^I,E^{\alpha}]=\alpha^IE^{\alpha}\,.
\end{equation}
In terms of Kac-Moody currents, these commutation relations are elevated to OPEs
\begin{equation}
    H^I(z)H^J(w)\sim\frac{k\,\kappa^{IJ}}{(z-w)^2}\,, \qquad  H^I(z)E^{\alpha}(w)\sim \frac{\alpha^I}{z-w}E^{\alpha}(w)\,,
\end{equation}
where $\kappa^{IJ}$ is the Killing form on the Cartan subalgebra which, by a convenient choice of basis, can be diagonalised, thus decoupling the different sectors of the Cartan subalgebra. Specifically, $\kappa^{IJ} = \text{diag}\left(-\frac{1}{2},-\frac{1}{2},-1\right)$ for the choice \eqref{eq:Cartan_basis}. In this case, there is an obvious generalization of spectral flow from $SL(2,\R)$ to $SU(2,2)$, which is most clearly expressed in terms of the modes of the currents \cite{Lerche:1989uy,Gaberdiel:2023nhb}
\begin{equation}
    \sigma_{{\omega}}\left(H^I_n\right)=H^I_n-\delta_{n,0}\,k\omega^I\,, \qquad  \sigma_{\mathbf{\omega}}(E^{\mathbf{\alpha}}_n)=E^{{\alpha}}_{n-{\alpha}\cdot{\omega}}\,,
\end{equation}
where $\vec{\omega}=(\omega_1,\omega_2,\omega_3)$ are spectral flow parameters associated with the three Cartan directions and $\vec{\omega}\cdot\vec{\alpha}=\omega_I\alpha^I$. Note that this is consistent with \eqref{eq:spectral_flow} since, for either $SU(2)$ subalgebra, $\omega^I = -\frac{\omega_I}{2}$ for $I=1,2$.

With this definition of spectral flow in mind, we can now state how to apply our methodology to this example. The spectral flow automorphism is readily realised using line defects. The natural generalisation from the $x_i=0$ case for $SL(2,\R)$ is to augment the contour integral factor in \eqref{eq: nucleation 2} or \eqref{eq: unifying j3 exponentials} as
\begin{equation}
    \exp\left(\frac{1}{2\pi i}\oint_{\p \mathcal{U}} \rd z\, J^3(z)\ln R(z)\right) \mapsto \exp\left(\frac{1}{2\pi i}\oint_{\p\mathcal{U}} \rd z \,\sum_{i=1}^n \vec{H}\cdot\vec{\omega}_i\ln\left(\frac{z-z_i}{z-u_i}\right)\right)\,.
\end{equation}
This may be written in a way that is perhaps more useful for calculations as
\begin{equation}
    \exp\left(\frac{1}{2\pi i}\oint_{\p \mathcal{U}}\rd z\,\sum_{I=1}^3 H^I\ln R_I(z) \right)\,,
\end{equation}
where
\begin{equation}
    R_I(z)=\prod_{i=1}^n\left(\frac{z-z_i}{z-u_i}\right)^{\omega_{i,I}}\,.
\end{equation}
In a basis where the Killing form $\kappa^{IJ}$ is diagonal, the three Cartan generators have trivial OPEs with each other and can be treated independently.

Our expectation is that only a subset of the Cartan generators will generate a non-trivial spectral flow automorphism (in the sense of mapping outside of the unflowed Hilbert space), whilst the others will map within the unflowed Hilbert space (as is the case for $SU(2)$). For example, in the case of $SU(2,2)$, we suspect that only the $U(1)$ generator $t^3$ will generate a non-trivial spectral flow, but this is not a point that we have investigated carefully.\footnote{For comparison, the notion of spectral flow introduced in \cite{Gaberdiel:2021jrv} utilises a single spectral flow parameter.}

By analogy with the $x$-basis discussed in the $AdS_3$ context, one could also introduce into the vertex operators a dependence on the spacetime coordinates $x^{\alpha\dot{\alpha}}$ of the boundary theory, where $\alpha,\dot{\alpha}=1,2$ are (anti-)chiral indices in the boundary spacetime. An obvious suggestion would be to use the zero mode of the conjugate momenta $P_{\alpha\dot{\alpha}}$ to define $x$-basis vertex operators
\begin{equation}
    V(x;z)=e^{x\cdot P_0}V(0;z)e^{-x\cdot P_0}\,,
\end{equation}
where $x\cdot P=x^{\alpha\dot{\alpha}}P_{\alpha\dot{\alpha}}$. We anticipate that the role of the operator $e^{xJ^+_0}J^3(z)e^{-xJ^+_0}=J^3(z)-xJ^+(z)$, which played a crucial role in our discussions of the $x$-basis correlators, would be played by $e^{x\cdot P_0}H^Ie^{-x\cdot P_0}$. The generalisation of the operator $J^3(z)-x(z)J^+(z)$, which featured in the $x(z)$-basis discussions, would also follow naturally. Indeed, we feel that much of the construction presented in this paper will have a natural generalisation to WZW models based on other non-compact  groups. Less clear is the generalisation to cosets, such as $AdS_5$. In order to discuss such spaces, one also needs to think about what remains of spectral flow once the quotient has been taken with the $SO(4,1)$ subgroup. With such coset generalisations in mind, the cigar geometry $SL(2,\R)/U(1)$ may provide a nice toy example for future investigation.

\section{Higher genus}\label{appendix:higher_genus}

We focused on genus zero correlators in the main text. Nevertheless, one could also study spectrally-flowed correlators at higher genus from the perspective of manipulating defect lines. We would like to sketch here how the implementation of this approach differs on worldsheets of higher genus. We will refer to the genus as $g$, which is not to be confused with the embedding field $g(z,\zbar)\in G$ of the WZW model.

Let us begin by recalling some useful definitions. We will only introduce the minimal amount of content required here; more extensive reviews can be found in, for instance, \cite{Nakahara:2003nw,Eynard:2018tcr} or specifically in a string theory context in \cite{DHoker:1988pdl,Polchinski:1998rq}.

The homology group $H_1(\Sigma,\Z)$ of a Riemann surface, $\Sigma$, of genus $g\geq 1$ is of dimension $2g$. A convenient basis is given by the $g$ $A$-cycles and $g$ $B$-cycles which we denote by $A_I$ and $B_I$ for $I=1, \dots, g$. These cycles have intersection numbers
\begin{equation}
    A_I\cap A_J = 0\,, \quad B_I\cap B_J = 0\,, \quad A_I\cap B_J = \delta_{IJ}\,.
\end{equation}
In order to construct integrals on $\Sigma$, we need to construct a basis of meromorphic one-forms, otherwise known as Abelian differentials. There exists a unique basis of one-forms, $\mathscr{W}_I$, such that
\begin{equation}
    \oint_{A_I}\mathscr{W}_J = \delta_{IJ}\,.
\end{equation}
The complex structure is then determined by specifying the periods around the $B$-cycles,
\begin{equation}
    \oint_{B_I}\mathscr{W}_J = \Omega_{IJ}\,,
\end{equation}
where $\Omega_{IJ}$ is symmetric and Im$\Omega$ is positive definite. For $g=1$, $\mathscr{W}_1 = \rd z$ and $\Omega_{11}=\tau$. We moreover need an object that plays the role of $(z-w)$ on $\Sigma$. This is known as a prime form, $E(z,w)$, which has a simple zero at $z=w$. The prime form is periodic around the $A$-cycles, while
\begin{equation}\label{monod}
    E(z+B_I,w) = E(z,w)\exp\left(-i\pi\Omega_{II}-2\pi i\int_{w}^z\mathscr{W}_I
\right)\,.
\end{equation}
One can use the prime form to construct the Green's function on $\Sigma$. It is given by
\begin{equation}\label{green}
G(z,w)=-\ln|E(z,w)|^2+2\pi\sum_{I,J} \text{Im} \left(\int_w^z\mathscr{W}_I\right)\big(\text{Im}(\Omega_{IJ})\big)^{-1}\text{Im}\left(\int_w^z\mathscr{W}_J\right)\,.
\end{equation}
$G(z,w)$ is periodic about the cycles of the Riemann surface but does not split naturally into holomorphic and anti-holomorphic sectors, unlike the prime form $E(z,w)$. The Green's function is, however, modular-invariant, whereas $E(z,w)$ is not.

The OPEs of vertex operators at higher genus are generalised to depend on the prime form\footnote{So that, for example, a chiral boson satisfies $\phi(z)\phi(w)\sim -\ln(E(z,w))$.}  $E(z,w)$, rather than $(z-w)$. At short distances, the prime form behaves as $E(z,w)\sim z-w$ so this fits with the expectation that the physics should not be able to distinguish between Riemann surfaces of different genus at short distances. Given this modification to the OPE, the natural generalisation of (\ref{eq: nucleation 2}) is
\begin{equation}\label{eq: nucleation g}
    \begin{aligned}
        V^w_{j,h}(z) = \exp\left(\frac{w}{2\pi i}\oint_{\partial U} \rd z'\, J^3(z')\ln\left(\frac{E(z'
        ,z)}{E(z',u)}\right)\,\right)E(z,u)^{m w} \\
        \times \lim_{\epsilon \rightarrow 0} V_{j,m}(z+\epsilon) \Omega_w(u-\epsilon)\,.
    \end{aligned}
\end{equation}
We can insert this inside a correlation function to generalise the tree-level $m$-basis result (\ref{eq:m_basis_canonical_form})
\begin{equation}\label{genus correlator}
\left\langle\prod_{i=1}^nV_{j_i,h_i}^{w_i}(z_i)\right\rangle=|\mathcal{P}|^2\left\langle\exp\left(\frac{1}{2\pi i}\oint_{\p\mathcal{U}}\rd z\,J^3(z)\ln R(z)\right)\prod_{i=1}^nV_{j_i,m_i}(z_i)\Omega_W(u)\right\rangle,
\end{equation}
to a Riemann surface of genus $g$. At genus zero ${\cal P}$, given by the prefactor in (\ref{eq:m_basis_canonical_form}), is a function of $(z_i-z_j)$ and $(z_i-u)$. At higher genus, a natural proposal is that the general form of (\ref{genus correlator}) still holds, except that ${\cal P}$ is replaced by ${\cal P}_g$, now a function of the appropriate prime forms $E(z_i,z_j)$ and $E(z_i,u)$, instead of $(z_i-z_j)$ and $(z_i-u)$. It is also necessary to generalise $R(z)$ from the genus zero case (\ref{eq:R(z)}) to
\begin{equation}
    R(z)=\prod_{i=1}^n\left(\frac{E(z,z_i)}{E(z,u)}\right)^{w_i}\,,
\end{equation}
where we have taken the coincident limit $u_i\to u$. This is not the only modification that is required at higher genus.

The monodromy (\ref{monod}) along the $B$-cycles means that
\begin{equation}\label{phase}
\frac{R(z+B_K)}{R(z)}=\prod_{i=1}^n\exp\left(2\pi iw_i\int_{u}^{z_i}\mathscr{W}_K\right)\,.
\end{equation}
Note that the right hand side of (\ref{phase}) is independent of $z$. As with genus zero, to compute the correlation function (\ref{genus correlator}), we pull the integral around $\p\mathcal{U}$ off the punctures $z_i$ and $u$ and evaluate it on the identity operator. Since the prime forms are not periodic around the $B$-cycles, we must insert a branch cut around each of the $g$ $A$-cycles. Deforming $\p\mathcal{U}$ off the punctures results in a set of $2g$ circular contours, comprised of two oppositely oriented circles, either side of each of the $g$ $A$-cycle branch cuts.  Evaluating the difference of the contour either side of the branch cut gives the following contribution for each $B$-cycle
\begin{eqnarray}\label{genus contribution}
{\cal S}_K:=\exp\left(\oint_{A_K}\frac{\rd z}{2\pi i}\Big(\ln\big[R(z)\big]J^3(z)- \ln\big[R(z+B_K)\big]J^3(z+B_K)\Big)\right)\nonumber\\
=
\exp\left(-\sum_{i=1}^nw_i\int_{u}^{z_i}\mathscr{W}_K\oint_{A_K}J^3\right),
\end{eqnarray}
where we have used (\ref{phase}) and assumed $J^3$ is periodic around the $B_K$ cycle. The full contribution in the correlation function will include a product of $g$ terms of the form (\ref{genus contribution}), one for each $B$-cycle. The contribution (\ref{genus contribution}) has no analogue at genus zero and we suspect that (\ref{genus contribution}) is necessary to ensure modular invariance of the spectrally-flowed correlation function. For the $m$-basis the correlation function takes the form
\begin{equation}
    \left\langle\prod_{i=1}^nV^{w_i}_{j_i,h_i}(z_i)\right\rangle_g=|{\cal P}_g|^2\left\langle\prod_{K=1}^g{\cal S}_K\prod_{i=1}^nV_{j_i,m_i}(z_i)\Omega_W(u)\right\rangle_g,
\end{equation}
where, as noted above, ${\cal P}_g$ is a rational function of prime forms and ${\cal S}_K$ is given by (\ref{genus contribution}).

\bibliographystyle{JHEP}
\bibliography{AdS3}

\end{document}